\documentclass[11pt,a4paper]{article}
\usepackage[T1]{fontenc}
\usepackage[margin=25mm]{geometry}
\usepackage{authblk}
\usepackage{amsthm}
\theoremstyle{plain}
\newtheorem{theorem}{Theorem}

\newtheorem{corollary}{Corollary}
\newtheorem{proposition}{Proposition}
\theoremstyle{definition}
\newtheorem{definition}{Definition}

\theoremstyle{remark}
\newtheorem{remark}{Remark}
\newtheorem*{remark*}{Remark}
\newenvironment{keywords}{\par\small\noindent\textbf{Keywords: }}{\par}

\usepackage{natbib}
\usepackage{graphicx} 
\usepackage{comment}%
\usepackage{bm}  
\usepackage{amsmath}
\usepackage[french,english]{babel}
\usepackage{url}
\usepackage{multirow} 
\usepackage{amsfonts}
\usepackage{graphicx}
\usepackage{color}
\usepackage[normalem]{ulem}  
\usepackage{tikz}
\usepackage{caption} 
\usepackage{mathrsfs}
\usepackage{dsfont}
\usepackage{enumitem}   
\usepackage{mathrsfs}
\usepackage{dsfont}
\usepackage{bm}
\usepackage{xcolor} 
\usepackage{soul}
\setstcolor{blue} 
\usepackage{float}
\usepackage{rotating}
\usepackage{tikz-cd}
\usepackage{amssymb} 
\usepackage{bbm}
\usepackage{chngcntr}

\allowdisplaybreaks
\setcitestyle{authoryear,open={(},close={)}} 
 
\newcommand{\PP}{\mathbb{P}}
\newcommand{\R}{\mathbb{R}}
\newcommand{\N}{\mathbb{N}}
\newcommand{\E}{\mathbb{E}}

\newcommand{\citea}[1]{\citeauthor{#1}}
\newcommand{\citeay}[1]{\citeauthor{#1}, \citeyear{#1}}

\usepackage{newtxtext}
\usepackage[subscriptcorrection]{newtxmath}

\graphicspath{{./}}

\usepackage[plain,noend]{algorithm2e}

\makeatletter
\renewcommand{\algocf@captiontext}[2]{#1\algocf@typo. \AlCapFnt{}#2} 
\def\@algocf@capt@plain{top}
\renewcommand{\algocf@makecaption}[2]{%
  \addtolength{\hsize}{\algomargin}%
  \sbox\@tempboxa{\algocf@captiontext{#1}{#2}}%
  \ifdim\wd\@tempboxa >\hsize
    \hskip .5\algomargin%
    \parbox[t]{\hsize}{\algocf@captiontext{#1}{#2}}
  \else%
    \global\@minipagefalse%
    \hbox to\hsize{\box\@tempboxa}
  \fi%
  \addtolength{\hsize}{-\algomargin}%
}
\makeatother

\usepackage[hidelinks]{hyperref}
\begin{document}

\title{Analysis of spatial extreme events via excursion set geometry}
\author[1]{Ryan Campbell}
\author[1,2]{Elena Di Bernardino}
\author[3]{Thomas Opitz}
\affil[1]{Laboratoire J.A. Dieudonn\'e, UMR CNRS 7351, Universit\'e C\^ote d'Azur, Parc Valrose, 06108 Nice, Cedex 2, France}
\affil[2]{Institut Universitaire de France}
\affil[3]{Biostatistics and Spatial Processes, INRAE, Avignon, 84914, France}
\affil[ ]{\texttt{ryan.campbell@univ-cotedazur.fr};\quad
\texttt{elena.di\_bernardino@univ-cotedazur.fr}\\
\texttt{thomas.opitz@inrae.fr}}
\renewcommand{\Affilfont}{\small}
\date{}
\maketitle

\begin{abstract} 
     In environmental data, interest often lies in characterising extreme behaviour on a spatial domain. 
     Environmental extremes are often associated with increasingly small and isolated exceedance clusters, leading to degenerate behaviour of standard limits from extreme-value theory. In this work, we investigate rescaling of spatial coordinates as a tool to characterise and estimate joint tail decay rates over spatial areas and reveal non-degenerate limits, leading to a framework that links spatial scaling to the extremal dependence structure of random fields. A \emph{scaling index}, which semi-parametrically quantifies the rate of spatial rescaling, is catalogued for various previously-studied random field models. 
     We propose two different estimation procedures for the scaling index using data over a regular-grid domain, both based on \emph{excursion sets}, defined as regions at which an exceedance of a predefined quantile threshold occurs.
     First,  we consistently estimate this index using the perimeter of excursion sets, allowing for the construction of a hypothesis test on its value, and therefore on the field's extremal dependence behaviour.
     Second, we estimate this index as a quantile regression coefficient with the extremal range, a local statistic that captures the spatial extent of extreme events in random fields, acting as the response variable. 
     Simulations demonstrate how the proposed methods accurately recover the underlying dependence structure via estimation of the scaling index. 
     We apply both approaches to hourly radar reanalyses of precipitation for continental France at 1km resolution, highlighting regional differences and revealing the influence of a seasonal covariate, in an effective and computationally efficient way.   
\end{abstract}


\section{Introduction} 
\subsection{Background on the geometry of excursion sets}
In extreme value analysis, the behaviour of rare and catastrophic events on a spatial domain is often considered for understanding phenomena and assessing risks.
Examples include  environmental processes such as precipitation, where understanding the most extreme events has direct implications on public safety \citep{davison2013geostatistics, HandbookExtremesCh25}.
Statistical modelling to capture the full complexity of spatial processes is challenging, particularly for large study domains with a high resolution of data locations.
Therefore, to extrapolate into the tails of the underlying spatial process, assumptions are often made on limits of the random field's extremal dependence. In doing so, inherent bias may be carried forward when estimating, for example, the occurrence probability of extreme events in a spatial domain.
Recently, the alternative approach of studying the geometry of \emph{excursion sets}, locations in the spatial domain at which values exceed a predefined threshold, has gained popularity \citep{di2024spatial, cotsakis2024spatial, Costakis2025Extent}.
In studying this geometry, it has been shown that one can infer the behaviour of the underlying random field in a computationally-efficient manner without making strong  assumptions on the form of its extremal dependence structure.

\begin{figure}[H]
    \includegraphics[scale=0.324, trim={30 0 30 0},clip]{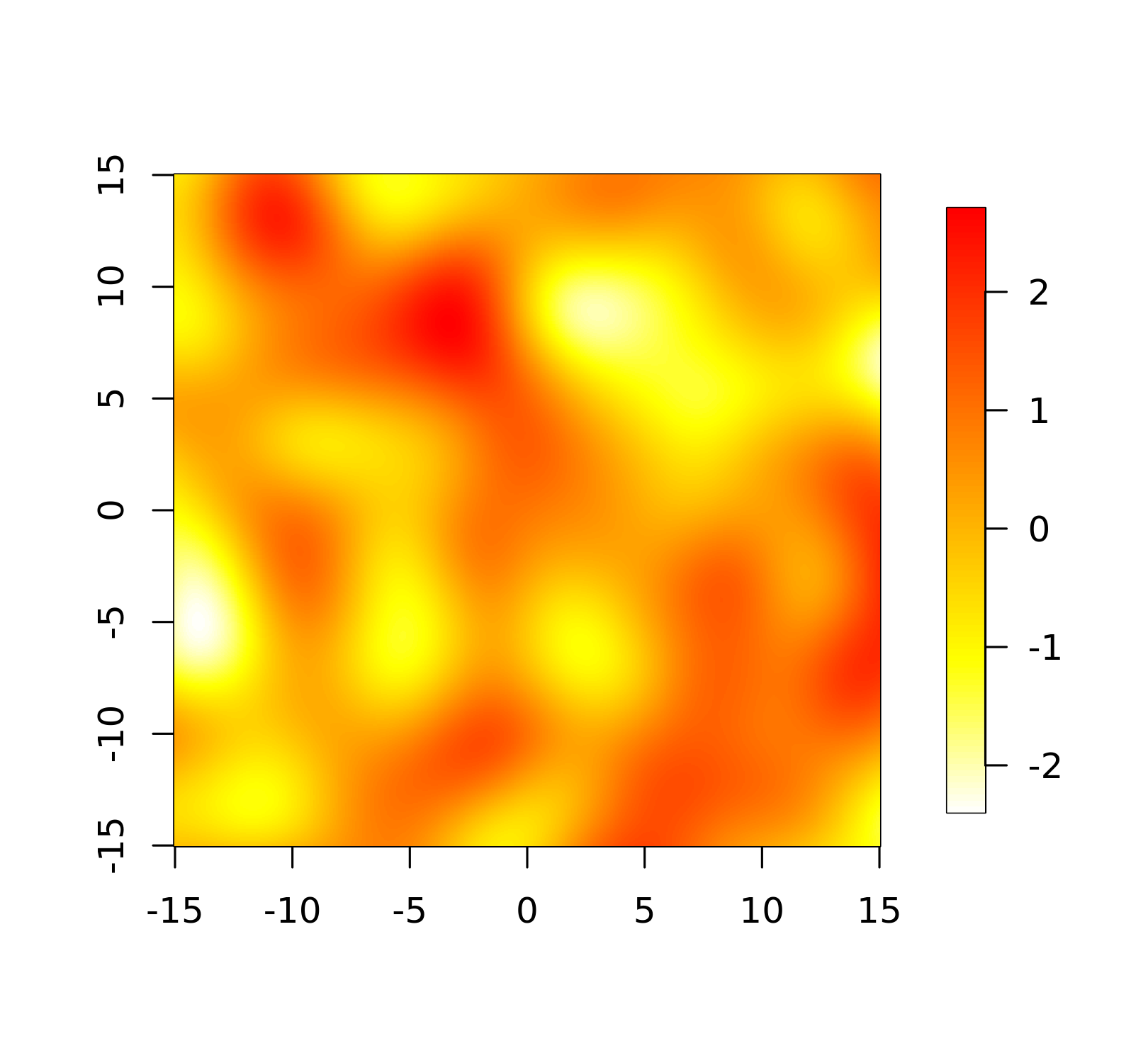}
    \includegraphics[scale=0.324,  trim={30 0 30 0},clip]{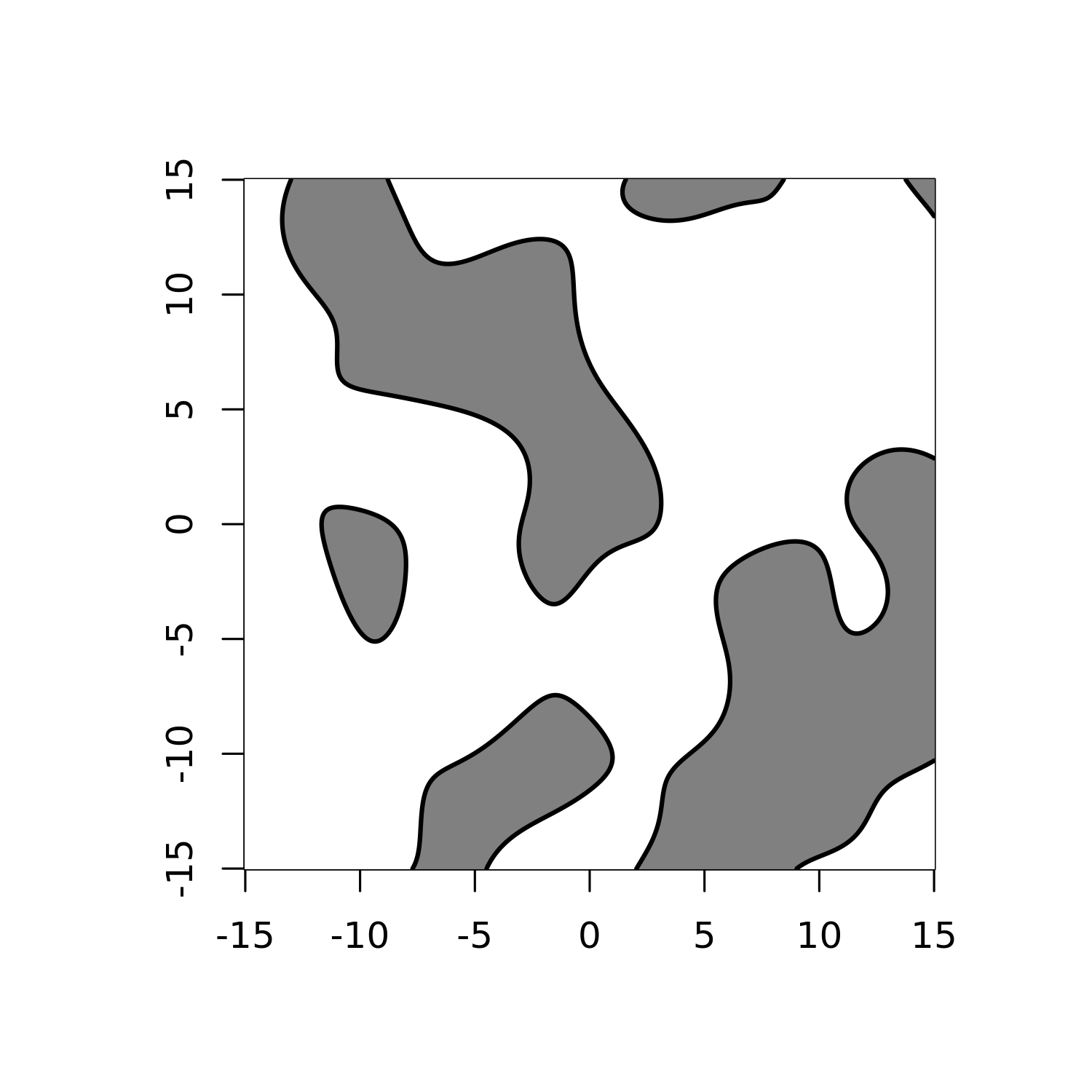}
    \includegraphics[scale=0.324,  trim={30 0 30 0},clip]{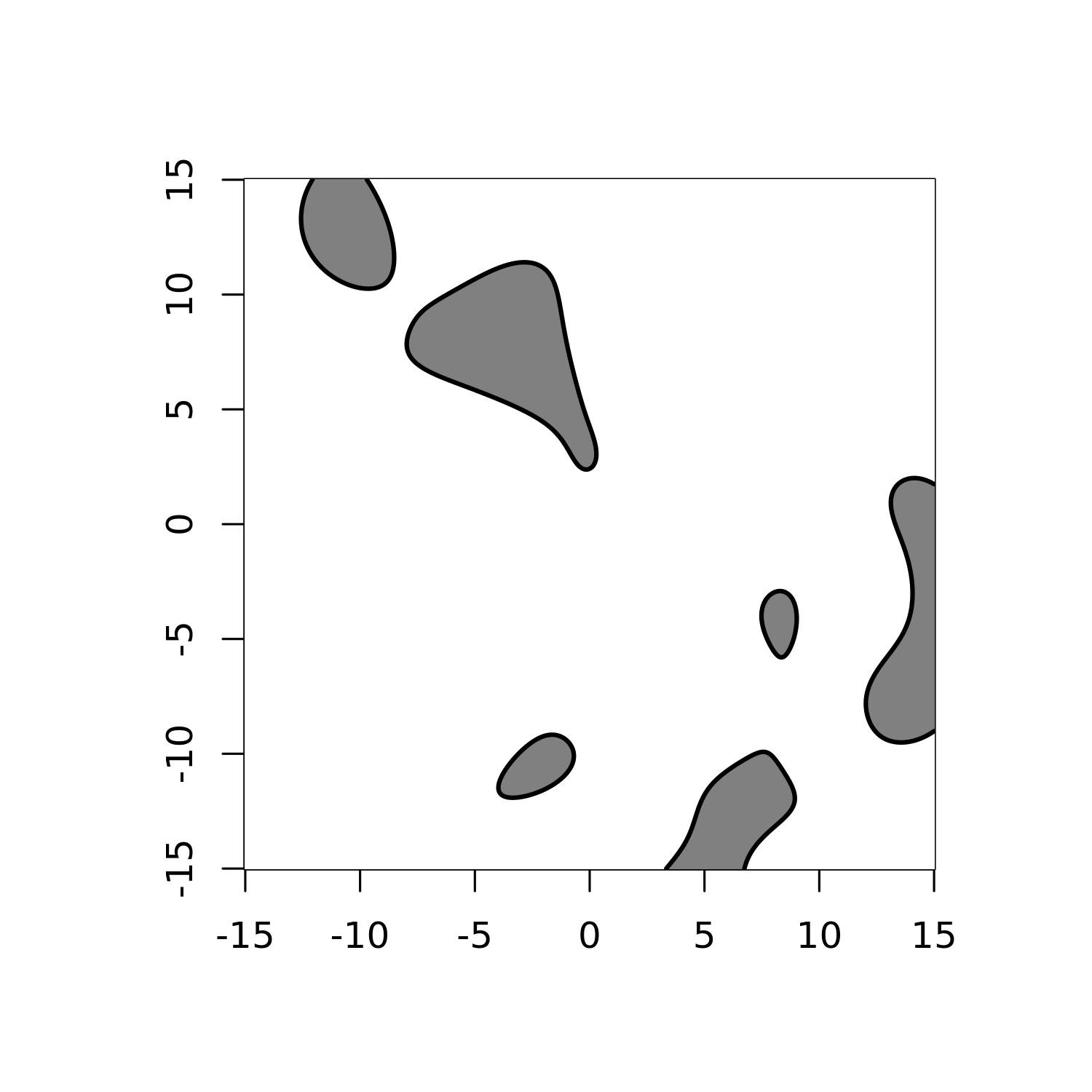}
    \includegraphics[scale=0.324, trim={30 0 30 0},clip]{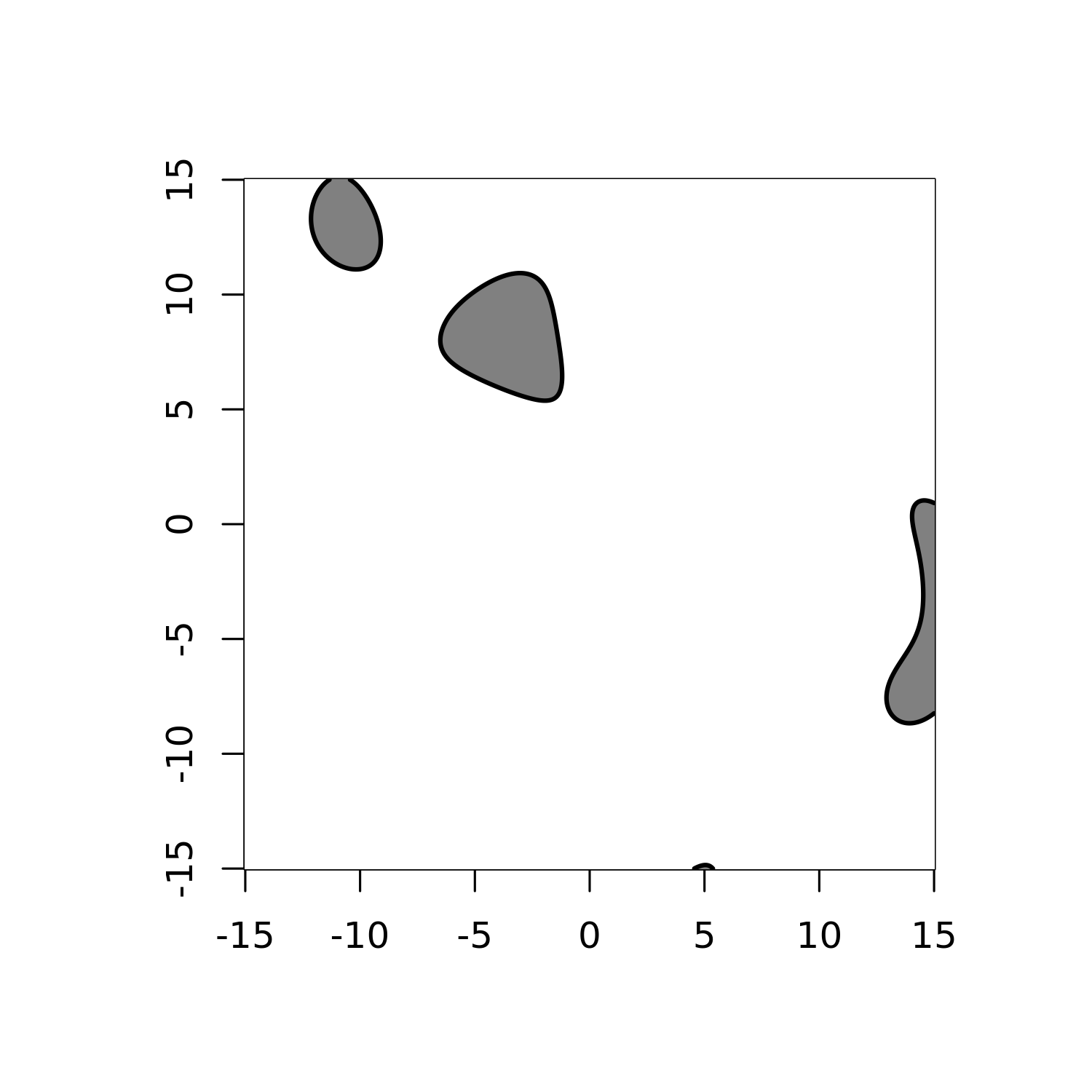}
    \vspace{-0.5cm}\caption{Left: An observation of a Gaussian random field possessing squared-exponential covariance function with scale parameter $\sigma^2=1$ and range parameter $l=3$ (see Equation \eqref{eq:sqexpcov}) on the domain $S=[-15,15]^2$. Centre-left, centre-right, and right: the corresponding excursion sets $E^X_p\cap  S$ (grey) and boundary $(\partial E^X_p)\cap S$ (black lines) for $p=0.70$, $0.90$, and $0.95$, respectively.}
    \label{fig:gaussEu_intro_illustration}
\end{figure}

Suppose $S \subset \R^2$ is a compact spatial domain. Let  $(\Omega, \mathcal{F}, \PP)$ be a probability space, and   $X : \Omega \times \mathbb{R}^2 \to \mathbb{R}$ 
  a random field.
Theoretical results are derived in this work by assuming stationarity of $X$ on $\mathbb{R}^2$ if not explicitly stated otherwise, with common continuous marginal distribution function denoted by $F$ (see Assumption \ref{A1} below). 
We define excursion sets through
\begin{equation}
    \label{eq:Eu-v3}
    E^X_p = \{{s}\in \mathbb{R}^2\,:\, X({s})\geq u_p\},
\end{equation}
where  the threshold is defined by the quantile $\smash{u_p=F^{-1}(p)}$ for some probability level $p\in(0,1)$ and   $F^{-1}$ the univariate quantile function of $X$.    Figure~\ref{fig:gaussEu_intro_illustration} demonstrates the evolution of $E^X_p$ in terms of  $p$ for a particular observation of a random field $X$.

Many theoretical results have been obtained for excursions of random fields, especially  Gaussian processes,  and include expected values of the so-called \emph{Lipschitz–Killing curvatures}
(LKCs), such as the area, the half-perimeter, and the Euler-Poincaré characteristic (equal to the number of connected components minus the number of holes, i.e., of connected components of the complement set) in two-dimensional Euclidean space. 
These scalar quantities offer convenient summary statistics of the geometric properties of $E^X_p$ in \eqref{eq:Eu-v3}. Studying properties of LKCs of excursion sets allows inference of behaviour of the underlying random field $X$. For example, for  Gaussian random
fields, the Euler-Poincaré characteristic is studied in  \cite{EL16} and \cite{del17},  the perimeter in \cite{berzin2021} and \cite{10.1214/19-AAP1508},  and the excursion volume is the subject of \cite{bulinski2012central} and \cite{pham}. \cite{Mu16} and \cite{Kratz2018}
jointly analyse LKCs, and  \cite{MESCHENMOSER,meschenmoser2013,shashkin2013}  establish
functional central limit theorems (CLTs) of LKCs.

To ensure regularity of the excursion sets and their boundaries, we use the following assumption:
\begin{enumerate}[label={\textcolor{blue}{($\mathcal{A}$0)}}]
\item\label{A1}   The random field $X:\Omega\times\mathbb R^2\to\mathbb R$
is stationary, with almost sure $C^2$ sample paths and
a continuous marginal distribution function $F$.
Furthermore, the gradient $\nabla X(s)$ and Hessian
$\nabla^2X(s)$ have non-degenerate distributions. For every fixed $p\in(0,1)$, the level $u_p$ 
is almost surely a regular value of $X$, that is,
$\mathbb P\!\left(
\exists s\in\mathbb R^2:
X(s)=u_p,\ \nabla X(s)=0
\right)=~0.$
\end{enumerate}
The 
regular value condition above ensures, by the implicit function theorem, that the excursion boundary is locally a $C^2$ curve, allowing the use of local tube formulas that underpin the calculations in the remainder of the paper. We  now explicitly define the half-perimeter and the area of the intersection $E^X_{p}\cap S$, which will be the two LKCs relevant in this work.

\begin{definition}[Excursion set half-perimeter and area] \label{def:LK}
Let $X$ be as in Assumption~\ref{A1}, and let $\partial S$ denote the boundary of $S$. Define,  for $p\in(0,1)$, the half perimeter $\mathcal{L}_{1}$ and the excursion area $\mathcal{L}_{2}$
of the excursion set $E^X_{p}$ as, respectively,
\begin{align}
\mathcal{L}_{1}(E^X_p, S) : &=  \frac{1}{2}   \mathcal{H}_1(S \cap \partial E^X_p)=\,\frac{1}{2} \lim_{\epsilon \to 0}\, \frac{1}{2\, \epsilon}\int_S \mathbbm{1}_{(u_p-\epsilon,u_p+\epsilon)}(X(s)) \left\|\nabla X(s)\right\|_2 \, ds\label{eq:UestCT1bis}  
\\
\mathcal{L}_{2}(E^X_p, S) : &=    \int_{S} \mathbbm{1}_{E_p^X}(s)\, ds,\label{eq:UestCT2}    
 \end{align}
where $\mathcal{H}_1$  is  the 1-dimensional Hausdorff measure, and $\mathbbm{1}_A(x)=1$ if $x\in A$, and 0 otherwise.
\end{definition}
\noindent 
The quantity $\mathcal{H}_1(S \cap \partial E^X_p)$ in~\eqref{eq:UestCT1bis} represents the perimeter of the excursion set $E^X_p$ within $S$, excluding the boundary of $S$ \citep[the solid black lines in Figure~\ref{fig:gaussEu_intro_illustration}; see][]{cotsakis2022perimeter,cotsakisAAP}. \vspace{-0.4cm}

\vspace{-0.4cm}

Throughout this work, we   consider increasing-domain asymptotics by assuming that compact spatial domains $S\subset\mathbb{R}^2$  tend to infinity in the Van Hove sense,
i.e., ${\mathcal{L}_1(S)}/{\mathcal{L}_2(S)}\rightarrow0$ as $S\nearrow\mathbb{R}^2$, the limit along any sequence of subsets $S$ of $\mathbb{R}^2$ growing to $\mathbb{R}^2$. This assumption is common (e.g., \citeay{bulinski2012central}), and is shortly referred to as VH-growing.   It is  useful to define the  normalised area and perimeter   and their associated densities.
\begin{definition}[Normalised half-perimeter  and area densities]\label{LKCdensities}
Let $E^X_p$ be the excursion set in \eqref{eq:Eu-v3} obtained from a random field $X$ satisfying Assumption~\ref{A1}. Define the normalized half-perimeter and area by
$C_i^{S}(E^X_p):=  \mathcal{L}_i(E^X_p, S)/{\mathcal{L}_2(S)},$ for $i=1,2,$ 
and, assuming they are finite, the associated  densities as 
\begin{align} \label{eq:defC1ast}
C^{\ast}_i(E^X_p) :=\lim_{S\nearrow\mathbb{R}^2} \E[C^{S}_i(E^X_p)],~\mbox{ for } i=1,2,
\end{align}  
where expected values are taken over the random field $X$.
\end{definition}

Due to Assumption~\ref{A1}, one can write without loss of generality
 $C^{\ast}_2(E^X_p)=\PP(X(0) \ge u_p)= 1-p$. Hence, no artificial contribution from $\partial S$ is added in Equation~\eqref{eq:UestCT1bis} and, by stationarity and the kinematic tube formulas, the expected half-perimeter is proportional to $\mathcal L_2(S)$, so that $C_1^*(E_p^X)$ is independent of $S$; see  \cite{cotsakisAAP}. LKCs and their densities have recently been used to provide local excursion set coefficients, which offer a parsimonious and informative summary of the behaviour of  $E^X_p$ around a \emph{reference location} $s_0$. Given a reference location ${s}_0$, we consider \emph{conditional excursion sets}:
   \begin{equation} \label{eq:Eu-cond}
    E^X_p ({s}_0)     :=E_p^X\mid\{s_0\in E_p^X\}.
\end{equation}   \cite{cotsakis2024spatial, Costakis2025Extent} formalise  the \emph{lower extremal range} (LER) coefficient as follows.

\begin{definition}[Lower extremal range (LER)]\label{def:LER}
Let $X$ be as in Assumption~\ref{A1} with the excursion set $E^X_p$  as in \eqref{eq:Eu-v3}, and let $s_0 \in E^X_p$. The lower extremal range (LER)  at $s_0$ is defined as the random variable
\begin{equation*} 
R^{(p)}_{s_0}
:=
\operatorname{dist}\!\left(
s_0,{\bigl(E_p^X(s_0)\bigr)^c}
\right)
=
\inf_{s\in\mathbb{R}^2\setminus E_p^X(s_0)}
\|s-s_0\|,
\end{equation*} 
i.e., the distance from $s_0$ to the closure of the complement of the conditional excursion set.
\end{definition} 
Though the complement $(E_p^X(s_0))^c$ is open, we bypass writing its closure in the distance definition of $R^{(p)}_{s_0}$. This is possible  under the well-behavedness of boundaries of excursion sets in our assumption~\ref{A1}, and we use this convention throughout the manuscript regarding complements of excursion sets.

The LER has an interesting connection to the half-perimeter and area densities from Definition~\ref{LKCdensities}. Indeed, 
\cite{cotsakis2024spatial} showed that 
\begin{equation}\label{eq:ler-limit-cdf}
    \PP\left(R^{(p)}_{s_0}\leq r\right)\sim 2\,\frac{C_{1}^\ast(E^X_p)}{C_{2}^\ast(E^X_p)}\,r, \quad \mbox {as } r\rightarrow0. 
\end{equation}
The LER in Definition \ref{def:LER} is a local variable, depending on a specific location $s_0$ exceeding a threshold. 
However, thanks to the stationarity Assumption  \ref{A1}, the asymptotic equivalence \eqref{eq:ler-limit-cdf} can be formulated with unconditional excursion sets through their half-perimeter and area densities (see Theorem 1 in \cite{cotsakis2024spatial}). 
An advantage of using conditional excursion sets is that the pairwise \emph{tail correlation coefficient} (see, \emph{e.g.}, \citeay{Coles99}, \citeay{beirlant2004book}, \citeay{davis2009extremogram}, and \citeay{strokorb2015tailcorrelation}) can be written in terms of the conditional excursion set:
\begin{align}\label{eq:chi-def}
\chi_p(s_0,s) = &  \, \PP\left(X(s)\geq u_p\middle|X(s_0)\geq u_p\right) =  \PP\left(s\in E^X_p(s_0)\right). 
\end{align}
In this work, we say that a random field is \emph{asymptotically dependent}
if $\chi_p(s_0,s)\rightarrow \chi(s_0,s)\in(0,1]$ for all $s\neq s_0$ such that $\left\|s-s_0\right\|<\infty$, as $p\rightarrow1$, and \emph{asymptotically independent} otherwise.

\subsection{Domain scaling for non-degeneracy}
 Complications arise from both mathematical and numerical points of view when examining  $E^X_p$ as  $p \to 1$, since it converges in probability to the empty set, provided that the supremum of $X$ is finite almost surely.  Rather than focusing on  $E^X_p$ in  \eqref{eq:Eu-v3}, we consider a scaled excursion set, defined by
\begin{equation}
    \label{eq:Eu-scaled}
    \tilde{E}^X_p = \left\{{s}\in \R^2 \,:\, X\left(\frac{{s}}{\sigma(p)}\right)\geq u_p \right\},
\end{equation}
and the scaled conditional excursion set  
\begin{equation*}
    \label{eq:Eu-cond-scaled}
    \tilde{E}^X_p ({s}_0)  
    :=\tilde{E}^X_p\mid\{s_0\in \tilde{E}^X_p\}
\end{equation*}
for some \emph{scaling function} $\sigma:[0,1]\rightarrow (0,\infty)$. Notice the relation between non-scaled and scaled excursion sets, $E^X_p={\tilde{E}^X_p}/{\sigma_p}$. The existence of a scaling function to ensure a non-degenerate limit set is formulated in the following assumption:
\begin{enumerate}[label={\textcolor{blue}{($\mathcal{A}$1)}}]
\item\label{A:scaling}
There exists a scaling function $\sigma:(0,1)\to(0,\infty)$
such that
$(\tilde E_p^X(s_0),
{(\tilde E_p^X(s_0))^c})$ 
converges in distribution, as $p\to1$, in the product Fell topology to
$(\tilde E^X(s_0),
{(\tilde E^X(s_0))^c})$,  
 where  $\tilde E^X(s_0)$ is a non-degenerate random closed set.
\end{enumerate}

A discussion on the topology for convergence in distribution in  Assumption \ref{A:scaling} is postponed to Remark \ref{topology}   of the  Supplementary
Material. 
We aim to study the properties of $\sigma(p)$, such that the scaled conditional excursion set will be non-degenerate for $p \to 1$.
Firstly, notice that stationary regularly varying random fields displaying asymptotic dependence \citep{HandbookExtremesCh16}  will allow choosing the scaling function as a constant $\sigma$. 
On the other hand, non degenerate stationary Gaussian random fields, displaying asymptotic independence, correspond to a scaling $\sigma(p)$ that increases with $p$. The intuition behind this is that 
$\chi_p\rightarrow0$ as $p\rightarrow1$ \citep{heffernan2000directory}.
This asymptotic degeneracy is illustrated in Figure~\ref{fig:GRF-ex} (in Appendix~\ref{supp:calculations} of Supplementary Material), where  $E^X_p(s)$ becomes arbitrarily small as $p \to 1$, and spatial scaling is needed to \emph{zoom in} on the random field to observe  non-degenerate behaviour in $\tilde E^X_p(s_0)$. In Assumption \ref{A:scaling}, in  addition to ensuring the convergence of rescaled excursion sets, we also ensure the convergence of their closed complements. This is to guarantee the well-behavedness of the LER of rescaled excursion sets as $p\to 1$ (see Definition~\ref{def:LER}) and proof of Theorem \ref{thm:LKCratio}).

Assumption \ref{A:scaling} is akin to previously-introduced asymptotic scaling assumptions, such as that of regular variation (see, for example, Chapter 6 of \citeay{resnick2007htp}, and the extensions to the functional setting in \citeay{hult2005extremal} and \citeay{wu2020regularly}). The key difference is that, rather than scaling the values of the random field, we apply scaling to the locations at which they are observed.  Furthermore, unlike functional regular variation, Assumption \ref{A:scaling} does not lead to asymptotic dependence between locations. Both asymptotically dependent and independent random fields can have Assumption \ref{A:scaling}, depending on the choice of scaling function $\sigma$. If the stationarity in Assumption \ref{A1} holds, then \ref{A:scaling} is an assumption on the entire random field (i.e., is independent of conditioning location $s_0$).  
 
\subsection{Structure of the paper} 
In Section \ref{sec:theory}, we analyse the behaviour of the scaling function $\sigma$ in \ref{A:scaling} for a large class of random fields. Furthermore,  we identify a regular variation index for the scaling function $\sigma$, and show that it can be consistently estimated and subsequently used to test for asymptotic spatial dependence (see Section \ref{sec:inference}). We then present simulations to analyse the numerical performance of our inference and testing procedures, followed by an application to  precipitation reanalysis data from France (Section \ref{sec:realdataSection}). Some future extensions are presented in Section \ref{ConclusionSection}.  
\vspace{-0.2cm}
\section{Extremal spatial dependence analysis  via excursion sets}\label{sec:theory}

\subsection{The scaling function and associated scaling  index}

Given a stationary random field, it is possible to use the distribution function approximation of $R_{s_0}^{(p)}$ in \eqref{eq:ler-limit-cdf}, involving the half-perimeter and area densities, to obtain the appropriate scaling function $\sigma$.  In Theorem~\ref{thm:LKCratio}, we show that the half-perimeter and area density ratio identifies the asymptotic order of the spatial scaling $\sigma$ appearing in Assumption~\ref{A:scaling}.

\begin{theorem}\label{thm:LKCratio} 
Suppose that Assumptions \ref{A1} and \ref{A:scaling} hold,  and there exists $p_0\in(0,1)$ such that for every $p\in(p_0,1)$  the rescaled local extremal radius
$\tilde R_{s_0}^{(p)}
=
\sigma(p)R_{s_0}^{(p)}
$ 
admits a density $f_{\tilde R_{s_0}^{(p)}}$ in a neighbourhood of
zero satisfying 
\begin{equation}\label{densityLER}
    \lim_{\delta\downarrow0}
\limsup_{p\to1}
\sup_{0\le r\le\delta}
\left|
f_{\tilde R_{s_0}^{(p)}}(r)
-
f_{\tilde R_{s_0}^{(p)}}(0)
\right|
=0,
\end{equation}
 
i.e., the densities \(f_{\tilde R_{s_0}^{(p)}}\) are uniformly equicontinuous at zero for $p \to 1$. 
Suppose also that
$\tilde R_{s_0} = \textup{dist}(
s_0,{\big(\tilde E^X(s_0)\big)^c})$ 
admits a density $f_{\tilde R_{s_0}}$ that is continuous at zero
and satisfies
$0<f_{\tilde R_{s_0}}(0)<\infty.$
Then, as $p\to1$,
\begin{equation}\label{eq:LKC-ratio-assumption}
\frac{C_1^\ast(E_p^X)}{C_2^\ast(E_p^X)}\sim
c_X \,\sigma(p), 
\end{equation}
where $c_X=\frac12 f_{\tilde R_{s_0}}(0)\in(0,\infty).
$
In particular, \eqref{eq:LKC-ratio-assumption} characterizes, up to a multiplicative constant, the scaling function $\sigma$ appearing in Assumption~\ref{A:scaling} through the ratio of the expected Lipschitz--Killing curvature densities of the excursion set.
\end{theorem} 
The proof of Theorem  \ref{thm:LKCratio}  is postponed to Section \ref{ProofsSection} of  Supplementary Material. Condition~\eqref{densityLER} ensures that the densities   remain close to their
respective values at zero in a neighbourhood whose size can be chosen
uniformly for $p$ sufficiently close to $1$. Numerical diagnostics for assessing this condition are provided in the Supplementary Material \ref{supp:assumption-check}.

\begin{remark}\label{rem:thm1}
The existence of a scaling function yielding a non-degenerate limit
for the rescaled excursion set is an important property.
\cite{Costakis2025Extent} use this spatial scaling principle to establish
the existence of summary statistics of excursion sets as $p\rightarrow1$.
\cite{strokorb2026domain} use a related spatial scaling principle to
define a new framework of regular variation, called
\emph{domain-scaled regular variation}, which retains the theoretical
conveniences of regular variation while accommodating the common
setting in which the unscaled excursion sets vanish as $p\to1$.\end{remark}

Assuming a semiparametric condition on the perimeter density of excursion sets allows for a useful tail characterisation of the scaling function, as is explained in the following corollary.

\begin{corollary}\label{cor:scaling}
Suppose that $X$ satisfies the conditions of
Theorem~\ref{thm:LKCratio}. Assume moreover that, for some
$\beta\in\mathbb{R}$,
\begin{equation}\label{eq:C1form}
C_1^*(E_p^X)
\sim
c_X(1-p)\{-\ln(1-p)\}^{\beta},
\qquad p\to1,
\end{equation}
where $c_X$ is the constant appearing in
Theorem~\ref{thm:LKCratio}. Then
$\sigma(p) \sim \{-\ln(1-p)\}^{\beta}$, $p\to1$.
In particular, $x\mapsto\sigma(1-e^{-x})$ is regularly varying
at infinity with index $\beta$. We call $\beta$ the \textit{scaling index} associated with the spatial rescaling in
Assumption~\ref{A:scaling}.
\end{corollary}

This regular-variation property of $\sigma(1-e^{-x})$ allows us to
study the extremal dependence structure of random fields through the
index $\beta$.  Namely, if $\sigma(1-e^{-x})$ is regularly varying with index $\beta$, the index $\beta$   can be used to describe the evolution of $\sigma$, and therefore the evolution of excursion sets.  
Recalling the $R_{s_0}^{(p)}$ from Definition~\ref{def:LER}, $\beta$ describes the rate of change of the LER at an arbitrary fixed location $s_0$ as $p$ evolves under Assumption \ref{A:scaling}. This can be seen by taking the ratio of the LER distribution function at two high thresholds $p_1$ and $p_2$,
\begin{align}\label{eq:LER-ratio}
    \frac{\PP\left(R_{s_0}^{(p_1)}\leq  r\right)}{\PP\left(R_{s_0}^{(p_2)}\leq r\right)}=\left[\frac{\ln(1-p_1)}{\ln(1-p_2)}\right]^{\beta}\left[1+\epsilon(p_1,p_2,r)\right],
\end{align} 
where $\epsilon(p_1,p_2,r)\to 0$ under a specified joint limiting regime for values of $p_1,p_2$ approaching 1 and $r$ approaching 0 (see Corollary~1 of \citeay{cotsakis2024spatial}). The value of $\beta$  is intrinsically linked to the asymptotic dependence structure of the underlying random field.
This is shown in the following proposition, whose proof is postponed to Section~\ref{ProofsSection}  of  Supplementary Material,  where $\sigma(p)\rightarrow\infty$ as $p\to 1$ (i.e., $\beta>0$) is shown to occur with asymptotic independence. 
\begin{proposition}\label{prop:scaling-AI}
Suppose that $X$ satisfies Assumptions~\ref{A1}
and~\ref{A:scaling}, and fix $s_0\in\mathbb{R}^2$.
Further assume that, for every compact set
$K\subset\mathbb{R}^2$ containing $s_0$ and every
$\epsilon>0$, there exists a compact set
$A\subset\mathbb{R}^2$ such that
\begin{equation}\label{ConfinementCondition}
\PP\left(
\sigma(p)\bigl((E_p^X(s_0)\cap K)-s_0\bigr)
\subset A
\right)\geq 1-\epsilon,
\end{equation} 
for all $p$ sufficiently close to $1$.
If $\sigma(p)\to\infty$ as $p\to1$, then
\[
\lim_{p\to1}\chi_p(s_0,s)=0,
\qquad
s\in\mathbb{R}^2\setminus\{s_0\}.
\]
By stationarity, this conclusion holds for every pair
of distinct locations in $\mathbb{R}^2$.
\end{proposition} Condition  \eqref{ConfinementCondition}   requires that, within every fixed compact neighbourhood of $s_0$, the entire conditional excursion set be contained, with arbitrarily high probability, in a ball centred at $s_0$ whose radius is of order $\sigma(p)^{-1}$. The containing compact set $A$ may depend on $K$ and $\epsilon$, but not on $p$. This confinement condition holds for instance for smooth Gaussian fields with nondegenerate gradient covariance and no perfect correlations at distinct locations, as well as for their strictly increasing marginal transformations (see, e.g., fields in Section~\ref{sec:examples}).   Numerical diagnostics for assessing condition~\eqref{ConfinementCondition} 
are provided in the Supplementary Material \ref{supp:assumption-check}.     Further note that if $\lim_{p\rightarrow1}\sigma(p)=c\in[1,\infty)$, then it is not guaranteed that $\lim_{p\rightarrow1}\chi_p(s_0,s) >0$.
  Indeed, finite limiting scaling suggests threshold stability in the random field, and it is possible to construct fields where we can find points that are always outside the limiting conditional excursion sets ($\lim_{p\rightarrow1}\chi_p(s_0,s)=0$).
    Examples of such random fields are given in Section \ref{supp:beta-dependence} of Supplementary Material.    Proposition~\ref{prop:scaling-AI} is closely related to Proposition~3.5 of \cite{strokorb2026domain}, where an analogous implication is
established under the stronger framework of domain-scaled regular
variation.\smallskip

Proposition~\ref{prop:scaling-AI} suggests that values of $\beta$ can give intuition on the limiting behaviour of random fields.
Indeed, suppose $\beta>0$, then the scaling $\sigma(p)$ grows with $p \to 1$. For an arbitrary location $s_0\in S$, this results in a zooming-in behaviour when applying the domain scaling in order to observe non-degeneracy of the scaled conditional excursion set $\tilde{E}^X_p(s_0)$.
This implies that excursion sets $E^X_p$ get smaller as $p\rightarrow1$, linked to a random field displaying more localised extremes as the threshold level $p$ increases (i.e., asymptotic independence). Under the asymptotic representation \eqref{eq:C1form}, a second possibility is \(\beta=0\), yielding \(\sigma(p)\sim1\). This indicates some form of threshold stability on the random field. Lastly, when $\beta<0$, the scaling function $\sigma(p)$ converges to $0$ as $p\rightarrow1$. Under the assumptions of Theorem~\ref{thm:LKCratio}, this corresponds to more extensive excursion sets, leading to the conclusion that the underlying random field displays increasingly widespread extremes as $p$ increases.  A natural question is how marginal transformations affect the scaling index $\beta$. In Section \ref{supp:calculations} in  Supplementary Material, we clarify, in a tractable Gaussian setting, that although $\beta$ is ultimately derived from the marginal quantile function, it entirely depends on the joint behaviour of the random field. \cite{Costakis2025Extent} distinguish between Gaussian
and regularly varying random fields through different values of
$\beta$. In next section, we extend this analysis by deriving the
constant $c_X$ and the index $\beta$ for a wide class of random fields.

\subsection{Examples where the asymptotic behaviour of the perimeter is as in Equation~\eqref{eq:C1form}}
\label{sec:examples}
Motivated by Theorem~\ref{thm:LKCratio}, we derive the form of the constant $c_X$ and the scaling function $\sigma$ for a wide variety of stationary random fields on a spatial domain in $\mathbb{R}^2$. Classes of considered random fields  are based on common choices in the literature, each satisfying Assumption~\ref{A1} and displaying different extremal dependence properties, appropriate for different modelling contexts. We also derive the scaling index $\beta$ in \eqref{eq:C1form}.
Lastly, note that changing a random field's covariance structure to affect the degree of asymptotic (in)dependence does not affect the scaling index $\beta$. We further provide the theoretical value of $c_X$ when available.

Henceforth, let $W=\{W(s)\}_{s\in S}$ be a stationary,
isotropic Gaussian random field, and let $W_1,\ldots,W_k$
be $k$ independent copies of $W$. We consider several classes of random fields:
Gaussian fields, $X(\cdot)=W(\cdot)$;
Gaussian-type random fields of the form
$X(\cdot)=g(W_1(\cdot),\ldots,W_k(\cdot))$,
where $g$ is a strictly monotone function;
location or scale mixtures $X(\cdot)=g(\Lambda,W(\cdot))$, where $\Lambda$
is a random variable independent of $W$;
and  max-stable random fields. 
 In Supplementary Material~\ref{supp:shot-noise}, we consider  shot noise random fields. The mathematical expressions are derived by using the Gaussian Kinematic Formula which  provides the mean LKCs of excursion sets of $X$ within a domain  $S$ (see, \textit{e.g.}, Theorem 15.9.5 in \cite{AT07} or Theorem 4.8.1 in \cite{AT11}).  Detailed calculations to obtain results below can be found in Section \ref{supp:calculations} in  Supplementary Material. 
\smallskip

\textbf{Gaussian-type random fields.} \textit{Stationary Gaussian random fields:}
Let $X$ be  a stationary isotropic Gaussian random field with  second spectral moment  $\lambda_2$. 
Then, for $p \to 1$, 
\begin{align*}
   C_1^\ast(E^X_p) =& c_X\, \varphi(u_p) \sim   c_X(1-p)(-\ln(1-p))^{1/2}, 
\end{align*} where $\varphi$  is  the univariate standard normal density and 
$c_X=\sqrt{\pi\lambda_2}/2$.  The scaling index in \eqref{eq:C1form} is therefore given by $\beta=1/2$. The same $\beta=1/2$ is obtained  for  affine anisotropic stationary Gaussian random field (see e.g., \citeay{cabana1987}),  $X(s)=W(As)$, $\forall s$, where $A$ is the affinity matrix   (see Section \ref{supp:calculations} in Supplementary Material).  
\smallskip

\textit{Standardised chi-squared random fields:} 
Let $X(s)= (2\kappa)^{-1/2}(\sum_{j=1}^\kappa W_j(s)^2-\kappa)$, for $s \in \R^2$ and  $\kappa \in \N^*$. Then, by using  Section 2.2.2 in 
 \cite{bierme2018lipschitz} and the form of marginal distribution function  for $p\rightarrow1$ (see   \citeauthor{NIST:DLMF}, Equation 8.11.2) we get,    $
    C_1^\ast(E_p^X) \sim c_X (1-p)\left[-\ln(1-p)\right]^{1/2}
$ with 
$c_X = \frac{\sqrt{\pi\lambda_2}}{2}$  
and  $\beta=1/2$.  \smallskip
   
\textbf{Random fields of the form \boldmath{$X(\cdot)=g(\Lambda,W(\cdot))$}.} 
Suppose   now   $X(s)=g(\Lambda,W(s))$, with     Gaussian random field $W$ and  an independent random variable $\Lambda$.  The following paragraphs are devoted to specific cases of the $\Lambda$  distribution and link function $g$, in which the calculations can be carried out explicitly. Theorem~\ref{thm:gauss-scale-mixture-beta} in Supplementary Material~\ref{supp:scale-mixtures} provides explicit expressions for the scaling function $\sigma$ and the scaling index $\beta$ for Gaussian scale mixtures $X(\cdot)=\Lambda W(\cdot)$, distinguishing between two different asymptotic regimes of the mixing variable $\Lambda$.
\smallskip
 
\textit{Laplace random fields \citep{Opitz.2016}:}  
Let $X(\cdot)=\sqrt{\Lambda}W(\cdot)$, with $\Lambda$ an exponential random variable with rate parameter  $1/2$.  
The limiting half-perimeter and area density ratio, as $p\rightarrow1$, is given by
$\frac{C_1^\ast (E^X_p)}{C_2^\ast (E^X_p)} \sim  c_X \sigma(p)$ as in \eqref{eq:LKC-ratio-assumption} with $c_X=\frac{1}{2}\sqrt{\frac{\pi\lambda_2}{2}}$ and $\sigma(p)=u_p^{1/2}$. Furthermore, we can prove that 
$C_1^\ast(E^X_p) =  c_X u_p^{1/2}e^{-u_p}\sim c_X(1-p)(-\ln(1-p))^{1/2}$, and so we have Equation \eqref{eq:C1form} with  scaling index  $\beta={1}/{2}$.\smallskip

\textit{Elliptical Student random fields \citep{Ma2013}:} 
Consider the stochastic representation of elliptical Student random fields,
$
    X(\cdot)=\Lambda W(\cdot)
$,
where $\Lambda\overset{\text{d}}{=}\sqrt{{k}/{\chi^2_k}}$.
Examining the half-perimeter and area density ratio, we get 
$
    \frac{C_1^\ast(E^X_p)}{C_2^\ast (E^X_p)}\sim c_X \,\sigma(p)
$
as $p\rightarrow1$ as in \eqref{eq:LKC-ratio-assumption} with 
$c_X= \frac{\sqrt{\pi\lambda_2}}{2}
    \frac{\Gamma(1+k/2)}{\Gamma((k+1)/2)}$ and $\sigma(p)=1$. In  this case,   we get Equation \eqref{eq:C1form}  with $\beta=0$.\smallskip


\textit{Sub-Gaussian random fields \citep{adler2010excursion}:} 
Suppose $\Lambda$ is such that $\mathbb{E}[e^{-t\Lambda}]~=~e^{-t^{\varrho/2}}$ (i.e., $\Lambda$ is positive ($\varrho$/2)-stable), then $X(\cdot)=\sqrt{\Lambda}W(\cdot)$ is called \emph{sub-Gaussian}.  Using the Gaussian Kinematic Formula, one can  obtain the  ratio of half-perimeter and area densities  
$\frac{C_1^\ast(E^X_p)}{C_2^\ast(E^X_p)}\sim c_X \, \sigma(p) $
as $p\rightarrow1$,
and we follow the form of \eqref{eq:LKC-ratio-assumption} with  
$c_X = \frac{\sqrt{\pi\lambda_2}}{2}\frac{\Gamma(1+\varrho/2)}{\Gamma((\varrho+1)/2)}$ 
and $\sigma(p)=1$. 
Furthermore,   it follows that
$C_1^\ast(E_p^X)\sim c_X (1-p)
$ as $p\rightarrow1$, meaning we follow the form of \eqref{eq:C1form}    and $\beta=0$.\smallskip

\textit{Gaussian-based threshold-stable processes \citep{di2024spatial}:} Recall that a scalar random variable $\Lambda$ is regularly-varying at infinity with index $\varrho$ (denoted $\Lambda\in\text{RV}_{\infty}(\varrho)$), if $\PP(\Lambda~>~\lambda)~\sim~L(\lambda)\lambda^{-\varrho}$ as $\lambda\to\infty$ for some slowly varying function $L$. 
With this, suppose that $X^{\textit{scal}}(s)=\Lambda W(s)$ for $\Lambda\in\text{RV}_{\infty}(\varrho)$, or $X^{\textit{loc}}(s)=\Lambda + W(s)$ for $e^{\Lambda}\in\text{RV}_{\infty}(\varrho)$.  In this setting, the half-perimeter and area density ratio is given by
\begin{center}
$
 \frac{C_1^{\ast}(E^{X^{\textit{scal}}}_p)}{C_2^{\ast}(E^{X^{\textit{scal}}}_p)}\sim \frac{1}{2}\pi^{1/2}\frac{\Gamma(1+\varrho/2)}{\Gamma((1+\varrho)/2)}\sqrt{\lambda_2}$ \quad  and \quad   $
 \frac{C_1^{\ast}(E^{X^{\textit{loc}}}_p)}{C_2^{\ast}(E^{X^{\textit{loc}}}_p)}\sim \frac{1}{2}\sqrt{\frac{\pi\lambda_2}{2}}\varrho
$, 
as $p\rightarrow1$. 
\end{center}
Therefore,  $\sigma(p)=1$ and~$\beta=0$. 

\textbf{Max-stable random fields (\cite{de1984spectral}):} An important random field construction for studying the tail behaviour of spatial processes arises with the use of \emph{max-stable random fields}, frequently defined through the spectral representation
$
    X(s)=\max_{i\geq 1}\xi_i f_i(s),
$
introduced in \cite{de1984spectral}.
In this representation, $\{\xi_i\}_{i\geq 1}$ are points of a Poisson point process on $[0,\infty)$ with intensity $\xi^{-2}d\xi$, and $\{f_i\}_{i\geq 1}$ is a collection of independent copies of a nonnegative random field $f$, independent of $\{\xi_i\}_{i\geq 1}$, each satisfying $\mathbb{E}[f(s)]=1$ for every $s\in\mathbb{R}^2$.  
Max-stable random fields do not have the almost sure $C^2$ sample path assumption in \ref{A1} which is needed to use the Gaussian kinematic formula; however, we bypass the need for this by obtaining expected perimeters and areas using methods from \cite{galerne2011computation}, demonstrated in Supplementary Material~\ref{supp:calculations}.
Stationary max-stable random fields of the given form have an \emph{extremal coefficient function} $\theta:\mathbb{R}^2\to[1,2]$, which describes the strength of asymptotic dependence at distance lag $h$ as
$
    \theta(h)=-\log\PP(X(0)\leq 1,X(h)\leq 1).
$
The precise form of $\theta(h)$ depends on the chosen spectral densities $f_i$, with several examples summarised in \cite{schlather2002models} and \cite{davison2012statistical}. 
 For example, the so-called \emph{Smith model} \citep{smith1990max} uses bivariate Gaussian densities with random location parameters and fixed covariance matrices $\sigma^2 I$ for $f_i$, with the resulting extremal coefficient function  given by $\theta(h)=2\Phi(\left\|h\right\|/(2\sigma))$. 
Examining the half-perimeter and area density ratio gives ${C_1^\ast(E_p^X)}/{C_1^\ast(E_p^X)}\sim (\pi c_\theta)/2$, where  $c_{\theta}:=\lim_{\,r\downarrow 0}r^{-1}\left(\theta(r\tilde{u})-1\right)$ for a direction $\tilde{u}\in\mathbb{R}^2$ such that $\left\|\tilde{u}\right\|=1$.
If such a limit exists, we recover Equation~\eqref{eq:C1form} with $\beta=0$. \smallskip
 
Table~\ref{tab:beta-vals} summarises the models studied above, where two cases arise in the setting of Theorem~\ref{thm:LKCratio}: $\beta=0$ or  $\beta=1/2$. Typically, in scale mixtures of the form $\Lambda W$, $\beta=0$ when the tails of $\Lambda$ are regularly varying, and thus decay at a slower rate than the Gaussian component $W$. Furthermore, $\beta=1/2$ for Gaussian fields and for scale mixtures when the tails of $\Lambda$ decay at a rate similar to that of the Gaussian component $W$, as is the case in the Laplace random field, for example. 

\begin{table}\centering
\begin{tabular}{|c|c|c|c|}
  \hline \hline  
 Random field $X$ &  Scaling  index $\beta$ & Scaling function  $\sigma$ &  Constant $c_X$ \\ \hline   \hline    
{\huge\strut}    Gaussian    & $1/2$ & $u_p/ \sqrt{2}$  &  $ \sqrt{\pi\lambda_2}/2$ \\ 
 {\tiny\strut} {\small \citep{bierme2018lipschitz}}   & &    &   \\    \hline   
{\huge\strut}   Anisotropic Gaussian  & $1/2$ & $u_p / \sqrt{2}$ & $\frac{\sqrt{\lambda_2}\,
P(\text{ellipse};a_1,a_2)}{4\sqrt{\pi}}$ \\ 
{\tiny\strut} {\small\citep{cotsakis2022perimeter}}&  &  &
\\  \hline   
{\huge\strut}  Standardised chi-squared  &  $1/2$ & $\sqrt{u_p}$ & $\sqrt{\pi\lambda_2}/2$\\
 {\tiny\strut}  {\small \citep{bierme2018lipschitz}} &  &  &   \\ \hline     
{\huge\strut}   Laplace  & $1/2$ & $\sqrt{u_p}$ & $\frac{1}{2}\sqrt{\frac{\pi\lambda_2}{2}}$ \\ 
{\tiny\strut} {\small   \citep{Opitz.2016}}  & &  & \\    \hline  
{\huge\strut}   Elliptical Student  & 0 & 1 & $ \frac{\sqrt{\pi\lambda_2}}{2}
    \frac{\Gamma(1+k/2)}{\Gamma((k+1)/2)}$ \\
{\tiny\strut} {\small  \citep{Ma2013}}  & &  & \\   \hline 
{\huge\strut}   Sub-Gaussian  & 0 & 1 & $\frac{\sqrt{\pi\lambda_2}}{2}\frac{\Gamma(1+\varrho/2)}{\Gamma((\varrho+1)/2)}$\\  
{\tiny\strut} {\small   \citep{adler2010excursion}}  & &  & \\    \hline   {\huge\strut}   Threshold-stable scale mix.  & 0 & 1 & $\frac{1}{2}\pi^{1/2}\frac{\Gamma(1+\varrho/2)}{\Gamma((1+\varrho)/2)}\sqrt{\lambda_2}$ \\  {\tiny\strut}  {\small   \citep{di2024spatial} }  & &  & \\    \hline   
{\huge\strut}   Threshold-stable loc. mix.  & 0 & 1 & $\frac{1}{2}\sqrt{\frac{\pi\lambda_2}{2}}\varrho$ \\  {\tiny\strut}  {\small   \citep{di2024spatial} }  & &  & \\    \hline   
{\huge\strut}   Max-stable  & 0 & 1 & $\frac{\pi}{2}c_{\theta}$ \\  {\tiny\strut}  {\small   \citep{de1984spectral} }  & &  & \\    \hline  
\hline  
\end{tabular} \vspace{-0.2cm}
\caption{Scaling indices $\beta$, scaling functions $\sigma$ and constants $c_X$ for several models  satisfying  the setting of  Corollary~\ref{cor:scaling} (see Section \ref{sec:examples}). Recall  that $u_p=F^{-1}_{X({0})}(p)$.}
\label{tab:beta-vals}
\end{table}

\subsection{Examples where the asymptotic behaviour of the perimeter differs from Equation~\eqref{eq:C1form}} \label{sec:COUNTERexamples} 

In this section,  we study cases when there is no analytic solution for $\beta$ because the half-perimeter density, $C_1^\ast(E^X_p)$, cannot be expressed in the form~\eqref{eq:C1form}.  Detailed calculations to obtain results below can be found in Section \ref{supp:calculationsBIS} in  Supplementary material. \smallskip

\textit{Student random fields:}\label{Tstudent} Let $X(s)=\sqrt{k-2}{W_0(s)}/\sqrt{\sum_{j=1}^k W_j(s)^2}$, where $k\geq3$ is an integer and $W_0,W_1,\ldots,W_k$ are independent copies of the stationary isotropic standard Gaussian field $W$.
Using Section 2.2.3 in \cite{bierme2018lipschitz}, one can derive the half-perimeter and area density ratio, as in \eqref{eq:LKC-ratio-assumption}, for $p\rightarrow1$, $\frac{C_1^\ast(E_p^X)}{C_2^\ast(E_p^X)}\sim c_X\,\sigma(p)$ with
$
    c_X=\frac{\sqrt{\pi\lambda_2}}{4}\frac{k}{\sqrt{k-2}}\frac{\Gamma(k/2)}{\Gamma((k+1)/2)}
$
and $\sigma(p)=u_p$.
Moreover, in this case, 
$C_1^\ast(E_p^X)\sim c_X(1-p)u_p\asymp(1-p)^{1-1/k}$
, and this random field's half-perimeter density does not follow the asymptotic form in~\eqref{eq:C1form}.
\smallskip 

\textit{Heavy-tail and Gaussian additive mixture:}
Suppose that $X(s)=\Lambda + W(s)$ for $\Lambda\in\text{RV}_{\infty}(\varrho)$ independent of $W$, with $\varrho>0$. Assume moreover that $\Lambda$ admits a density $f_\Lambda$ satisfying $f_\Lambda(u)\sim\varrho\bar F_\Lambda(u)/u$ as $u\to\infty$.
For this random field, $\chi_p(s_0,s)\rightarrow1$ as $p\to1$, consistently with the power-law case of Proposition~1 in \cite{krupskii2018factor}.
\cite{engelke2019extremal} derived the coefficient of tail dependence \citep{ledford1996statistics,ledford1997modelling} for $X$ using various definitions of $\Lambda$. One can prove that the half-perimeter and area density ratio is given by $\frac{C_1^\ast(E^X_p)}{C_2^\ast(E^X_p)}\sim c_X \sigma(p)$, as $p\rightarrow1$, and we follow the form of \eqref{eq:LKC-ratio-assumption} with $c_X=\frac{\sqrt{2\pi\lambda_2}}{4}\varrho$ and $\sigma(p)=u_p^{-1}$. 
Composing this scaling function with the standard exponential distribution function gives $\sigma(1-e^{-x})\rightarrow0$ as $x\rightarrow\infty$. More precisely, regular variation of the marginal tail implies $\sigma(1-e^{-x})=\exp\{-x/\varrho+o(x)\}$ as $x\to\infty$. Thus, $x\mapsto\sigma(1-e^{-x})$ is rapidly varying with index $-\infty$ and does not admit a finite scaling index $\beta$ in the framework of Corollary~\ref{cor:scaling}. Under the assumed sample continuity of $W$, the conditional excursion sets $E_p^X(s_0)$ contain every fixed compact set $K\subset S$ containing $s_0$ with probability tending to one as $p\to1$.
Moreover, $C_1^\ast(E^X_p)\sim c_X(1-p)u_p^{-1}$, which, by the preceding asymptotic behaviour of $u_p^{-1}$, does not follow the logarithmic-power form in Equation~\eqref{eq:C1form}.

Several of the functions $\sigma$ obtained in Sections~\ref{sec:examples} and~\ref{sec:COUNTERexamples} are graphically illustrated in Figure~\ref{fig:sigmaplot} in Supplementary Material Section \ref{supp:SSfigs}.

\section{Statistical inference for $\beta$ and a   spatial   dependence testing
hypothesis }\label{sec:inference}

We have seen how the scaling index is a useful tool for describing the asymptotic behaviour of a stationary random field. A natural question is how, given observations of a random field, one can estimate $\beta$.
We now propose two strategies; one using estimates of the perimeter  and the other using the estimates of the LER of observed excursion sets. Each of these strategies has its own benefits and drawbacks.
The perimeter approach has built-in asymptotic normality, allowing us to develop a hypothesis test on the value of our estimator for $\beta$. Convergence to the normal distribution depends on our spatial domain $S$ growing in size.  
Boundary effects in perimeter estimation could potentially lead to inaccuracies in the estimation of $\beta$.  
The LER approach is a more indirect method, using regression on point estimates of the LER as a response variable. Estimation of the LER does not suffer as severely from  boundary effects, and can be accurately obtained on a relatively coarse grid, saving on computation time.  
\cite{Costakis2025Extent} developed a linear-regression procedure for the estimation of $\beta$ using LER estimates as the response variable. Here, we propose a related quantile regression-based procedure, and include the possibility to use covariates, demonstrating how $\beta$ can behave under temporal non-stationarity, for example. The principal drawback  of this approach is that it is difficult to establish asymptotic normality   with explicit theoretical variance.
 
\begin{figure}[ht!] \centering
\begin{tikzpicture}[scale=0.5]
\def\n{5}
\draw[thick] (0,0) rectangle (\n,\n);
\foreach \i in {0,...,\n}{
    \draw (\i,0) circle (0.09);
    \draw (\i,\n) circle (0.09);
    \draw (0,\i) circle (0.09);
    \draw (\n,\i) circle (0.09);
}
\foreach \i in {1,...,4}{
    \foreach \j in {1,...,4}{
        \draw (\i,\j) circle (0.09);
    }
}

\foreach \p in {(2,4),(4,4),(2,2),(4,2)}{
    \draw[blue, thick] \p circle (0.09);
}
\foreach \p in {(0,0),(0,2),(0,4)}{
    \draw[blue, thick] \p circle (0.09);
}
\foreach \p in {(2,0),(4,0)}{
    \draw[blue, thick] \p circle (0.09);
}
\node[below left]  at (0,0) {\scriptsize{$s_{0,0}$}};
\node[below]       at (1,0) {\scriptsize{$s_{1,0}$}};
\node[below]       at (2,0) {\scriptsize{$s_{2,0}$}};
\node[left]        at (0,1) {\scriptsize{$s_{0,1}$}};
\node at (1,1.3) {\scriptsize{$s_{1,1}$}};
\node[left]        at (0,2) {\scriptsize{$s_{0,2}$}};
\node[left] at (0,5) {$S$};
\draw[<->, thick] (5.4,3) -- (5.4,4)
node[midway,right=6pt] {$\delta$};
\draw[<->, thick] (0,5.4) -- (5,5.4)
node[midway,above=2pt] {$2t$};
\end{tikzpicture} \vspace{-0.2cm}
 \caption{An illustration of the quantities defined in Definition~\ref{def:grid}. The positions of the elements of $\mathbb{G}(\delta,S)$  in $\R^2$ are shown as circles, and the subset   $I^{(2,\delta,S)}$ is highlighted in blue. Here, $M=6$, and the side length of $S$ is $(M-1)\delta = 5\delta$.} \label{fig:sampling}
 \end{figure}
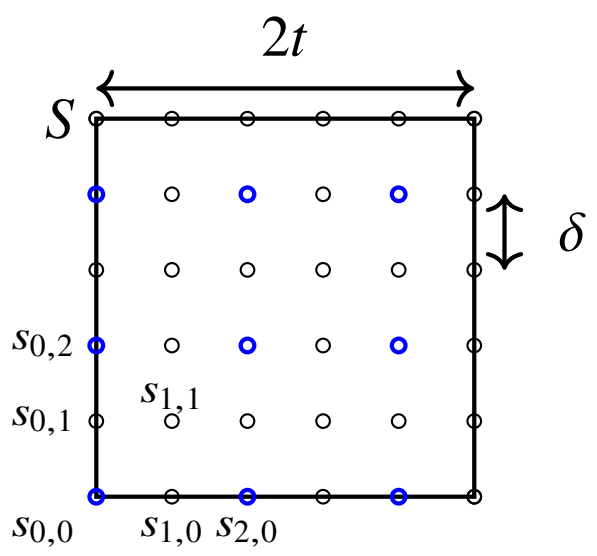

\smallskip
In practice, the estimates of the perimeter and the LER will be done on a pixelized  spatial domain. For the sake of simplicity, we assume that $S$ is a square region within $\mathbb{R}^2$. We introduce a grid of points and its associated \textit{subgrid}.
\begin{definition}\label{def:grid}
\textit{Suppose  $S= [-t,t]^2 \subset \R^2$ for some $t>0$ and fix the pixel width $\delta > 0$. Define a square grid  as
\begin{center} $\mathbb{G}(\delta,S) := \big\{s_{i,j}: i,j\in\N_0\big\}\cap S$  with $s_{i,j} := (-t+i\delta, -t+j\delta)\in\R^2.$ \end{center}   For $M$ the number of rows (and the number of columns) of $\mathbb{G}(\delta,S)$, define the index set $I^{(\delta, S)} := \{0,\ldots,M-1\}\subset \N$. For $m\in \N\setminus \{0\}$, consider the associated subgrid as $ I^{(m, \delta, S)} := \{i\in I^{(\delta, S)}: i \equiv 0\ (\mathrm{mod}\ m)\}.$
}\end{definition}
\noindent An illustration of $\mathbb{G}(\delta,S)$ is given in Figure~\ref{fig:sampling}, where the elements with indices in $ I^{(m,\delta,S)}$, for subwindows with $m=2$ pixels on each side, are highlighted in blue.  

\subsection{The perimeter estimation  approach}\label{sec:perim-approach}

\noindent
\textbf{Building an estimator for \boldmath{$\beta$} using the perimeter.} Taking inspiration from Equation  \eqref{eq:C1form},  we propose a plug-in estimator of the scaling index $\beta$ based on a consistent estimator of the perimeter and derive its asymptotic normality property.
\cite{cotsakisAAP}, \cite{bierme:hal-02793752}, and \cite{abaach2024pixel} consider an estimator for the excursion set perimeter   that is constructed with the $\ell_{1}$-norm. However, by the nature of pixelisation, this estimator is biased (see \cite{Baccelli, cotsakisAAP, bernardino:hal-05679352}).
We instead consider the unbiased $\ell_{2}$-norm-based estimator leading to an estimate having a multigrid convergence property (see \cite{cotsakis2022perimeter}).

\begin{definition}[A consistent estimator of $C^*_1$]\label{def:estimator}
Let   $\zeta_s(p) =  $ $\mathbbm{1}_{[u_p,\infty)}(X(s))$ where $u_p=F_{X(0)}^{-1}(p)$.    Define the estimate of the  half-perimeter density  $C^*_1(E^X_p)$ in  \eqref{eq:defC1ast}  as 
\begin{equation}\label{eqn:C1hat}
    \hat{C}_{1}^{(m,  \delta,  S)}(X, p)  :=\frac{\delta}{{2\,\mathcal{L}_2(S_n)}}\sum_{a,b\in I^{(m, \delta, S)}} \big\|\big(N^{a,b}_{X,h}(p), N^{a,b}_{X,v}(p)\big)\big\|_2,
\end{equation} 
where the components in the $\ell_2$-norm in \eqref{eqn:C1hat} are given by 
\begin{equation*}
    N^{a,b}_{X,h}(p) :=  \sum_{i = a}^{(a+m-1) \wedge (M-1)}\ \sum_{j = b}^{(b+m-1) \wedge (M-2)}
    |\zeta_{s_{i,j}}(p) - \zeta_{s_{i,j+1}}(p)|,
\end{equation*}
\begin{equation*}
    N^{a,b}_{X,v}(p) :=  \sum_{i = a}^{(a+m-1) \wedge (M-2)}\ \sum_{j = b}^{(b+m-1) \wedge (M-1)} |\zeta_{s_{i,j}}(p) - \zeta_{s_{i+1,j}}(p)|.
\end{equation*}
 
\end{definition} 

Notice that $\hat{C}_{1}^{(m,  \delta,  S)}(X, p)$  uses only the information contained in the random binary matrix  $\zeta$, and is obtained by separating $\zeta$ into sub-matrices of length-size $m$. In each sub-matrix, the length of the line segment that separates the observed  ones from the zeroes  is computed using the $\ell_{2}$-norm. For the asymptotic growing-domain analysis, we introduce the VH-growing  \emph{dilation} $S_n$ of a fixed initial domain $S$,   $S_n:= \{ns:s\in S\}$, for $n\in\mathbb{N}$.  In addition, suppose that $p_1 \neq p_2\in[p_{\min},1)$ are two fixed probability levels used to define the excursion sets, where $p_{\min}$ is close to 1. Then $\hat{C}^{(m,\delta,S_n)}_1(X,p_1)$ and $\hat{C}^{(m,\delta,S_n)}_1(X,p_2)$ as in \eqref{eqn:C1hat} are unbiased estimators for $C_1^*(E^X_{p_1})$ and $C_1^*(E^X_{p_2})$ with asymptotic Gaussian fluctuations  (see  Theorem 2 in~\cite{cotsakis2022perimeter}).   As a consequence of Equation \eqref{eq:C1form}, 
 the ratio of $\hat{C}^{(m,\delta,S_n}_1(X,p_1)$ and $\hat{C}^{(m,\delta,S_n)}_1(X,p_2)$   can be used 
 to build the following plug-in estimator:
\begin{align}\label{eq:beta-C1}
\hat \beta(p_1,p_2):=\hat \beta(p_1,p_2;X,m,\delta,S_n) &=
\frac{\ln\left(
{\hat{C}^{(m,\delta,S_n)}_1(X,p_1)}/{\hat{C}^{(m,\delta,S_n)}_1(X,p_2)}\right) - \ln\left(\frac{1-p_1}{1-p_2}\right)
}{\ln\left({\ln(1-p_1)}/{\ln(1-p_2)}
\right)}.
\end{align}
We now prove a CLT for \eqref{eq:beta-C1} for a sequence of increasing (\textit{resp.} decreasing) $m$ (\textit{resp.} $\delta$) values. \smallskip

\textbf{CLT for the \boldmath{$\hat \beta(p_1,p_2)$}  and hypothesis testing.}  
The $\hat \beta(p_1,p_2)$-estimator  in  \eqref{eq:beta-C1} is a differentiable function of the $\hat{C}_{1}^{(m,  \delta,  S)}(X, p)$ estimator in Definition \ref{def:estimator}; therefore, we can use the delta method to establish asymptotic normality, as $n\rightarrow\infty$.  

\begin{theorem}\label{thm:CLTbeta}
Suppose the random field $X$ satisfies Assumption~\ref{A1}, is strongly mixing (see Definition \ref{stronglyMIXING}) and has no local extrema on $\partial S$.   Let
$p_1\neq p_2\in[p_{\min},1)$ be two fixed probability levels, and let
$u_{p_j}=F^{-1}_{X(0)}(p_j),$  $j=1,2.$ Suppose $X$ has no critical points at the levels $u_{p_1}$ and $u_{p_2}$ inside $S\subset\mathbb{R}^2$.
Let $(S_n)_{n\geq1}$ be a VH-growing sequence of domains, and suppose
that the sequences $(m_n)_{n\geq1}$ and $(\delta_n)_{n\geq1}$ satisfy
the conditions required for the multilevel perimeter CLT
simultaneously at the levels $u_{p_1}$ and $u_{p_2}$ (see Theorem 2 in \cite{cotsakis2022perimeter}). Define
\[
\beta(p_1,p_2)
:=
\frac{
\ln\left\{
C_1^*(E^X_{p_1})/C_1^*(E^X_{p_2})
\right\}
-
\ln\left\{
(1-p_1)/(1-p_2)
\right\}
}{
\ln\left\{
\ln(1-p_1)/\ln(1-p_2)
\right\}
}.
\]
Then,  it holds that 
$
\sqrt{\mathcal{L}_2(S_n)}
\left[
\hat{\beta}
(p_1,p_2;X,m_n,\delta_n,S_n)
-
\beta(p_1,p_2)
\right]
\xrightarrow{d}
\mathcal{N}\left(0,\upsilon^2(p_1,p_2)\right),
$ for $ n\to\infty,
$ 
where $\hat \beta(p_1,p_2)$ as in \eqref{eq:beta-C1} and 
\[
\upsilon^2(p_1,p_2)
=
\left(
\ln\left(
\frac{\ln(1-p_1)}{\ln(1-p_2)}
\right)
\right)^{-2}
\left(
\frac{\nu(p_1,p_1)}
     {C_1^*(E^X_{p_1})^2}
+
\frac{\nu(p_2,p_2)}
     {C_1^*(E^X_{p_2})^2}
-
2\frac{\nu(p_1,p_2)}
       {C_1^*(E^X_{p_1})C_1^*(E^X_{p_2})}
\right),
\]
and $\nu(p_i,p_j)$ denotes the asymptotic covariance between the
half-perimeter density estimators at levels $p_i$ and $p_j$.
\end{theorem}

\begin{remark}[Convergence to the scaling index $\beta$]
\label{rem:beta_convergence}
The quantity $\beta(p_1,p_2)$ appearing in
Theorem~\ref{thm:CLTbeta} is a finite-threshold version of the
asymptotic scaling index $\beta$. Indeed, suppose that
\[
C_1^*(E_p^X)
=
c_X(1-p)\{-\ln(1-p)\}^{\beta}\{1+r(p)\},
\qquad r(p)\rightarrow 0,
\]
as $p\to1$, as in Equation~ \eqref{eq:C1form}. Then
\[
\begin{aligned}
\beta(p_1,p_2)
&=
\frac{
\ln\left\{
C_1^*(E^X_{p_1})/C_1^*(E^X_{p_2})
\right\}
-
\ln\left\{
(1-p_1)/(1-p_2)
\right\}
}{
\ln\left\{
\ln(1-p_1)/\ln(1-p_2)
\right\}
}
\\
& =
\beta+
\frac{
\ln\{1+r(p_1)\}-\ln\{1+r(p_2)\}
}{
\ln\left\{
\ln(1-p_1)/\ln(1-p_2)
\right\}
}.
\end{aligned}
\]
Consequently, if $p_1,p_2\to1$ in such a way that
$\frac{-\ln(1-p_1)}{-\ln(1-p_2)}
\rightarrow
\lambda\in(0,\infty)\setminus\{1\},
$  
then
$\beta(p_1,p_2)\rightarrow~\beta$.  In Figure \ref{fig:betap1p2} in Supplementary material Section  \ref{supp:SSfigs}, we display the values of  
 $\beta(p, (p+1)/2)$ along with $\beta$ for several choices of  $p_1$ for the  Gaussian, standardised chi-squared, and Smith  models (see Section \ref{sec:examples}, and note that for the anisotropic Gaussian field,  the curve is exactly identical to that of the isotropic Gaussian field).
 \end{remark}

In practice, for fixed large $n$, the empirical variability of the estimator $\hat{\beta}(p_1,p_2;X,m_n,\delta_n,S_n)$ can be large. To remedy this, a median version of Theorem \ref{thm:CLTbeta}  is given in the following result.

\begin{corollary}[Median version of Theorem~\ref{thm:CLTbeta}]
\label{corr:beta-median} 
Let $\hat{\beta}^{(i)}_n
=
\hat{\beta}^{(i)}
(p_1,p_2;X,m_n,\delta_n,S_n),
$ for $i=1,\ldots,k,$ 
be $k$ independent replicates of the estimator in
Theorem~\ref{thm:CLTbeta}, each computed from an independent
realization of the random field $X$ over the same VH-growing domain
$S_n$. Define
\[
\tilde\beta_{n,k}(p_1,p_2)
:=
\operatorname{median}_{i=1,\ldots,k}
\hat\beta_n^{(i)}.
\]

For every fixed odd integer $k\geq 1$, as $n\to\infty$,
$
\frac{\sqrt{\mathcal L_2(S_n)}}
{\upsilon(p_1,p_2)}
\left\{
\tilde\beta_{n,k}(p_1,p_2)
-
\beta(p_1,p_2)
\right\}
\xrightarrow{d}
M_k,
$ 
where $M_k$ is the sample median of $k$ independent standard
Gaussian random variables. The limiting random variable $M_k$ has  variance
\[
\tau_k^2
=
\frac{k!}
{\left\{\left(\frac{k-1}{2}\right)!\right\}^{2}}
\int_{-\infty}^{\infty}
x^2
\Phi(x)^{(k-1)/2}
\{1-\Phi(x)\}^{(k-1)/2}
\phi(x)\,dx.
\]
Moreover,
$\tau_k^2\sim\frac{\pi}{2k},
$ for $ k\to\infty.$
\end{corollary}

The proofs of both Theorem \ref{thm:CLTbeta}  and Corollary \ref{corr:beta-median} are postponed to Section \ref{ProofsSection}  of  Supplementary Material. Note that the theoretical expression of $\upsilon^2(p_1,p_2)$ involves
quantities that are difficult to evaluate in practice. We therefore estimate
the asymptotic variance empirically from independent replicates of the
median estimator. More precisely, we use
\begin{equation}\label{eq:emp-beta-var}
    \hat{\eta}_k^2(p_1,p_2)
    =
    \mathcal{L}_2(S_n)\,
    \hat{\mathbb{V}}\left\{
        \tilde{\beta}_{n,k}(p_1,p_2)
    \right\},
\end{equation}
where $\hat{\mathbb{V}}$ denotes the empirical variance computed
from independent replicates of $\tilde{\beta}_{n,k}(p_1,p_2)$. 
\smallskip

\textbf{A test for  asymptotic spatial dependence.}   
Given the CLT in Theorem~\ref{thm:CLTbeta}, we  consider the hypothesis test
\begin{equation*}
    \text{H}_0: \beta{(p_1,p_2)} \leq 0 \quad \mbox{ versus } \quad \text{H}_1: \beta{(p_1,p_2)}>0.
\end{equation*} 
Under the asymptotic representation~\eqref{eq:C1form}  and the threshold regime of Remark~\ref{rem:beta_convergence}, a positive value of $\beta(p_1,p_2)$ at sufficiently high thresholds provides evidence for $\beta>0$. In this case, $\sigma(p)\to\infty$, and Proposition~\ref{prop:scaling-AI} implies asymptotic independence. 
Under the assumptions of Proposition~\ref{prop:scaling-AI} and with the counterexamples in Supplementary Material~\ref{supp:beta-dependence} in mind, rejection of $\text{H}_0$ at sufficiently high thresholds $p_1$, $p_2$ provides evidence of asymptotic independence of the random field, while failure to reject $\text{H}_0$ is not sufficient to establish the tail dependence behaviour of the random field. 

In order to test the null hypothesis, we use the residual
$$Z^{H_0}=\sqrt{\frac{\mathcal L_2(S_n)}
{\hat\eta_k^2(p_1,p_2)}}
\,\tilde\beta_{n,k}(p_1,p_2)= \sqrt{\frac{1}
{\hat{\mathbb{V}}\{
        \tilde{\beta}_{n,k}(p_1,p_2)\}}}
\,\tilde\beta_{n,k}(p_1,p_2).$$
Assuming that
$\hat\eta_k^2(p_1,p_2)
\xrightarrow{\mathbb P}
\upsilon^2(p_1,p_2)\tau_k^2\in(0,\infty),$ $ n\to\infty,$
Corollary~\ref{corr:beta-median} and Slutsky's theorem yield,
at the boundary $Z^{H_0}\xrightarrow{d}{M_k}/{\tau_k}.$  The null hypothesis is rejected if
$Z^{H_0}>q_{1-\alpha,k}$, where $q_{1-\alpha,k}$ denotes
the $(1-\alpha)$-quantile of $M_k/\tau_k$.
These quantiles can be computed using $\Phi(M_k)~\sim~
\operatorname{Beta}\!\left((k+1)/2,(k+1)/2\right)$. 
The Monte Carlo estimate of the rejection probability
at significance level $\alpha\in(0,1)$, based on $N$
replicates of the residual, is
$\hat{\mathbb P}(\text{reject }H_0) =\frac1N\sum_{i=1}^N
\mathbbm{1}_{(q_{1-\alpha,k},\infty)}(Z^{H_0}_{\,i})$.
For large $k$, the standard Gaussian quantile
$\Phi^{-1}(1-\alpha)$ provides an approximation to
$q_{1-\alpha,k}$. We follow this normal approximation in the rest of the paper.
\smallskip

\textbf{Numerical studies for 
\boldmath{$\tilde\beta_{n,k}(p_1,p_2)$}.} 
Let $W$ be a stationary Gaussian random field using the squared-exponential covariance function
\begin{equation}\label{eq:sqexpcov}
    C(s_0,s) = \sigma^2 \exp\left(-\frac{1}{2 l^2}\left\|s-s_0\right\|^2\right),
\end{equation}
where $\sigma^2>0$ and $l$ are the scale and range parameters, both fixed to 1. With this covariance function, the second spectral moment of $W$ is given by $\lambda_2=\sigma^2 l^{-2} =1$. 
We perform simulation studies on a square region $S=[-15,15]^2$, with associated  gridded version $S^\delta$ on spacing $\delta=0.0293$.  We set the number of pixels on the sides of the subwindow used for unbiased perimeter estimation to be $$m=\Big\lfloor\frac{1}{3}\left(\frac{\sqrt{\mathcal{L}_2(S)}}{{\hat{\mathcal{L}}}_0(E^X_p\cap  S)}\right)^{1/3}\delta^{-2/3}\Big\rfloor\approx \Big\lfloor 52.61\left({\hat{\mathcal{L}}}_0(E^X_p\cap  S)\right)^{-1/3}\Big\rfloor,
$$ as defined in \cite{cotsakis2022perimeter}. The value ${\hat{\mathcal{L}}}_0(E^X_p\cap  S)$ is an estimate of the Euler-Poincaré characteristic of the set $E^X_p\cap S$, obtained by taking the difference between the number of connected components and the number of holes in observations of $E^X_p\cap S$.

To do so in a computationally-efficient manner, we employ the circulant embedding method via the fast Fourier transform, with code adapted from   Section~12.2.2 of \cite{Kroese2015simulation}.
Using this procedure, we generate observations from the Gaussian, anisotropic Gaussian, and standardised chi-squared random fields. The affinity matrix of the anisotropic Gaussian field set to $A=\left(2, 1 ; 1 , 1.5\right)$, and the chi-squared field uses $\kappa=3$ degrees of freedom.
    Each of these random fields has positive true scaling index values, as derived in Section~\ref{sec:examples}.
    In order to study an example with $\beta=0$, we generate observations from the Smith model with $\sigma^2=1$. Observations from this model are obtained using the \texttt{rmaxstab} function of the \texttt{SpatialExtremes} package within the \texttt{R} computing language \citep{Ribatet2022}.

Results presented in Figure~\ref{fig:C1boxplots-supp} in  Supplementary Material (Section \ref{supp:SSfigs}) show that, in the Gaussian, anisotropic Gaussian,  chi-squared and   Smith settings, excursion set perimeters are accurately estimated across   a broad range of levels $p$.    Therefore, following Theorem  \ref{thm:CLTbeta},  we opt for very high values of $p$ to better capture the tail behaviour of the respective random fields.

\begin{figure}[htpb]\centering 
    \includegraphics[width=0.24\linewidth]{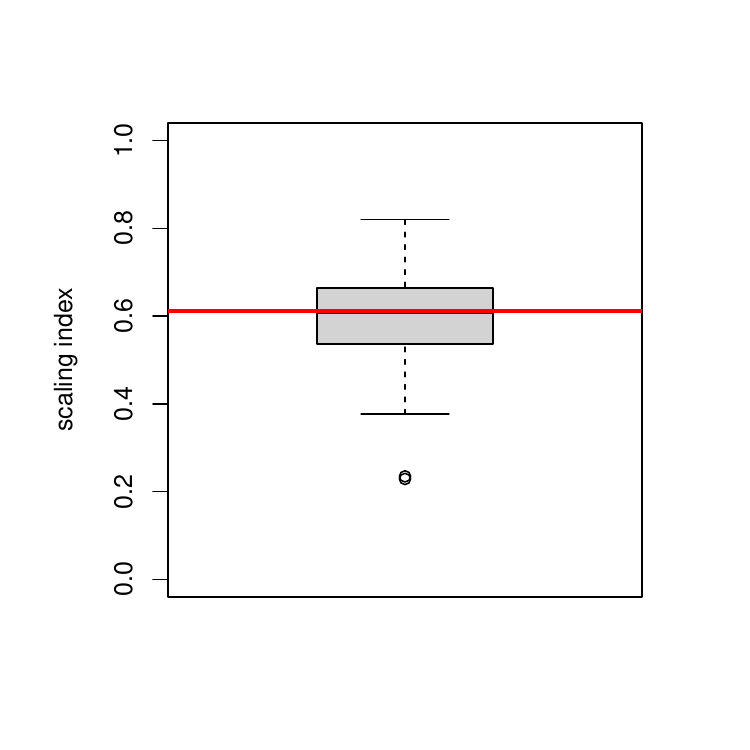}
    \includegraphics[width=0.24\linewidth]{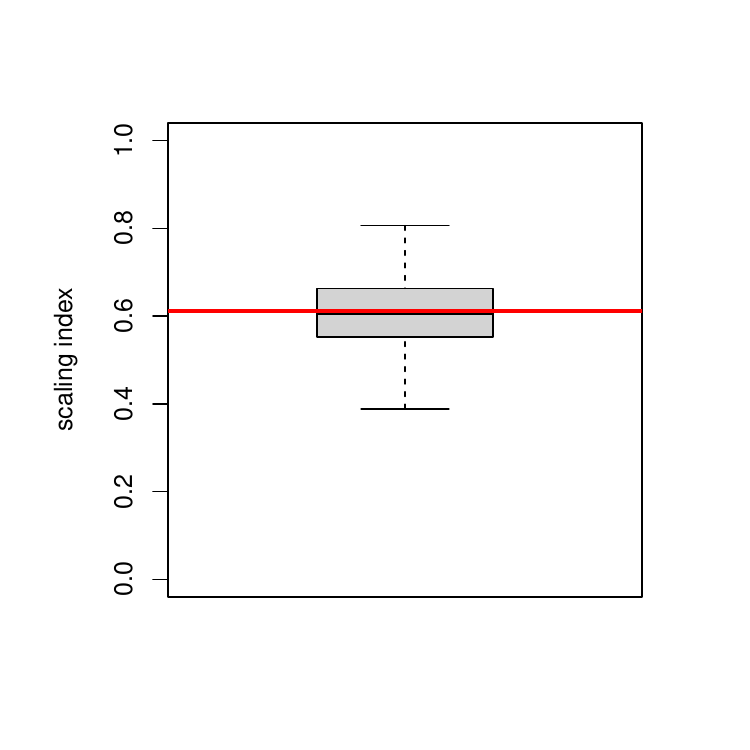}
    \includegraphics[width=0.24\linewidth]{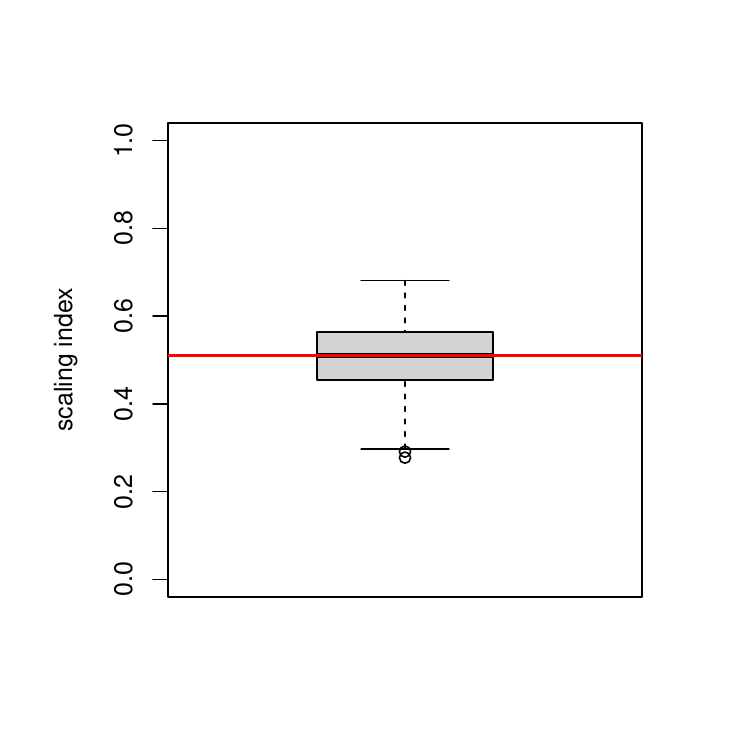} 
    \includegraphics[width=0.24\linewidth]{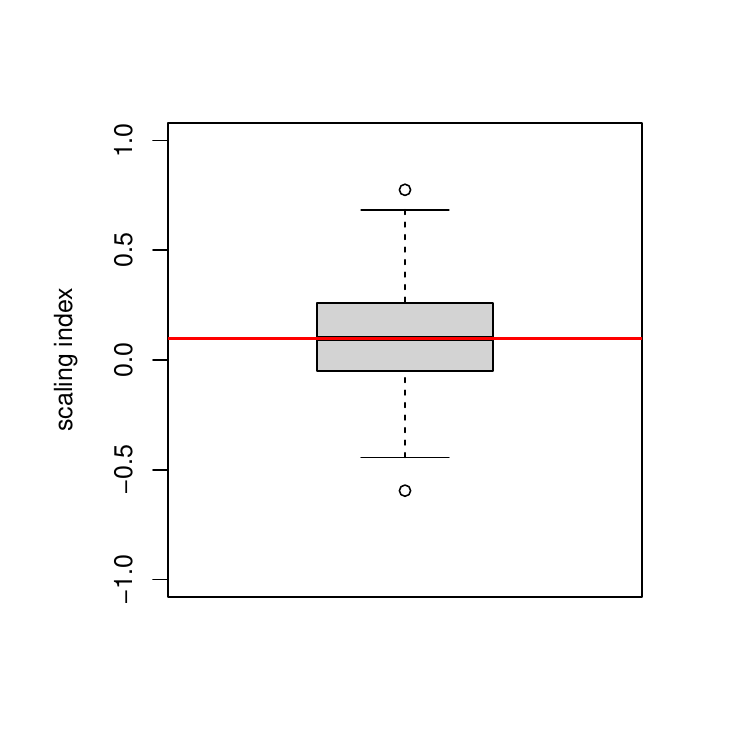}
    \vspace{-0.5cm}
    
    \includegraphics[width=0.24\linewidth]{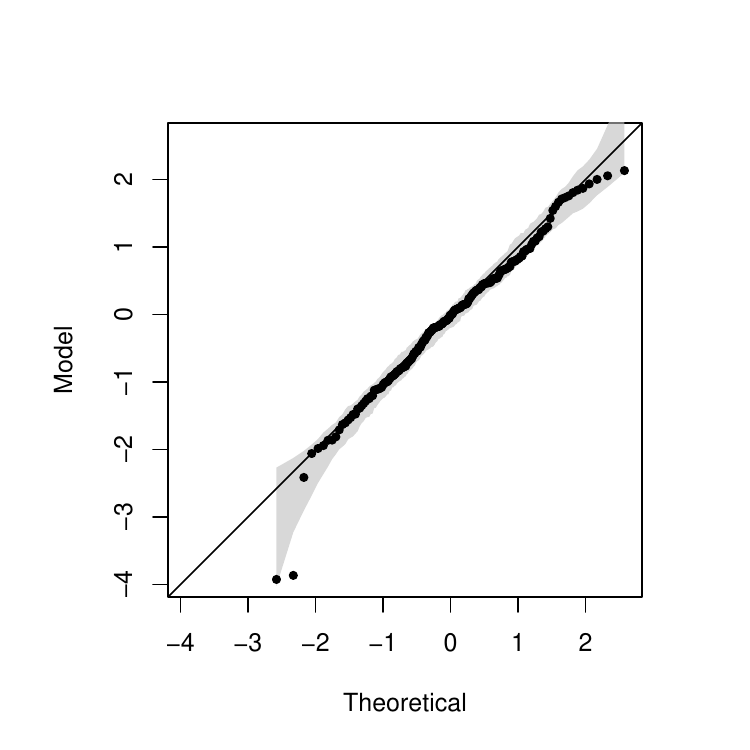}
    \includegraphics[width=0.24\linewidth]{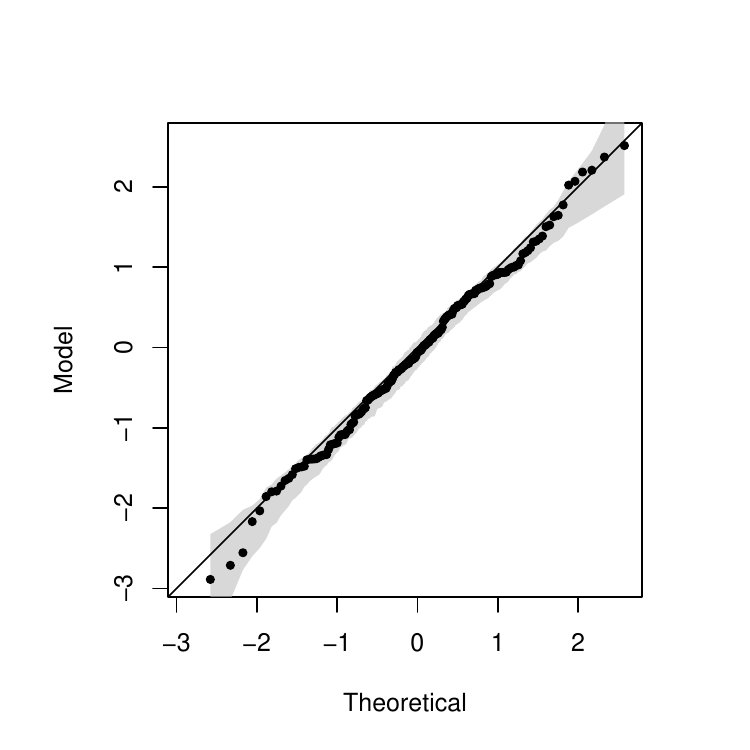}
    \includegraphics[width=0.24\linewidth]{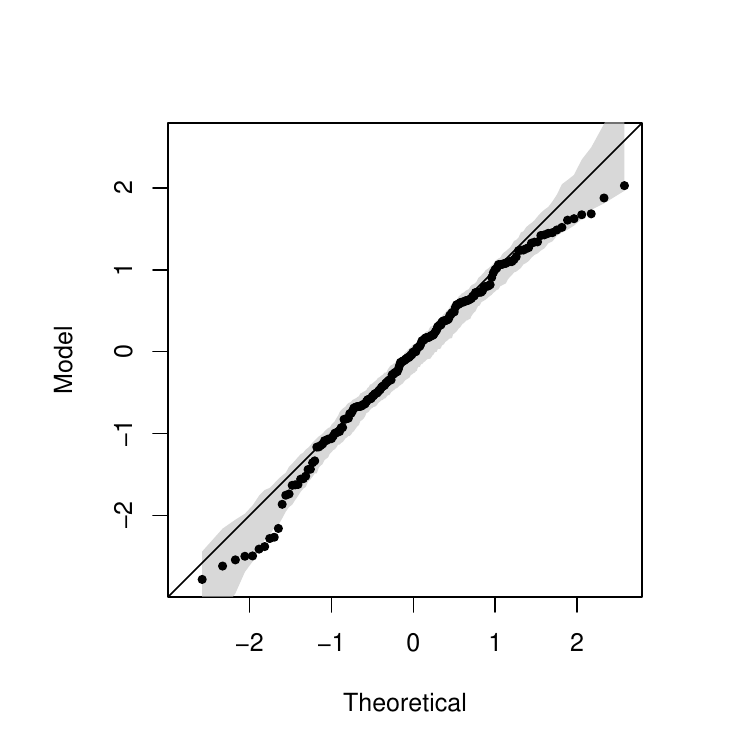}
    \includegraphics[width=0.24\linewidth]{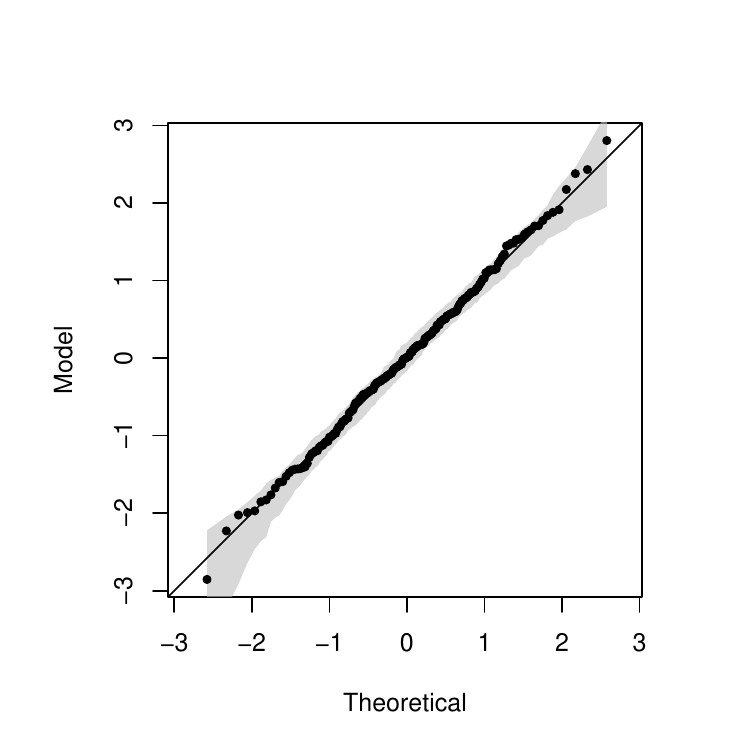}
    \caption{Boxplot (top row) of estimated $\beta(p_1,p_2)$ using the perimeter-based approach, with true values indicated by the red line, with the associated QQ plot (bottom row) for $\beta(p_1, p_2)$ for the Gaussian, anisotropic Gaussian, standardised chi-squared, and Smith random fields (left to right).    Median values $\tilde{\beta}_{n,k}$ are taken using groups of $k=31$   values of the estimate $\hat{\beta}_n$.} 
    \label{fig:SSboxplots-perim}
\end{figure}

For each of the four random fields above, we generate 6000 realisations on $S^\delta$. At each realisation, the perimeters are estimated at levels $p_1$ and $p_2$, resulting in 6000 estimates  
$\hat \beta(p_1, p_2)$ in \eqref{eq:beta-C1}. We take medians of groups of $k=31$ estimates, resulting in roughly 200 observations of the reduced-variance estimator $\tilde \beta$, displayed in boxplots in Figure~\ref{fig:SSboxplots-perim} for each of the four underlying random fields at the 
pair of threshold levels $(p_1,p_2)$ provided in Table~\ref{tab:SSperim-prH0}. These thresholds are fixed to $p_1=0.93$ and $p_2=0.94$, and  were chosen to not be too large so as to accurately estimate excursion set perimeters, but still large enough to capture estimates of $\beta(p_1,p_2)$ close to the limiting value $\beta$.  

 In Figure~\ref{fig:SSboxplots-perim}, good agreement is shown between estimated and true values of $\beta(p_1, p_2)$. Furthermore, the finite-sample Gaussian approximation of the median estimator \(\tilde\beta_{n,k}\) is assessed through quantile-quantile (QQ) plots. These are obtained by repeatedly computing $\tilde \beta_{i, k}(p_1,p_2;m,\delta,S^\delta)$ for observations $i~\in~\{1,\dots,N\}$. For each estimate, we compute the residual
$$
    Z_i = \sqrt{\frac{\mathcal{L}_2(S^\delta)}{\hat{\eta}_k^2(p_1,p_2)}}\left[\tilde \beta_{i, k}(p_1,p_2;m,\delta,S^\delta)-\beta(p_1, p_2)\right],
$$
where $\hat{\eta}_k^2(p_1,p_2)$ is empirical variance of the estimated scaling indices, presented in    \eqref{eq:emp-beta-var}. 
The ordered residuals $Z_{(1)}\leq\dots\leq Z_{(N)}$ can be plotted against the standard Gaussian quantiles.  
In Figure~\ref{fig:SSboxplots-perim}, we see that while all four random fields have QQ plots in general agreement with the diagonal, albeit with some slight deviation in the tails, and some skewness in the chi-squared model.  In Table~\ref{tab:SSperim-prH0}, the Gaussian, anisotropic Gaussian, and chi-squared random fields, for which $\beta(p_1,p_2)>0$, reject $H_0$ with high probability, indicating high empirical power in these settings. For the Smith model, which lies on the boundary of the null hypothesis that $\beta(p_1,p_2)=0$,  the empirical rejection probability is above 0.05, indicating a small bias with respect to the nominal type 1 error appearing at the 0.05 significance level.  
\begin{table}[H] \small
    \centering
    \begin{tabular}{c|c|c|c|c}
    \hline \hline  
         Random field &  $(p_1, p_2)$ & $\beta(0.93,0.94)$  & $\beta$ & $\hat{\PP}(\text{reject H}_0)$  \\ \hline
         Gaussian & $(0.93, 0.94)$ & $0.612$ & $0.50$   & $1.00$ \\
         Anisotropic Gaussian & $(0.93, 0.94)$ &  $0.612$ & $0.50$ & $1.00$ \\
         Standardised chi-squared & $(0.93, 0.94)$ & $0.511$ &  $0.50$ & $1.00$ \\
          Smith  & $(0.93, 0.94)$ & $0.096$  & $0$ & $0.115$  \\
      \hline \hline   \end{tabular}
    \vspace{-0.23cm}
    \caption{Estimated probability of rejecting   hypothesis   $\text{H}_0 : \beta{(p_1, p_2)}\leq 0$, using 200 estimated  values  $\tilde \beta_{i, k}(p_1,p_2;m,\delta,S^\delta)$, at the $0.05$ significance level.}
    \label{tab:SSperim-prH0}
\end{table}    

\subsection{The quantile regression approach based on the LER}\label{sec:qr-beta}  
 
As an alternative to the perimeter-based approach presented in Section \ref{sec:perim-approach}, we also consider an extension to the estimation approach for $\beta$ presented in \cite{Costakis2025Extent} based on observations of the LER.
Letting $q_{a}(R_{s_0}^{(p)}) = \inf \{r>0\,:\,\PP(R_{s_0}^{(p)}<r)=a \}$ be the $a$-quantile of the LER at threshold level $p$ and location $s_0\in S$, Theorem 1 of \cite{Costakis2025Extent} states that
\begin{equation}
    \label{eq:eq-cond-v2}
    \ln q_{a}(R_{s_0}^{(p_2)})\sim \ln  q_{a}(R_{s_0}^{(p_1)}) - \beta\left[\ln(-\ln(1-p_2))-\ln(-\ln(1-p_1))\right]
\end{equation}
as $\min\{p_1,p_2\}\rightarrow 1$.
By fixing $p_1$ to be moderately high and allowing $p_2$ to vary, Equation \eqref{eq:eq-cond-v2} can be reformulated into a linear regression  procedure
\begin{equation}
    \label{eq:lin-fn}
    \ln q_{a}(R_{s_0}^{(p)})=\beta_0 - \beta\ln(-\ln(1-p)),
\end{equation}
thus making it possible to estimate the scaling index $\beta$ through the estimated coefficients in the linear function in \eqref{eq:lin-fn}.
\cite{Costakis2025Extent} estimate  $\beta$ by taking several observations of $R_{s_0}^{(p)}$ for a range of threshold levels $p$, obtaining response observations for $\ln q_{a}(R_{s_0}^{(p)})$ by taking the $a$ quantile at each $p$, then solving the linear regression problem in \eqref{eq:lin-fn}.
In this work, we instead directly model quantiles of $R_{s_0}^{(p)}$ by performing quantile regression at the $a$ level via minimization of the check function,
\begin{equation}\label{eq:qr-lik}
    L(\bm{\beta};\cdot) = \sum\limits_{i : r_i(s_0;p)>0}\left[\ln r_i(s_0;p) - f(\cdot;\bm{\beta})\right]\left[a - \mathbbm{1}_{(-\infty,0)}(\ln r_i(s_0;p) - f(\cdot;\bm{\beta}))\right],
\end{equation}
where $r_i(s_0;p)$ are observed values of the LER at location $s_0\in S$ and threshold level $p$,   and $f(\cdot;\bm{\beta})$ is a linear predictor of the response.  This  minimization can be done efficiently using the \texttt{rq} function in the \texttt{quantreg} package \citep{quantregRpackage}, and is advantageous because it estimates the conditional quantile relationship from the individual LERs rather than first reducing the data to a collection of empirical quantiles. The standard choice is given by the right-hand side of \eqref{eq:lin-fn}; however, another advantage of our quantile regression procedure is the ease with which additional covariates and interaction terms can be introduced, depending on the available dataset.  

\smallskip
\textbf{Numerical studies for \boldmath{$\beta$} based on the LER.}  For each of the stationary random fields described in Section~\ref{sec:examples}, we implement the quantile regression procedure in order to estimate $\beta$ on observations from the four models described in the perimeter-based simulations in Section~\ref{sec:perim-approach}. 
To ensure non-trivial LER observations at threshold level $u_p$, we simulate from conditional random fields $\{X(s)\mid X(s_0)>u_p\}_{s\in S}$ using \texttt{R} code accompanying \cite{Costakis2025Extent} for the Gaussian and chi-squared random fields. 
This simulation techniques involves direct computation of the Gaussian covariance matrix via the eigenvalue decomposition. 
This is computationally-expensive; therefore, we use a larger pixel width of $\delta= 0.2479$ on the discretisation of the domain $S=[-15,15]^2$.  In order to simulate from the conditional Smith model at high quantiles, we adopt a local approach. At threshold level $u_p$, the Poisson point process representation of the Smith model indicates that the number of contributions (or \emph{storms}) $\xi_i f(0;s_i,\sigma^2 I)$ at location 0 exceeding $u_p$ follows a Poisson distribution with rate ${1}/{u_p}$. Therefore, for $p$ approaching 1, we only need to generate a small number of storms exceeding $u_p$ at the origin. To do so, we first draw an observation $n_{u_p}$ from $\text{Poisson}({1}/{u_p})$. We then simulate locations $s_i$ from $N(0,\sigma^2 I)$ followed by magnitudes $\xi_i$ from  $\text{Pareto}({u_p}/{f(0;s_i,\sigma^2 I)},1)$  for $i=1,\dots,n_{u_p}$. Using this, we finally obtain an observation $X(s)=\max_{i\in\{1,\dots,n_{u_p}\}}\xi_i f(s;s_i,\sigma^2 I)$ that ensures $X(0)>u_p$.

For the Gaussian random field used to generate Gaussian, anisotropic Gaussian, and chi-squared observations, we keep the same scale parameter $\sigma^2$ of 1 as in the perimeter-based simulation studies in Section~\ref{sec:perim-approach}, but we increase the range parameter $l$ of the squared-exponential covariance function from 1 to 9 to avoid pixelisation effects for conditional excursion sets at high threshold probability levels $p$. Similarly, we increase the marginal variance $\sigma^2$ of the bivariate Gaussian density in the Smith model from $1$ to 9. This results in sensible LER values, rather than observations of $E_p^X(0)$ being a single pixel at the origin.

\begin{figure}[ht!]
    \centering    \includegraphics[width=0.5\linewidth]{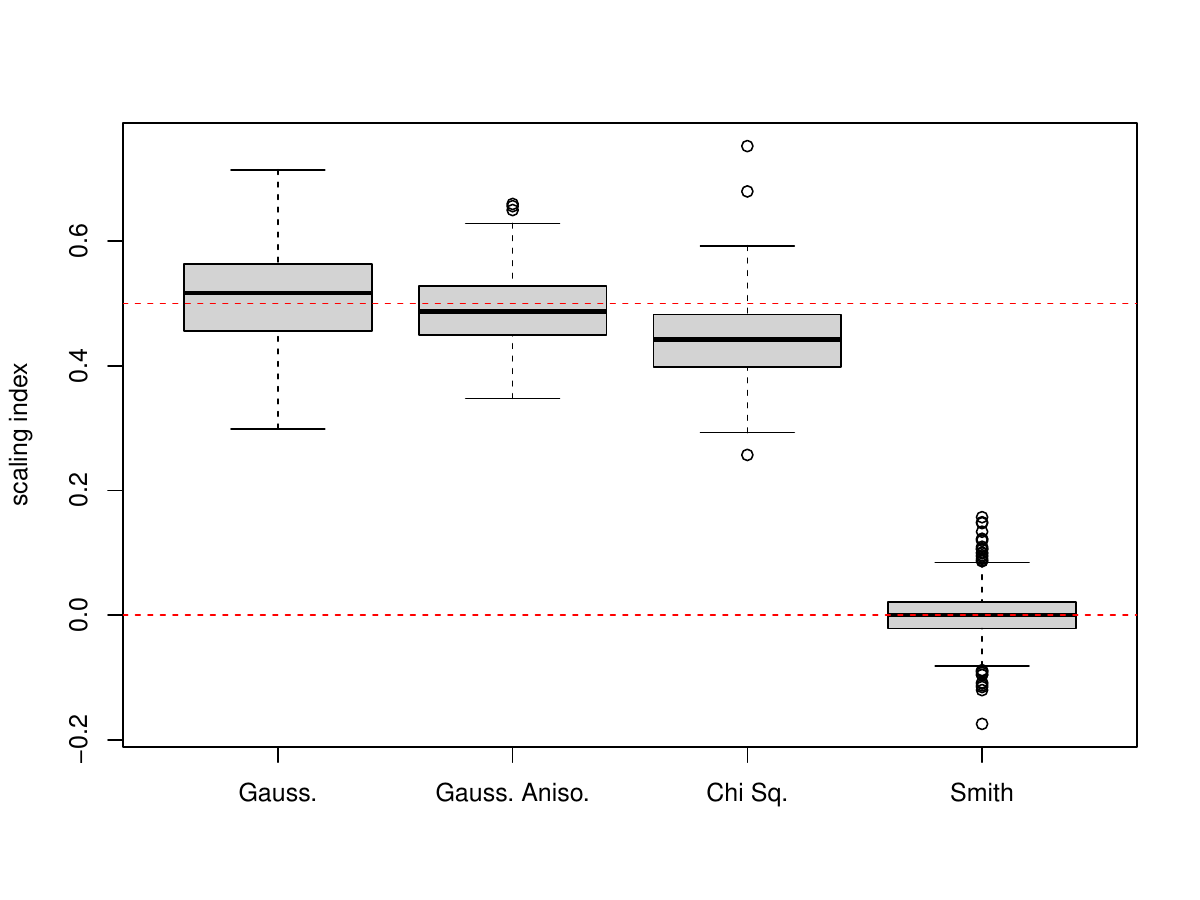}
\vspace{-0.6cm}
    \caption{Boxplot of estimated  $\beta$ using the LER method, with $\beta=0$ and $\beta=0.5$ highlighted by the dashed lines. }
    \label{fig:SSboxplotsLER}
\end{figure}

For a mesh of values $p$ of length 50, we simulate 50 observations of $X\mid \{X(s_0)>u_p\}$ on  $S^{\delta}$ at each threshold $p$.
The reference site is fixed to $s_0=(0,0)$, and the quantile  level in   \eqref{eq:eq-cond-v2} is set to $a=0.50$. We sample 2500 values of the pair $(-\ln(-\ln(1-p)),\ln r(s;p))$ and we estimate $\beta$ using the quantile regression procedure at level  $a$.
To assure equal separation of the covariate $-\ln(-\ln(1-p))$, the values of $p$ are generated  using $p_j=1-\exp\{-\exp\{x_j\}\}$, where $x_1,\dots,x_{50}$ are equally-spaced values between 1.10 and 2.44. This is repeated 200 times, and the 200 resulting estimates of $\beta$ are shown  in Figure~\ref{fig:SSboxplotsLER}.  We see general agreement with the perimeter-based results displayed in Figure~\ref{fig:SSboxplots-perim}; with some  bias in estimates of $\beta$ for the chi-squared random field, but with very good accuracy for the Smith model.     

\section{Application to French precipitation data}\label{sec:realdataSection}

We investigate the index $\beta$ to describe the locality of heavy rainfall across France by applying the perimeter- (see  Section~\ref{sec:perim-approach}) and LER-based techniques (see Section \ref{sec:qr-beta}) on COMEPHORE  
data \citep{Caillaud2019_COMEPHORE}. This reanalysis dataset, provided by M\'{e}t\'{e}o France, combines rain gauge observations and radar-sensed precipitation to give hourly measurements (in units $10^{-1}\text{mm}$) across  metropolitan France $S^\delta$ at grid size $\delta=1\,\text{km}$. We consider the period  between 1 August 2024 and 1 April 2026  resulting in data points at $n=14588$ time steps over a grid of size $1536\times1536$; by abuse of language, we subsequently refer to these data as observations. 
 
\begin{figure}[H]
    \centering
    \includegraphics[width=0.24\linewidth,trim={50 30 30 50},clip]{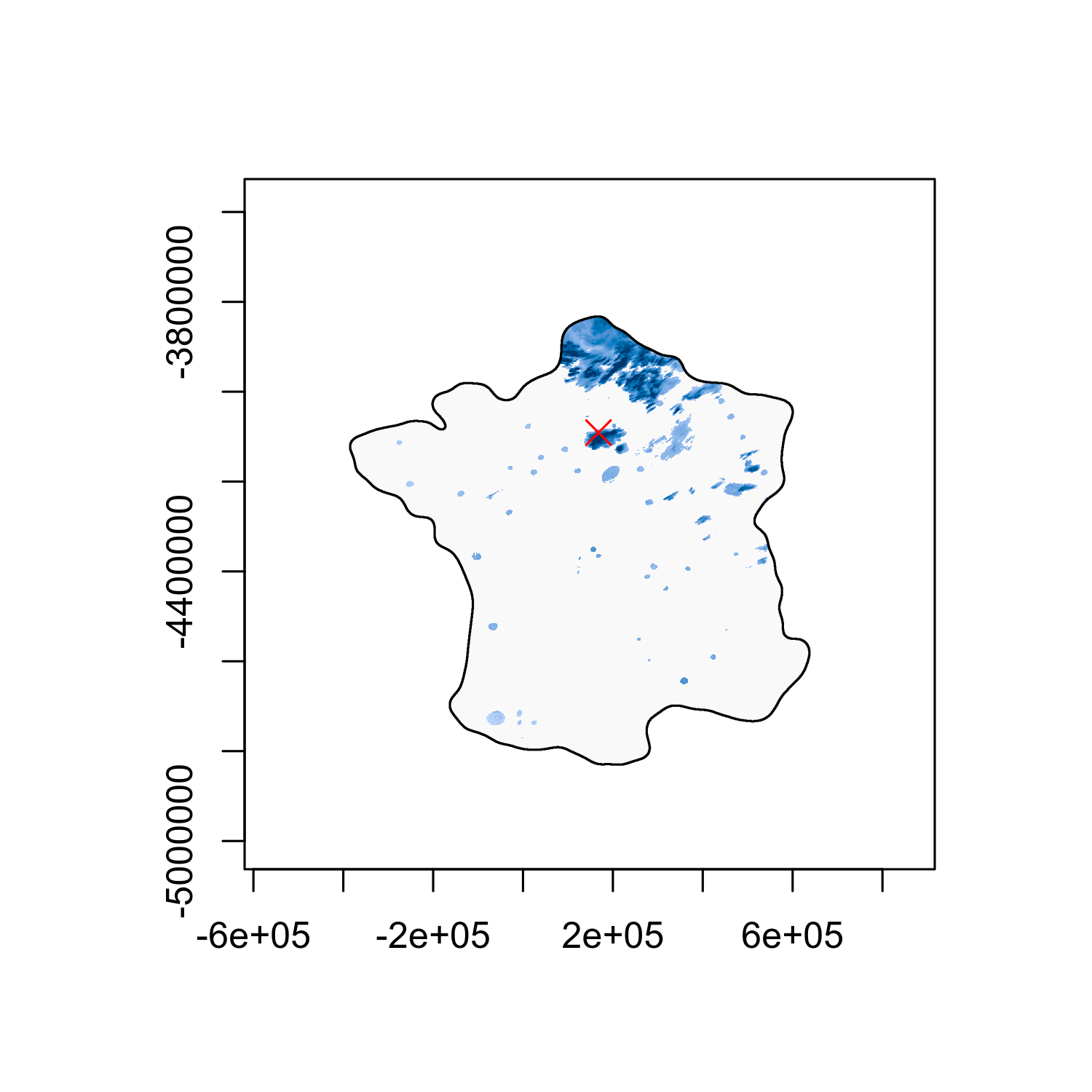}
    \includegraphics[width=0.24\linewidth,trim={50 30 30 50},clip]{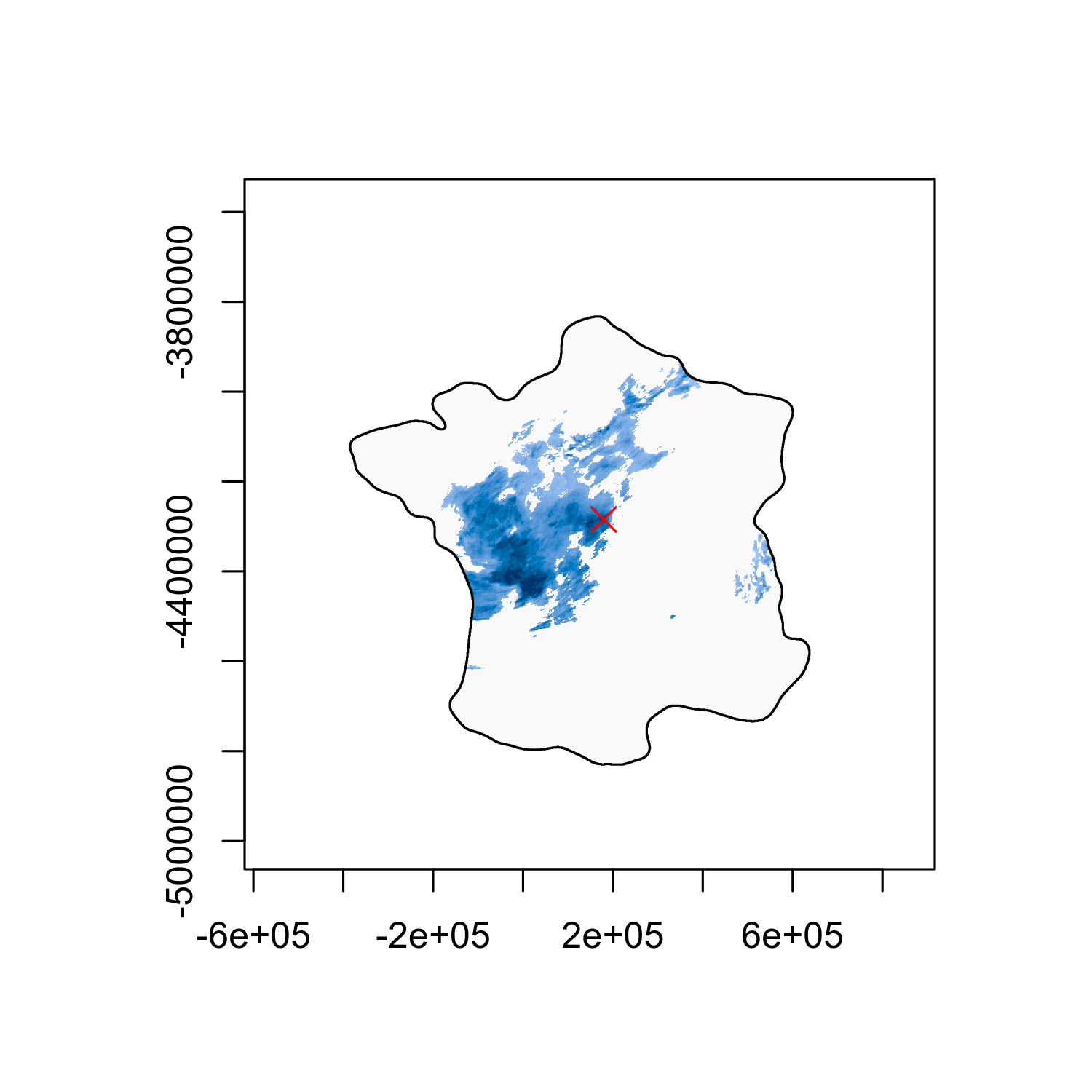}
    \includegraphics[width=0.24\linewidth,trim={50 30 30 50},clip]{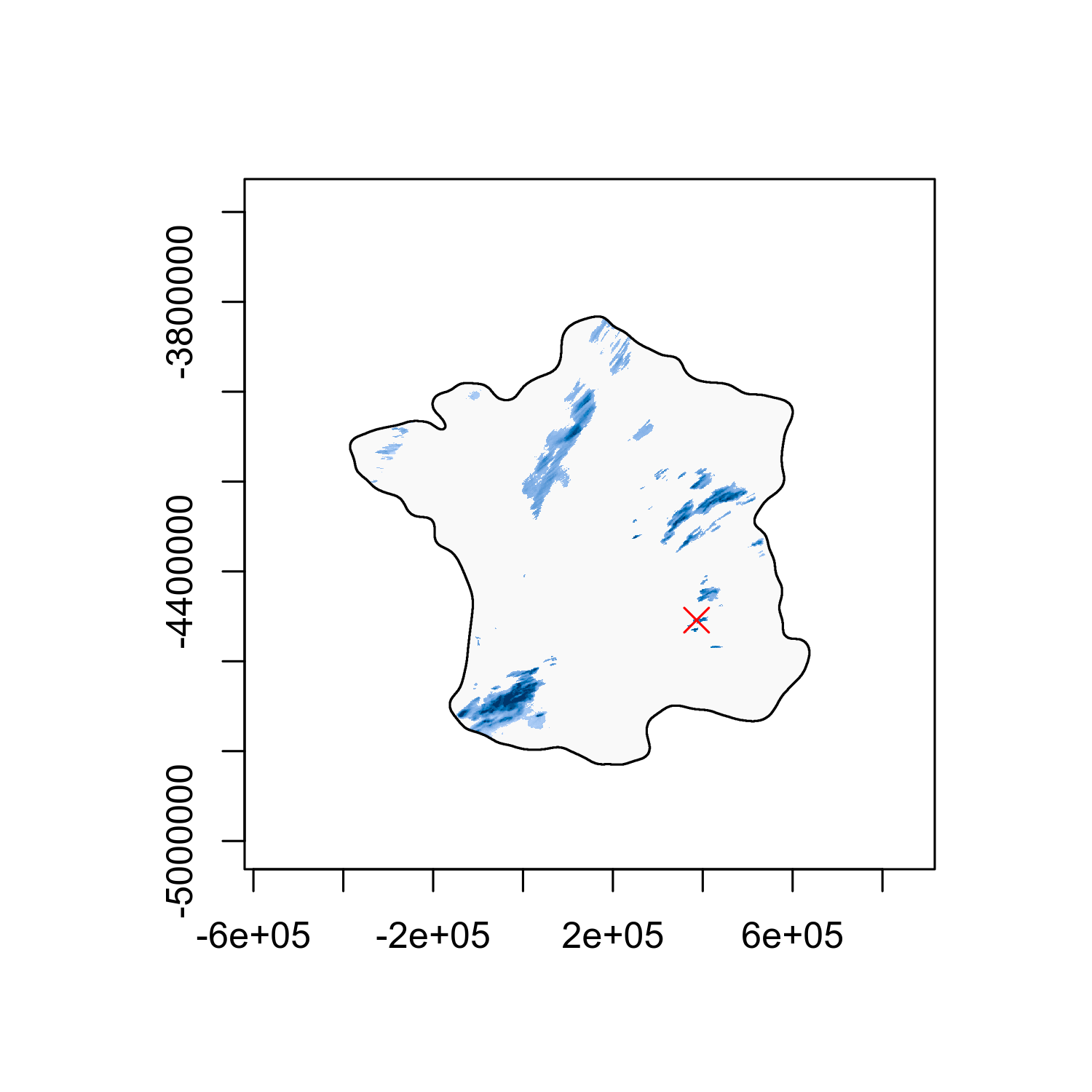}
    \vspace{-0.3cm}
    \caption{Observations of hourly precipitation conditioning on locations $s_0$ of interest exceeding the  $p=0.995$ threshold for Paris, Bourges, and Valence (left to right). The ``$\times$'' symbol represents the locations of interest, and values  are standardized to exponential margins, with darker colour indicating higher precipitation.}
    \label{fig:french-precip-data}
\end{figure}

Denote observations on the original scale at grid point $s_j$ by $X_o(s_j)$.
The data is standardised pointwise to have unit-exponential margins (above some relatively low threshold)  using the probability integral transform $X(s_j)=-\ln(1-F_{o,j}(X_o(s_j)))$, where  $F_{o,j}$ is the distribution function associated with $X_o(s_j)$. Values below the threshold represent absence of rainfall; we take care to conduct the excursion-set analysis using only threshold levels $u_p$ above this threshold.   Due to the unavailability of the true $F_{o,j}$, we approximate it using the empirical counterpart $\hat{F}_{o,j}(X_o(s_j)$  with
$\hat{F}_{o,j}(x)=n^{-1}\sum_{i=1}^n \mathbbm{1}_{(-\infty,x]}(x_i)$.
By the dense nature of the gridded domain $S^\delta$, standard tools for assessing extremal spatial dependence in $\{X(s)\}_{s\in S}$ , such as through the coefficient $\lim_{p\rightarrow1}\chi_p(s_0,s)$, become less feasible.  We highlight the computational advantages of excursion-based statistics and focus on  three pixels representing our locations (cities) of interest: $s_0 \in \{\text{Paris, Bourges, and Valence}\}$. 
Three observations of the standardised data fields are displayed in Figure~\ref{fig:french-precip-data}, each representing an exceedance above the $p=0.995$ threshold for a different conditioning site $s_0$.
By the geophysical nature and the size of France, the stationarity assumption in \ref{A1} is no longer plausible.  Therefore, we distinguish the scaling index for each $s_0$ and adapt estimation to subdomain-specific $\beta(s_0)$. This intuitively describes the rate of change of excursion sets conditioning on location $s_0$ lying in the excursion set.  The order of estimates will give intuition as to which region has more localised extreme rainfall and which are more wide-reaching.
\vspace{-0.2cm}

Perimeter-based estimation of $\beta(s_0)$ for this dataset
uses the perimeter of
the connected component of the excursion set containing the location $s_0$.
It is important to note that while the perimeter-based
approach in Section~\ref{sec:perim-approach} is constructed
for the unconditional setting, we show that the ratio of conditional
and unconditional expected perimeters of the full
excursion set converges to one on expanding domains,
for each fixed $p$, under suitable spatial
decorrelation conditions. 
The theoretical details are provided in 
Supplementary Material~\ref{supp:cond-perims}, where the  stationarity 
is crucial. Because stationarity is not assumed in this application,
we use a local conditional adaptation of the
perimeter-based estimator in
Section~\ref{sec:perim-approach},
based on the connected excursion component
containing the conditioning location $s_0$.

In practice, we compute the conditional scaling index $\hat{\beta}(s_0)$ by obtaining the perimeter of excursion sets conditioned on $X(s_0)>u_p$ via
\begin{align*}
    \hat{\beta}_{(p_1,p_2)}(s_0):=  &
\frac{\ln\left(
{\hat{C}^{(m,\delta,S^\delta)}_1(s_0,X,p_1)}/{\hat{C}^{(m,\delta,S^\delta)}_1(s_0,X,p_2)}\right) - \ln\left(\frac{1-p_1}{1-p_2}\right)
}{\ln\left({\ln(1-p_1)}/{\ln(1-p_2)}
\right)},
\end{align*} 
with the term $\hat{C}^{(m,\delta,S^\delta)}_1(s_0,X,p)$ obtained through the estimated perimeter of the connected component of $E_p^X(s_0)$ containing $s_0$.
Without knowledge of the true value of $\beta$, it is difficult to know which set of thresholds $(p_1,p_2)$ to use. To remedy this, we consider an equally-spaced mesh of $p$ values in the range $[0.900,0.999]$. For every pair $(p_1,p_2)$ in this mesh, we estimate a total of 105 values $\hat{\beta}_{(p_1,p_2)}(s_0)$, with Figure~\ref{fig:beta-hist-comephore} in Section~\ref{supp:SSfigs}    of Supplementary Material displaying histograms of all such estimates.
The variability of estimates $\hat{\beta}(s_0)$ displayed in Figure~\ref{fig:beta-hist-comephore} in Supplementary Material further motivates taking the median across the 105 values. The final median estimates are displayed in Table~\ref{tab:perim-beta-est-allFrance}.


In addition to the perimeter-based approach, LER-based estimates of $\beta$ are obtained via the quantile regression procedure outlined in Section~\ref{sec:qr-beta}.
For the same regular mesh of values $p\in[0.900,0.999]$ used in the perimeter-based analysis, we compute the empirical values of the LER, $r(s_0;p)$, at each of the three locations of interest  $s_0$. 
Table~\ref{tab:perim-beta-est-allFrance} shows the estimates obtained when minimizing the check loss \eqref{eq:qr-lik} with the linear function $f$ and the quantile set to $a=0.5$, meaning we model the median LER values. The estimated scaling indices  in Table~\ref{tab:perim-beta-est-allFrance} are positive across all locations and estimation methods, providing evidence of increasingly localised extreme rainfall events at high thresholds, consistent with an asymptotically independent regime. 
This is an expected feature of rainfall data, as is shown in  \cite{davison2013geostatistics}, for example. All confidence intervals are located entirely to the right of the value $0.50$, so that the spatial scale of excursion sets decreases at rates systematically faster than in the setting of Gaussian or other common asymptotically independent random fields (cf. Table~\ref{tab:beta-vals}). 
We further observe that the ordering of estimated scaling indices across estimation methods indicates that extreme rainfall events in Paris and Bourges show the strongest tendency towards localisation, followed by Valence, where extreme events are more widespread at the same high threshold.

 \begin{table}[h!]
    \centering
    \begin{tabular}{c|c|c}
        $s_0$ & $\hat{\beta}_{(p_1,p_2)}(s_0)$ (perim.) & $\hat{\beta}(s_0)$ (LER) \\ \hline
        Paris & 1.17 \ \ (0,92,1.61) & 1.65  \ \ \small{(1.36,1.94)} \\
        Bourges & 1.22  \ \ \small{(0.77,1.58)}  & 1.42  \ \ \small{(1.18,1.67)} \\
        Valence & 0.88  \ \ \small{(0.62,1.60)}  &0.94  \ \ \small{(0.66,1.23)}
    \end{tabular}
    \caption{Estimates  for the perimeter-based and the LER-based inference regimes, with 95\%  confidence intervals based on a 3-day block bootstrap approach.
    The estimates for the perimeter-based approach are obtained by taking the median across estimates obtained by all pairs of $(p_1,p_2)$ selected in a mesh between $0.900$ and $0.999$.
    }
    \label{tab:perim-beta-est-allFrance}
\end{table}

Additionally, we may include covariates to the linear model in the LER-based scaling index estimation procedure. For instance, suppose we wish to study the role that seasonality plays in the excursion set evolution of our precipitation data (keeping marginal distributions stationary, i.e., we capture all seasonal differences in the excursion geometry). To do so, we  minimize  \eqref{eq:qr-lik} with linear function
$
    f(p,t;\beta_0,\beta_1,\beta_2) = \beta_0 - \beta_1\ln(-\ln(1-p)) + \beta_2 t,
$ 
where $t$ is a binary covariate with 0 indicating winter (1 October -- 31 March) and 1 indicating summer (1 April -- 30 September).
Table~\ref{tab:beta-est-allFrance} shows the resulting parameters for this model setup.  We see that the extent of extreme rainfall events is generally lower in the summer for the northern locations of Paris and Bourges while in the south, this seasonal difference is less pronounced. Overall, Valence shows the least variation in excursion geometry across different threshold levels $p$ and seasons. Linear predictors with interaction between time and scaling index were also tested, but the interaction effect was statistically insignificant at all locations.
\begin{table}[H]
    \centering
    \begin{tabular}{c|c|c|c}
        $s_0$ & Intercept & $\hat{\beta}(s_0)$ (LER) & Seasonal effect \\ \hline
        Paris & 10.64 \small{(10.31,10.97)}   &    1.53 \small{(1.24,1.80)}  &    $-0.40$ \small{($-0.54$,$-0.31$)} \\
        Bourges & 10.50 \small{(10.24,10.81)} &      1.32 \small{(1.12,1.56)} &    $-0.48$ \small{($-0.58$,$-0.37$)}  \\
        Valence & 9.86 \small{(9.51,10.20)}   &     0.89 \small{(0.61,1.19)}  &    $-0.21$ \small{($-0.30$,$-0.07$)} 
    \end{tabular}
    \caption{Quantile regression-based estimates for a linear predictor with log-LER as response and an indicator of season (summer, encoded as $1$, vs. winter, encoded as $0$) as covariate, with 95\% 3-day block bootstrap confidence intervals.  
    }
    \label{tab:beta-est-allFrance}
\end{table}

\begin{figure}[ht!]     \includegraphics[width=0.33\linewidth,trim={10 10 10 10},clip]{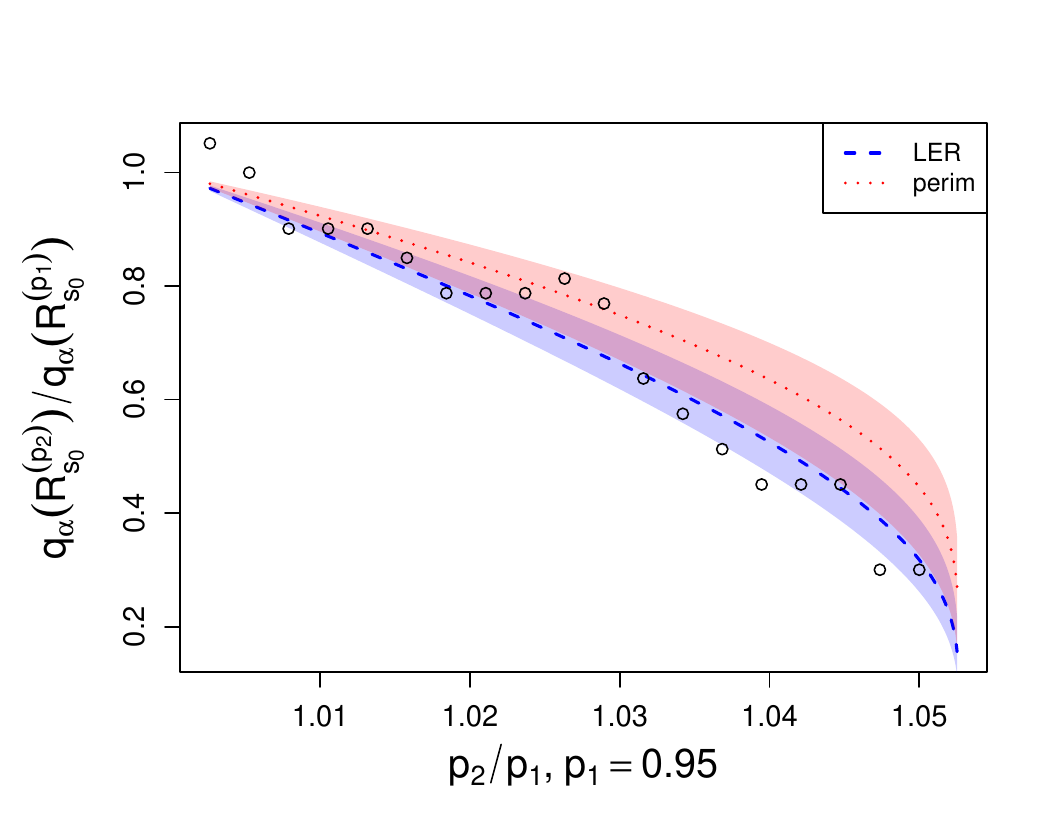}    \includegraphics[width=0.33\linewidth,trim={10 10 10 10},clip]{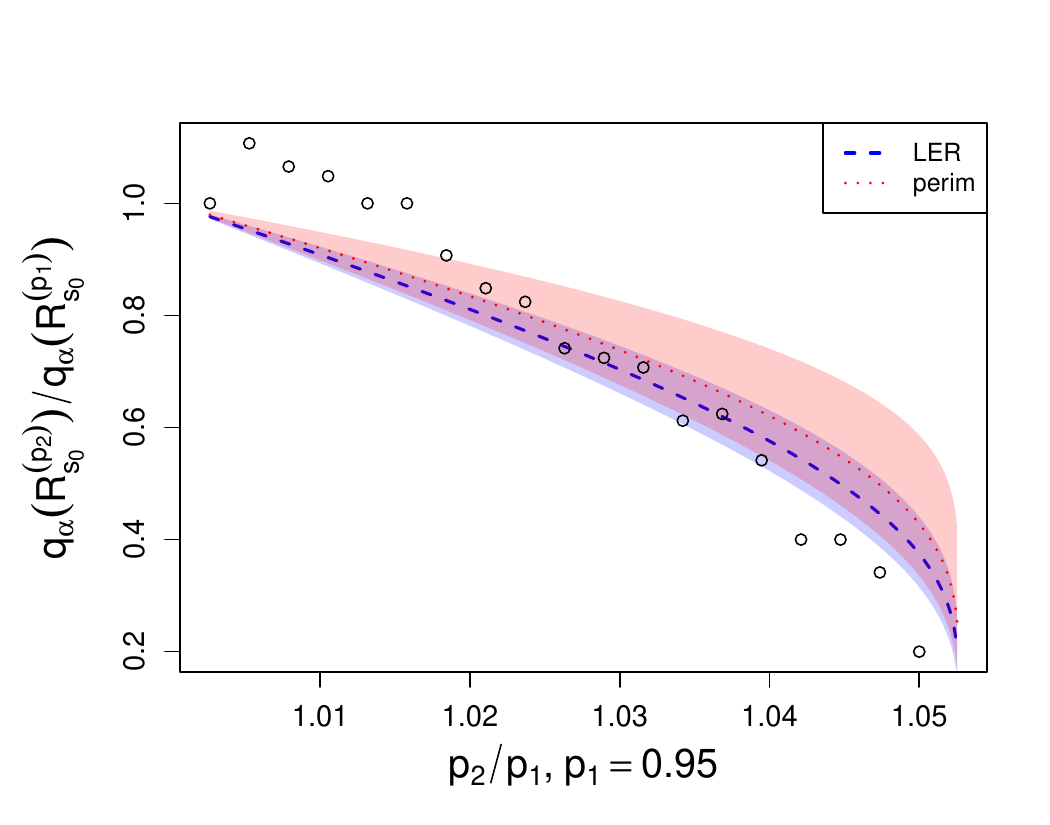}    \includegraphics[width=0.33\linewidth,trim={10 10 10 10},clip]{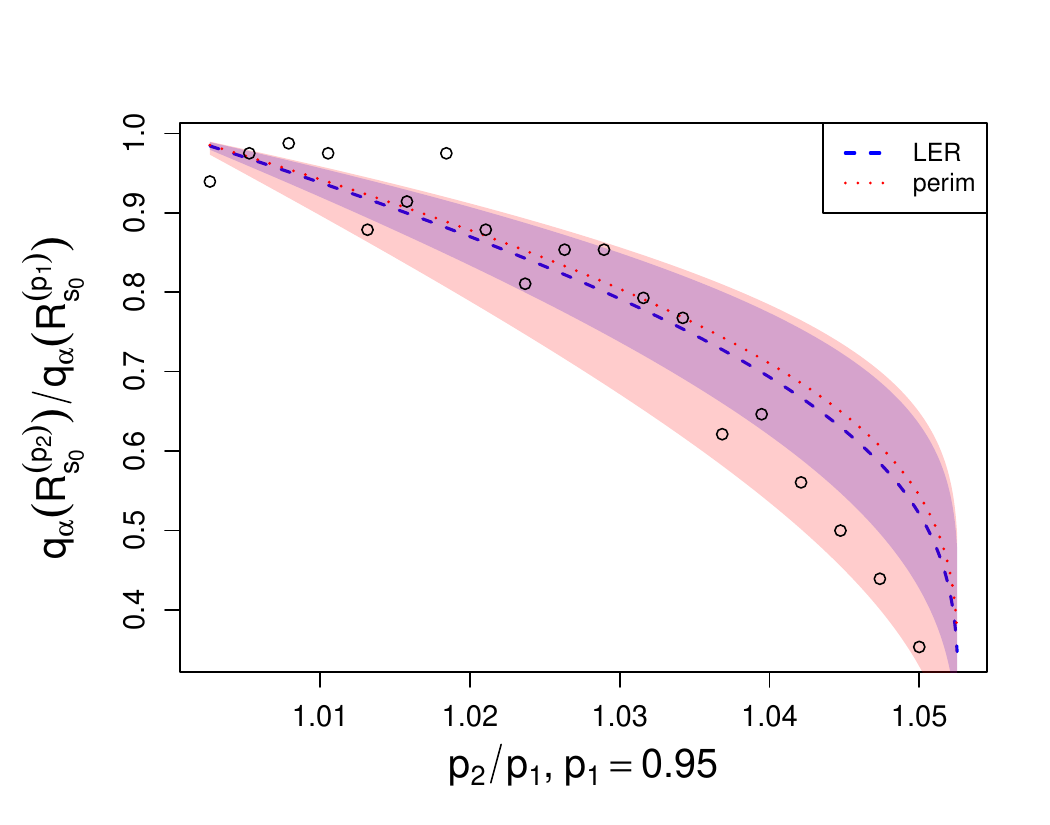}
  \vspace{-0.52cm}
    \caption{ 
    Extrapolation of the median LER up to the $p_2=0.9999$ threshold on COMEPHORE data using estimated values of $\beta(s_0)$ given in Table~\ref{tab:perim-beta-est-allFrance}, with dotted line corresponding to perimeter-based estimates, and dashed line corresponding to the quantile-regression approach. Conditioning locations $s_0$ are Paris, Bourges, and  Valence (left to right). Points represent empirical estimates obtained by taking the median of LER values on a mesh of thresholds between $0.95$ and $0.999$. Shaded regions correspond to 95\%  confidence intervals using a 3-day block bootstrap approach.}
    \label{fig:Comephore-LER-extrapolation}
\end{figure}

Estimates of $\beta$ can be used to extrapolate the behaviour of excursion sets, and of related coefficients satisfying distance scaling \citep{Costakis2025Extent}, to high yet unseen levels of the threshold. Consider the median LER as an example. The asymptotic equivalence in \eqref{eq:eq-cond-v2} shows that, at a fixed but high level of $p_1$, one could extrapolate its value to higher unseen threshold values $p_2$. Fixing a level $p_1=0.95$, we plot in Figure~\ref{fig:Comephore-LER-extrapolation} a selection of observed pairs  
$
    \big(p_2/p_1,  q_{a}(R_{s_0}^{(p_2)})/q_{a}(R_{s_0}^{(p_1)})\big)
$
for each of the  three locations of interest $s_0$. 
We also plot the predicted values  
$\big(p_2/p_1, \left(\ln(1-p_2)/\ln(1-p_1)\right)^{-\beta} \big)$  
on a curve for estimates of $\beta$ from each of the two procedures.
Using this, we extrapolate to the threshold $p_2=0.9999$, well beyond any observed value of rainfall at each of the  three locations of interest in the dataset. Both the LER and perimeter-based approach show generally good agreement between empirical values and model-based curves and envelopes, particularly at higher levels of extrapolation. 

 \vspace{-0.75cm}
\section{Conclusion}\label{ConclusionSection}
Assessing extremal spatial dependence becomes increasingly challenging as the size and pixel resolution of the domain grows. To address this, we introduced a scaling index based on the decay rate of the size of excursion sets. The proposed index provides a parsimonious description of the localisation of extremes across space, allowing regions with more concentrated extremal behaviour at increasingly high thresholds to be distinguished from those exhibiting more diffuse extremes.
When estimated through excursion set perimeters, the scaling index further admits an asymptotically normal estimator, facilitating formal inference and the construction of a hypothesis test for asymptotic independence. Through simulation studies, we demonstrated that the proposed methodology accurately recovers the scaling behaviour in both asymptotically independent and asymptotically dependent settings.
Our application to precipitation data illustrates that the scaling index captures meaningful spatial variation in local extremal behaviour and can be used to extrapolate geometric characteristics of excursion sets beyond observed thresholds. The large scaling indices we estimate indicate that tail decay rates are faster than those of commonly used statistical models for spatial extremes, e.g. Gaussian copula models \citep[see the review of stochastic precipitation generators by][]{HandbookExtremesCh25}.

Several directions remain for future work.  Theorem \ref{thm:CLTbeta} concerns increasing-domain asymptotics for fixed threshold levels and therefore provides inference for the finite-threshold quantity \(\beta(p_1,p_2)\). Establishing asymptotic normality directly around \(\beta\) under a simultaneous regime \(p_{1,n},p_{2,n}\to1\) and \(\mathcal L_2(S_n)\to\infty\) would require a triangular-array extension of the multilevel perimeter CLT, together with uniform control of the threshold-dependent covariance terms and of the remainder in \eqref{eq:C1form}, and represents a challenging direction for future theoretical work. Instead of predicting  a specific quantile in the  regression  proposed for estimating the scaling index through the lower extremal range, we could extend inference from the conditional median to the full distribution. With the asymptotic Assumption \ref{A:scaling}, the scaling function would be proportional to a scale parameter of the response distribution, such that covariate effects can be flexibly incorporated. This extension would require identifying an appropriate response distribution. It could be formulated with survival regression to accommodate right censoring induced by edge effects at the domain boundary and to apply artificial left-censoring to reduce bias from discretisation effects for small responses. It could also provide more precise diagnostic procedures to assess asymptotic assumptions, specifically whether the chosen scaling function indeed yields non-degenerate excursion-set limits. Establishing such model checks could further strengthen the practical  applicability.
 \vspace{-0.5cm}

\section*{Supplementary Material}
In  Supplementary Material, Section \ref{ProofsSection} includes proofs of the results in the main text.   Section \ref{AppenidxDEF} contains further definitions, figures and auxiliary results. Detailed calculations for Section \ref{sec:examples} (\emph{resp.}  Section \ref{sec:COUNTERexamples}) are given in Section \ref{supp:calculations} (\emph{resp.}  Section \ref{supp:calculationsBIS}). In Section~\ref{supp:shot-noise}, we consider  shot noise random fields. The relationship between scaling index and tail dependence is discussed in Section \ref{supp:beta-dependence}, and 
  the tail behaviour of a specific type of random field (Gaussian-based scale mixture) and its effect on the scaling index is provided in Section \ref{supp:scale-mixtures}. Additional figures are postponed to Section~\ref{supp:SSfigs}.  Empirical assessment of Assumption~\ref{A:scaling} is presented in Section~\ref{supp:assumption-check}.  Finally, in Section~\ref{supp:cond-perims}, we provide  sufficient conditions for the ratio of the conditional and unconditional expected perimeters of excursion sets to converge to one
as the domain tends to  $\mathbb{R}^2$.\vspace{-0.5cm}

\section*{Declaration of the use of generative AI and AI-assisted technologies}
During the preparation of this work, the authors used generative AI tools (OpenAI ChatGPT GPT-5.6 Sol, high reasoning effort, September 2026) for the development of Assumption~\ref{A:scaling} and the proof of Theorem~\ref{thm:LKCratio}. The authors reviewed and edited the content as necessary and take full responsibility for the content of the publication.

\section*{Acknowledgement and code availability}  This work has been supported by the  French Agence Nationale de la Recherche (ANR), under grant ANR-22-EXIR-0002 (project COCHAIR) within PEPR Risques, PEPR Platforme (ANR-22-EXIR-0008) and project LessBig (ANR-25-CE40-1607). Thomas Opitz further acknowledges support from the projects EXSTA (ANR-23-CE40-0009) and CLIMATHS (ANR-23-EXMA-0003, one of the target projects of PEPR Maths VivES), as well as from the Chaire Geolearning funded by Andra, BNP-Paribas, CCR and SCOR Foundation (https://chaire-geolearning.org).
Code for reproducing the results and figures in this manuscript can be found at \hyperlink{https://github.com/ryancampbell514/ExcursionDependence}{https://github.com/ryancampbell514/ExcursionDependence}.

 \vspace{-0.5cm}
 
\clearpage
  \appendix
\providecommand{\theHproposition}{}
\renewcommand{\theHproposition}{supp.\arabic{proposition}}
\providecommand{\theHtheorem}{}
\renewcommand{\theHtheorem}{supp.\arabic{theorem}}
\providecommand{\theHlemma}{}
\renewcommand{\theHlemma}{supp.\arabic{lemma}}
\providecommand{\theHcorollary}{}
\renewcommand{\theHcorollary}{supp.\arabic{corollary}}
\providecommand{\theHdefinition}{}
\renewcommand{\theHdefinition}{supp.\arabic{definition}}
\providecommand{\theHremark}{}
\renewcommand{\theHremark}{supp.\arabic{remark}}
\setcounter{remark}{0}
\renewcommand{\theHequation}{supp.\arabic{equation}}
\renewcommand{\theHfigure}{supp.\arabic{figure}}
\renewcommand{\theHtable}{supp.\arabic{table}}

\section*{\large Supplementary Material of ``Analysis of spatial extreme events via excursion set geometry''}
\label{SM} 

\begin{center}
    Ryan Campbell, Elena Di Bernardino, Thomas Opitz
\end{center}

\counterwithout{figure}{section}
\counterwithout{table}{section}
\counterwithout{equation}{section}
\counterwithout{proposition}{section}
\counterwithout{theorem}{section}
\counterwithout{lemma}{section}
\counterwithout{corollary}{section}
\counterwithout{definition}{section}

\setcounter{figure}{0}
\setcounter{table}{0}
\setcounter{equation}{0}
\setcounter{proposition}{0}
\setcounter{theorem}{0}
\setcounter{lemma}{0}
\setcounter{corollary}{0}
\setcounter{definition}{0}

\renewcommand{\thefigure}{S.\arabic{figure}}
\renewcommand{\thetable}{S.\arabic{table}}
\renewcommand{\theequation}{S.\arabic{equation}}

\renewcommand{\theproposition}{S.\arabic{proposition}}
\renewcommand{\thetheorem}{S.\arabic{theorem}}
\renewcommand{\thelemma}{S.\arabic{lemma}}
\renewcommand{\theremark}{S.\arabic{remark}}
\renewcommand{\thecorollary}{S.\arabic{corollary}}
\renewcommand{\thedefinition}{S.\arabic{definition}}

\section{Proofs}\label{ProofsSection}
\begin{proof}[Proof of Theorem  \ref{thm:LKCratio}] 
By Assumption~\ref{A:scaling} and continuity of the distance functional
for the Fell topology,
$\tilde R_{s_0}^{(p)}
=\sigma(p)R_{s_0}^{(p)}
\ \xrightarrow{d}\
\tilde R_{s_0}$, when $p\to1.$
Hence, for every continuity point $r>0$ of the distribution function of
$\tilde R_{s_0}$,
$\mathbb P\big(\tilde R_{s_0}^{(p)}\le r\big)
\rightarrow
\mathbb P\big(\tilde R_{s_0}\le r\big).
$ 
The equicontinuity hypothesis at zero of the densities
$f_{\tilde R_{s_0}^{(p)}}$, together with the continuity of $f_{\tilde R_{s_0}}$ at zero, then yields $f_{\tilde R_{s_0}^{(p)}}(0)
\rightarrow
f_{\tilde R_{s_0}}(0)$.  For every fixed $p\in(0,1)$, Theorem~1 in
\cite{cotsakis2024spatial} under Assumption~\ref{A1} gives
$$
\lim_{r\downarrow0}
\frac{1}{r}
\mathbb P\left(R_{s_0}^{(p)}\le r\right)
=
2\,\frac{C_1^\ast(E_p^X)}
        {C_2^\ast(E_p^X)}.
$$
Since $\tilde R_{s_0}^{(p)}=\sigma(p)R_{s_0}^{(p)}$, it follows that
$ f_{\tilde R_{s_0}^{(p)}}(0)
=
\frac{2}{\sigma(p)}
\frac{C_1^\ast(E_p^X)}
     {C_2^\ast(E_p^X)}.
$ Therefore, for $p \to 1$
$
\frac{1}{\sigma(p)}
\frac{C_1^\ast(E_p^X)}
     {C_2^\ast(E_p^X)}
\rightarrow
\frac12 f_{\tilde R_{s_0}}(0)
=:c_X.
$ Hence the result.  
\end{proof}

\begin{proof}[Proof of Theorem \ref{thm:CLTbeta}]
For notational convenience, we  write for $j=1,2$,  
$C_{p_j}:=C_1^*(E^X_{p_j}),$ $
\hat C_{p_j,n}
:=
\hat C_1^{(m_n,\delta_n,S_n)}(X,p_j).$   Under Assumption~\ref{A1}, the multilevel central limit
theorem (see Theorem 2 in \cite{cotsakis2022perimeter}) for the half-perimeter estimators yields
\[
\sqrt{\mathcal{L}_2(S_n)}
\begin{pmatrix}
\hat C_{p_1,n}-C_{p_1}\\
\hat C_{p_2,n}-C_{p_2}
\end{pmatrix}
\xrightarrow{d}
\mathcal{N}_2
\left(
\begin{pmatrix}0\\0\end{pmatrix},
\Sigma(p_1,p_2)
\right),
\]
where
\[
\Sigma(p_1,p_2)
=
\begin{pmatrix}
\nu(p_1,p_1) & \nu(p_1,p_2)\\
\nu(p_1,p_2) & \nu(p_2,p_2)
\end{pmatrix},
\]
where for $i,j\in\{1,2\}$, 
$
\nu(p_i,p_j)
:=
\frac{1}{4}
\int_{\mathbb{R}^2}
H_s(p_i,p_j)\,{\rm d}s,
$ 
with 
\[
\begin{aligned}
H_s(p_i,p_j)
={} 
f_{X(0),X(s)}&(u_{p_i},u_{p_j})
\,
\mathbb{E}\!\left[
\|\nabla X(0)\|_2
\|\nabla X(s)\|_2
\,\middle|\,
X(0)=u_{p_i},\,X(s)=u_{p_j}
\right]
\\
&-
f_{X(0)}(u_{p_i})
f_{X(s)}(u_{p_j})
\,
\mathbb{E}\!\left[
\|\nabla X(0)\|_2
\,\middle|\,
X(0)=u_{p_i}
\right]
\mathbb{E}\!\left[
\|\nabla X(s)\|_2
\,\middle|\,
X(s)=u_{p_j}
\right].
\end{aligned}
\]
Consider the function $h:\mathbb{R}_+^2\to\mathbb{R}$ defined by
\[
h(x,y)
=
\frac{
\ln(x/y)
-
\ln\left\{(1-p_1)/(1-p_2)\right\}
}{
D(p_1,p_2)
},
\qquad
D(p_1,p_2)
:=
\ln\left\{
\frac{\ln(1-p_1)}{\ln(1-p_2)}
\right\}.
\]
By definition,
$
h(\hat C_{p_1,n},\hat C_{p_2,n})
=
\hat{\beta}(p_1,p_2;X,m_n,\delta_n,S_n),
$ whereas
$h(C_{p_1},C_{p_2})=\beta(p_1,p_2).
$
Moreover,
\[
\nabla h(x,y)
=
\frac{1}{D(p_1,p_2)}
\begin{pmatrix}
x^{-1}\\
-y^{-1}
\end{pmatrix}.
\]

The multivariate delta method therefore gives
\[
\sqrt{\mathcal{L}_2(S_n)}
\left[
\hat{\beta}(p_1,p_2;X,m_n,\delta_n,S_n)
-\beta(p_1,p_2)
\right]
\xrightarrow{d}
\mathcal{N}
\left(
0,
\nabla h(C_{p_1},C_{p_2})^\top
\Sigma(p_1,p_2)
\nabla h(C_{p_1},C_{p_2})
\right).
\]
Finally,
\[
\nabla h(C_{p_1},C_{p_2})^\top
\Sigma(p_1,p_2)
\nabla h(C_{p_1},C_{p_2})
=
\frac{1}{D(p_1,p_2)^2}
\left(
\frac{\nu(p_1,p_1)}{C_{p_1}^2}
+
\frac{\nu(p_2,p_2)}{C_{p_2}^2}
-
2\frac{\nu(p_1,p_2)}{C_{p_1}C_{p_2}}
\right),
\]
which is precisely $\upsilon^2(p_1,p_2)$ in  Theorem \ref{thm:CLTbeta}. 
\end{proof}

\begin{proof}[Proof of Corollary \ref{corr:beta-median} ]
By Theorem~\ref{thm:CLTbeta}, for each $i\in\{1,\ldots,k\}$,
\[
Z_n^{(i)}
:=
\frac{\sqrt{\mathcal L_2(S_n)}}
{\upsilon(p_1,p_2)}
\left\{
\hat\beta_n^{(i)}
-
\beta(p_1,p_2)
\right\}
\xrightarrow{d} Z_i,
\]
where $Z_i\sim\mathcal N(0,1)$. Since the $k$ replicates are independent, for every fixed $k$,
$\left(Z_n^{(1)},\ldots,Z_n^{(k)}\right)
\xrightarrow{d}
\left(Z_1,\ldots,Z_k\right),$ 
where $Z_1,\ldots,Z_k$ are independent standard Gaussian random
variables. By the continuous mapping theorem,
$\operatorname{median}_{i=1,\ldots,k} Z_n^{(i)}
\xrightarrow{d}
\operatorname{median}_{i=1,\ldots,k} Z_i
=:M_k$.  Since $k$ is odd, $M_k$ is the $((k+1)/2)$th order statistic of
$k$ independent standard Gaussian random variables, 
$\operatorname{Var}(M_k)
=
\frac{k!}
{\left\{\left(\frac{k-1}{2}\right)!\right\}^{2}}
\int_{-\infty}^{\infty}
x^2
\Phi(x)^{(k-1)/2}
\{1-\Phi(x)\}^{(k-1)/2}
\phi(x)\,dx
=
\tau_k^2.
$ 
Finally, the classical asymptotic variance of the sample median gives
$\tau_k^2\sim \frac{\pi}{2k},
 $ $ k\to\infty$.
\end{proof}

\begin{proof}[Proof of Proposition~\ref{prop:scaling-AI}]
By stationarity, take $s_0=0$. Fix
$s\in\mathbb R^2\setminus\{0\}$ and choose a compact set $K$
containing $0$ and $s$. Let $\epsilon>0$, and let $A$ be
a compact set provided by the assumption for this $K$ and
$\epsilon$.

Since $\sigma(p)\to\infty$ as $p\to1$, it follows that
$\|\sigma(p)s\|\to\infty$. Since $A$ is compact, there exists
$p_{\epsilon}\in(0,1)$ such that $\sigma(p)s\notin A$
for all $p>p_{\epsilon}$. Consequently, for such $p$,
\[
\{s\in E_p^X(0)\}
\subset
\left\{
\sigma(p)\bigl(E_p^X(0)\cap K\bigr)
\not\subset A
\right\}.
\]
It follows that, for $p$ sufficiently close to $1$,
\begin{align*}
\chi_p(0,s)
&=\mathbb P\left(s\in E_p^X(0)\right)\le
\mathbb P\!\left(
\sigma(p)\bigl(E_p^X(0)\cap K\bigr)
\not\subset A
\right)
\le\epsilon.
\end{align*}
Since $\epsilon>0$ is arbitrary, $\chi_p(0,s)\to0$.
Finally, stationarity gives
$\chi_p(t_0,t)=\chi_p(0,t-t_0)$,
so the result holds for every $t_0\ne t$.
\end{proof}

\section{Further definitions and auxiliary results} \label{AppenidxDEF}

 \begin{remark}[Topology for convergence in distribution in Assumption~\ref{A:scaling}]\label{topology}
Excursion sets $\tilde E_p^X(s_0)$ conditioned on exceedance at $s_0$
are random closed subsets of $\mathbb{R}^2$, and we have to define an
appropriate topology for convergence to a limit set. Because we define
exceedances of the form ``$\geq$'' and assume continuous sample paths
(see Assumption~\ref{A1}), all excursion sets $\tilde E_p^X(s_0)$ are
closed, and we can use the Fell topology on
\[
\mathcal{F}(\mathbb{R}^2)
=
\{A\subset\mathbb{R}^2 : A \text{ is closed}\}.
\]
As $p\to1$, $\tilde E_p^X(s_0)$ must intersect every open set
intersecting $\tilde E^X(s_0)$ while avoiding every compact set
disjoint from $\tilde E^X(s_0)$
(see Definition C.1 of \cite{molchanov2017theory} and
Definition 2.1.1 of \cite{schneider2008stochastic}).
Equipping $\mathcal{F}(\mathbb{R}^2)$ with its Borel $\sigma$-algebra
$\mathcal{B}(\mathcal{F}(\mathbb{R}^2))$ allows for the convergence
\[
\mathbb{P}\bigl(\tilde E_p^X(s_0)\in\cdot\bigr)
\xrightarrow{w}
\mathbb{P}\bigl(\tilde E^X(s_0)\in\cdot\bigr)
\]
on $\mathcal{F}(\mathbb{R}^2)$ as $p\to1$, where $\mathbb{P}$ refers
to the probability law when conditioning on $X(s_0)\geq u_p$
(by abuse of notation since the law changes with $p$) and
$\xrightarrow{w}$ denotes the weak convergence of probability measures.

Therefore, by Definition 1.7.1 of \cite{molchanov2017theory}, the
assumption of the existence of the convergence in distribution in
Assumption~\ref{A:scaling} is understood as the weak convergence of probability
laws of random closed sets on
$\bigl(\mathcal{F}(\mathbb{R}^2),
\mathcal{B}(\mathcal{F}(\mathbb{R}^2))\bigr)$. Notice that the spatial scaling Assumption~\ref{A:scaling} is related to
domain-scaled regular variation (DSRV, \cite{strokorb2026domain})
in that both assume location-scaling to obtain non-degenerate behaviour
of the random field. However, DSRV is stronger than Assumption~\ref{A:scaling}
since it assumes convergence of the entire rescaled random field,
while Assumption~\ref{A:scaling} solely assumes convergence of excursion sets,
a particular random-set functional.
\end{remark}

\begin{figure}[H]
    \centering
\includegraphics[height=110pt, trim={40 40 40 40},clip]{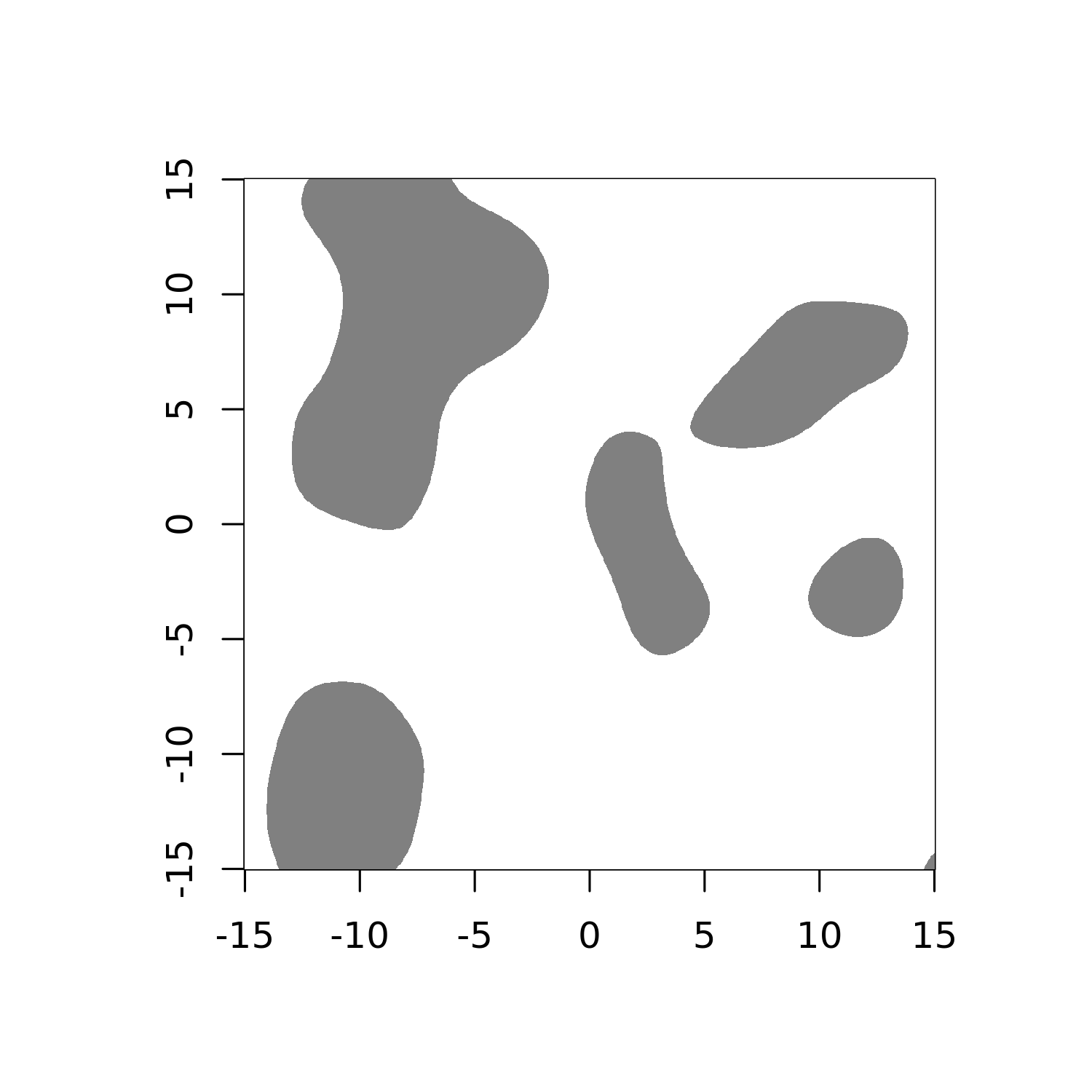}
    \includegraphics[height=110pt, trim={40 40 40 40},clip]{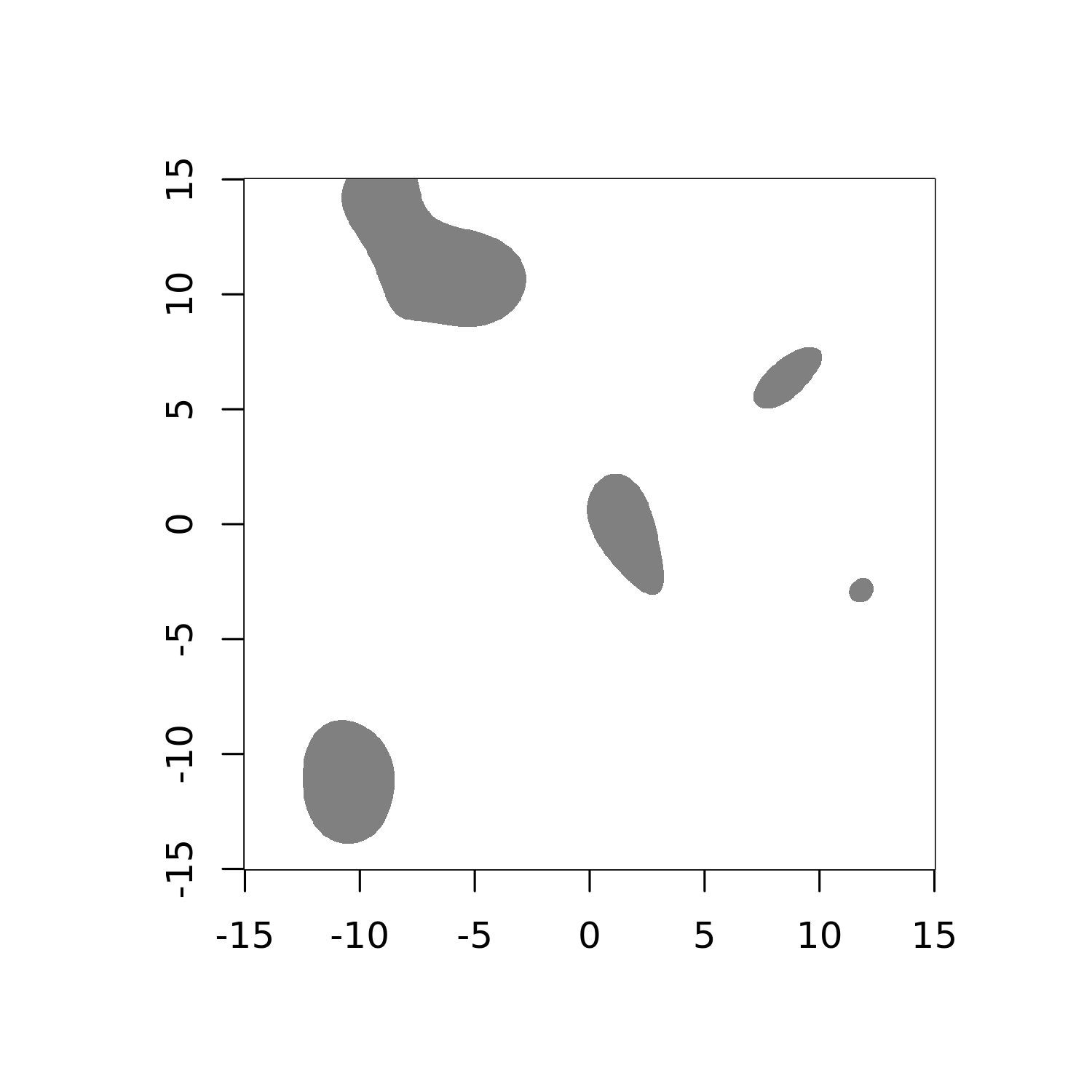}
    \includegraphics[height=110pt, trim={40 40 40 40},clip]{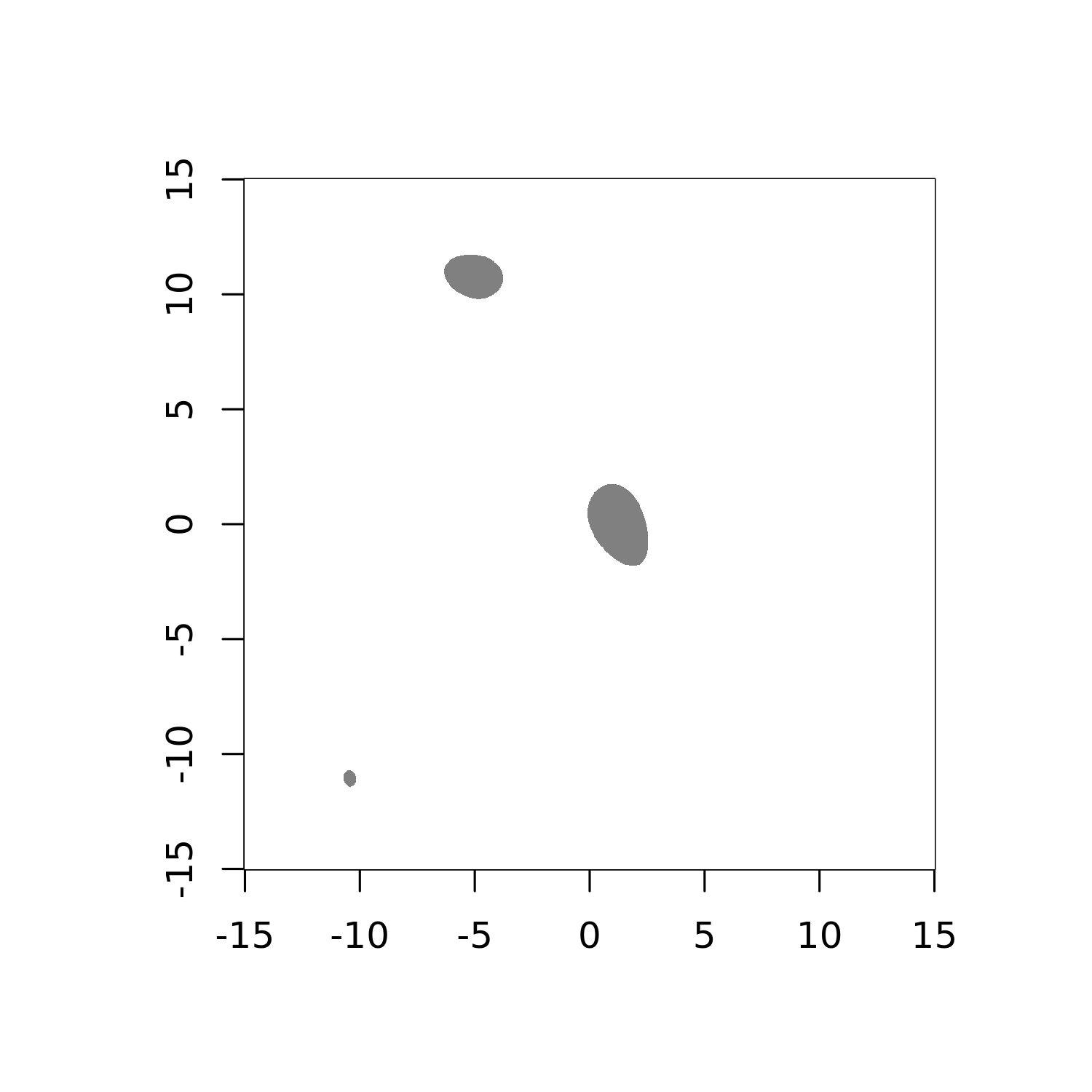}

    \includegraphics[height=110pt, trim={40 40 40 40},clip]{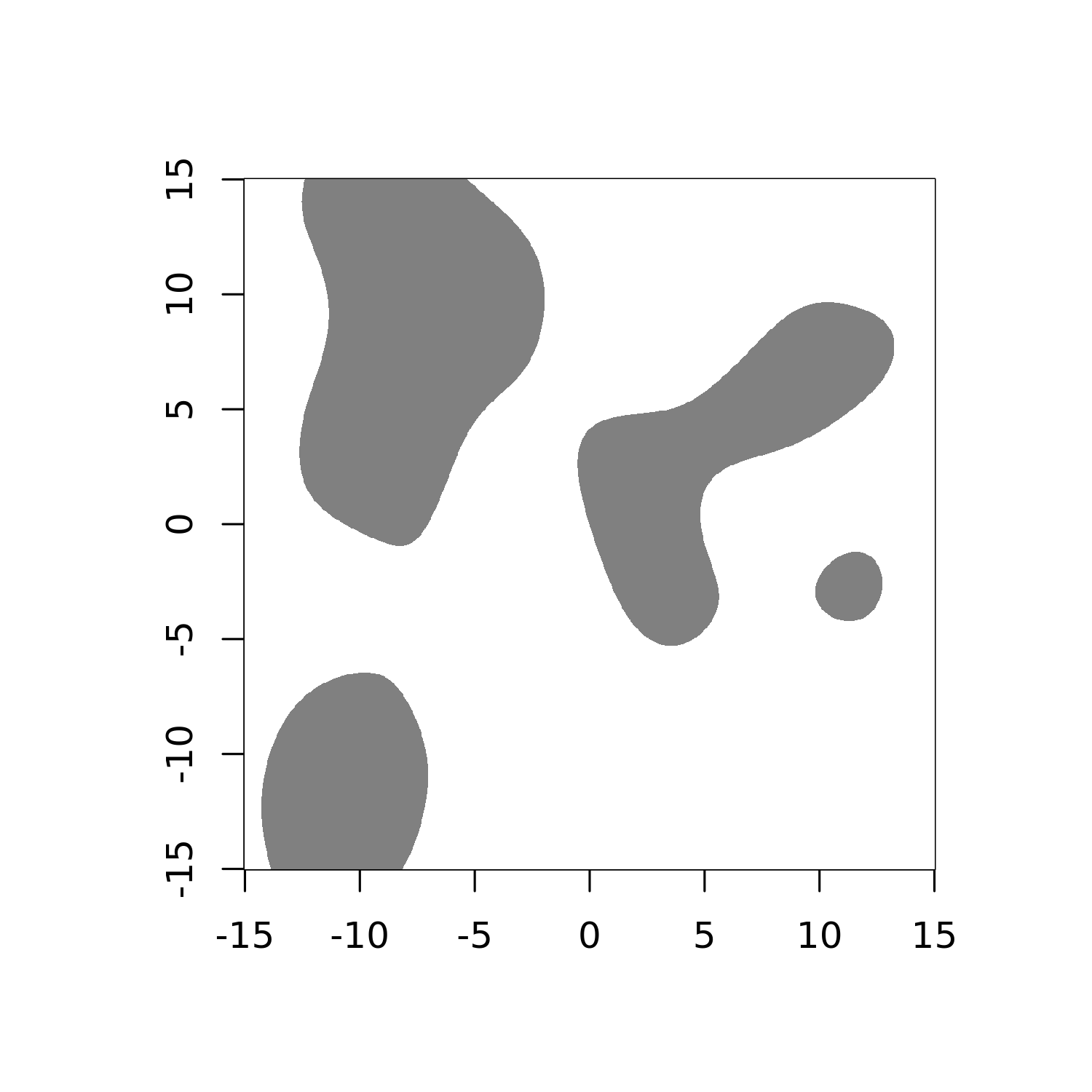}
    \includegraphics[height=110pt, trim={40 40 40 40},clip]{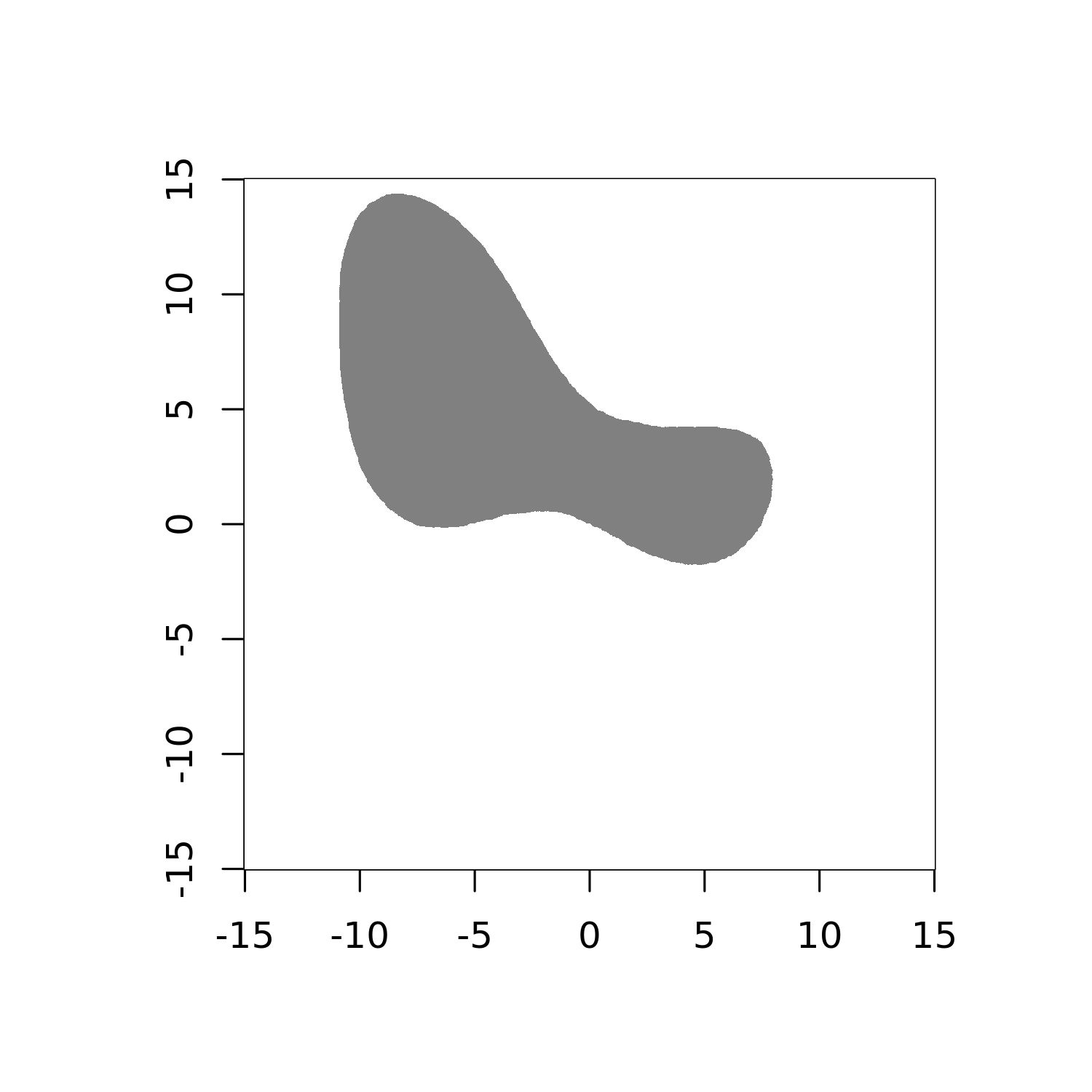}
    \includegraphics[height=110pt, trim={40 40 40 40},clip]{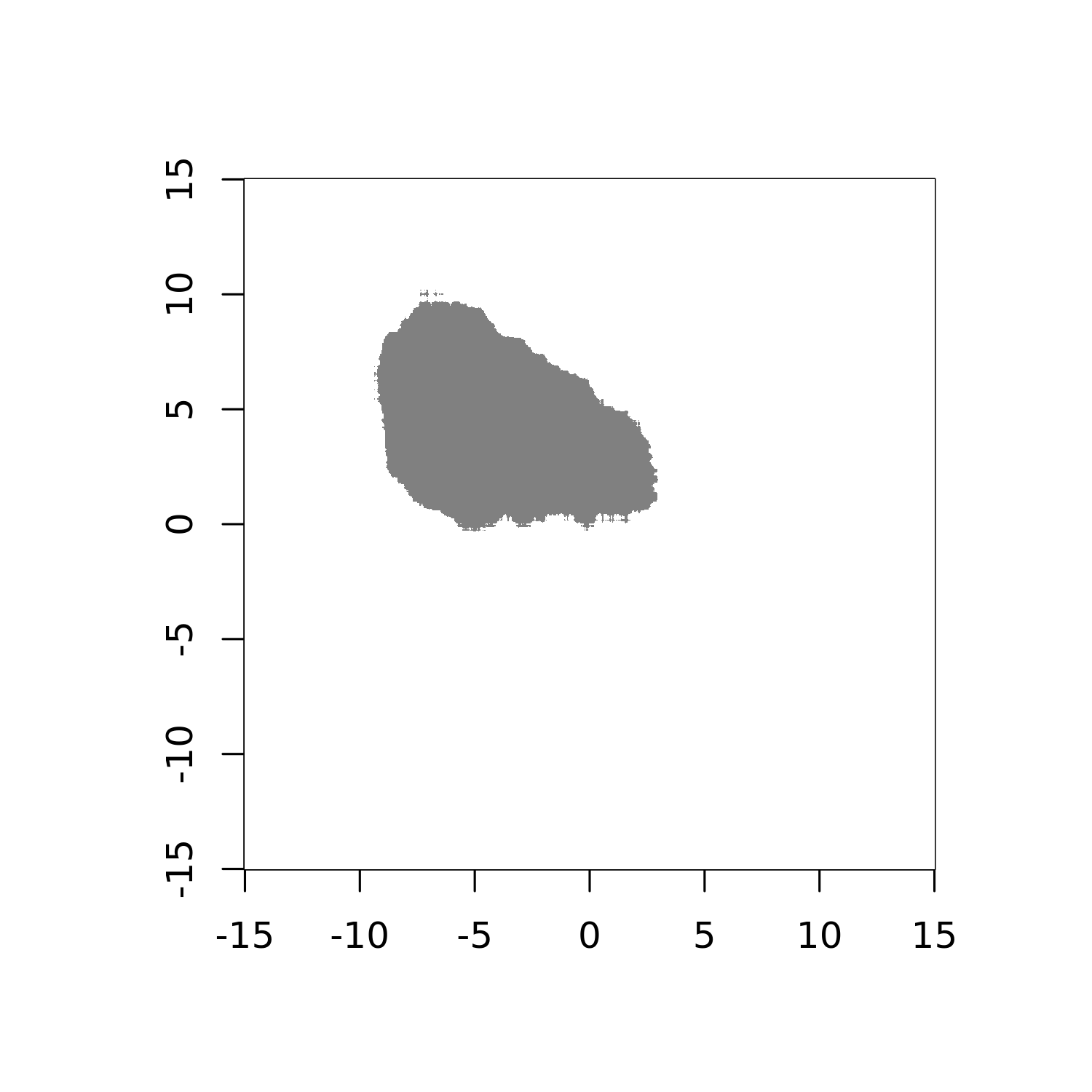}
    \caption{
    First row: The conditional excursion set $E^X_p(s_0)$ corresponding to an observation of a stationary isotropic Gaussian random field for $p=1-10^{-1}$, $1-10^{-2}$, and $1-10^{-3}$ (left to right) with conditioning site $s_0=(0,0)$. Second row: three observations of the corresponding scaled excursion set $\tilde{E}^X_p(s_0)$ using $\sigma(p)=\Phi^{-1}(p)$ for $p=1-10^{-1}$, $1-10^{-2}$, and $1-10^{-3}$ (left to right).}
    \label{fig:GRF-ex}
\end{figure}

In Figure   \ref{fig:GRF-ex} we display the the conditional excursion set $E^X_p(s_0)$  and the corresponding scaled excursion set $\tilde{E}^X_p(s_0)$  corresponding to   a stationary isotropic Gaussian random field    using $\sigma(p)=\Phi^{-1}(p)$,  where $\Phi$ is the univariate standard normal distribution function.  Indeed, in  this Gaussian case, we can provide, via the half-perimeter and area densities derived in Section 2.2.1 of \cite{bierme2018lipschitz} and 
Proposition 2 of \cite{cotsakis2024spatial},  $\tilde{E}^X_p(s_0)$ is the excursion set for locations scaled by $\sigma(p)=u_p=\Phi^{-1}(p)$. The interested reader is also refereed to \cite{hashorva2021sojourn}, which derives the scaling behaviour for Gaussian and other Gaussian-related asymptotically independent random fields.

 \begin{definition}[Strongly mixing random field]\label{stronglyMIXING}  
    A random field  $X$ is said to be {strongly mixing}, or {uniformly mixing}, if there exists a function $\psi(\rho):\R^+\rightarrow\R^+$ tending to 0 as $\rho\rightarrow\infty$, such that for any two measurable sets $S_1,S_2\subset\R^2$ that satisfy $\inf\{||s_1-s_2||_2:s_1\in S_1, s_2 \in S_2\} =: \rho > 0$, and for any events $A_1$ and $A_2$ in the  sigma fields generated by $\{X(s):s\in S_1\}$ and $\{X(s):s\in S_2\}$ respectively, it holds that $|\PP(A_1\cap A_2)-\PP(A_1)\PP(A_2)| < \psi(\rho)$.   
 \end{definition}

\section{Detailed calculations for Section \ref{sec:examples}}\label{supp:calculations}
\textbf{Stationary Gaussian-type random fields.}
Let $X$ be a stationary isotropic Gaussian random field
with standard normal margins.
Then one gets the half-perimeter and area density ratio
\begin{equation}\label{eq:Gauss-LKCratio}
    \frac{C_1^{\ast}(E^X_p)}{C_2^{\ast}(E^X_p)}
    = \frac{1}{4}
    \frac{\sqrt{2\pi\lambda_2}\varphi(u_p)}{1-\Phi(u_p)}
    \sim \frac{1}{4}\sqrt{2\pi\lambda_2}\,u_p,
\end{equation}
as $p\rightarrow1$, where $\varphi$ and $\Phi$ are the
univariate standard normal density and distribution functions,
respectively, and $\lambda_2$ is the second spectral moment
of $X$, obtained via
$\text{Var}(\nabla X(0))=\lambda_2\,\text{diag}(1,1)$
(see, e.g., Section 2.2.1 in \citeay{bierme2018lipschitz}).
Therefore, we follow the form in
\eqref{eq:LKC-ratio-assumption} as $p\rightarrow1$ with
$c_X=\sqrt{\pi\lambda_2}/2$
and $\sigma(p)=u_p/\sqrt{2}$.
Furthermore, for $u_p=\Phi^{-1}(p)$,
\begin{align*}
    C_1^\ast(E^X_p)
    &= \frac{c_X}{\sqrt{2}}\,\varphi(u_p)
    \sim \frac{c_X}{\sqrt{2}}
    (1-\Phi(u_p))u_p= \frac{c_X}{\sqrt{2}}(1-p)u_p
    \sim c_X(1-p)(-\ln(1-p))^{1/2}.
\end{align*}
Then the scaling index in \eqref{eq:C1form} is therefore
given by $\beta=1/2$.
We can also consider the affine anisotropic stationary Gaussian
random field (see e.g., \citeay{cabana1987}),
$X(s)=W(As)$, $\forall s$, where $W$ is a stationary isotropic Gaussian
random field with standard normal margins,
and the affinity matrix $A$
is invertible and has singular values
$a_1$ and $a_2$.
The associated excursion set is denoted by $E^X_p$.
Using stationarity and Equation~\eqref{eq:Gauss-LKCratio},
we have
\begin{equation*}
    \frac{C_1^{\ast}(E^X_p)}{C_2^{\ast}(E^X_p)}
    \sim
    \frac{P(\text{ellipse};a_1,a_2)}{\sqrt{2\pi}}
    \frac{1}{4}\sqrt{\lambda_2}\,u_p,
\end{equation*}
where $\lambda_2$ is the second spectral moment of
$W$ and
$P(\text{ellipse};a_1,a_2)$ is the perimeter of an ellipse
with semi-major and semi-minor axes $a_1,a_2$.
Therefore, we follow the form of
\eqref{eq:LKC-ratio-assumption} with
\[
c_X=\frac{\sqrt{\lambda_2}\,
P(\text{ellipse};a_1,a_2)}{4\sqrt{\pi}}
\qquad\text{and}\qquad
\sigma(p)=\frac{u_p}{\sqrt{2}}.
\]
Figure~\ref{fig:GRF-ex} shows the effect that the scaling
function $u_p$ (proportional to $\sigma(p)$)
has on excursion sets at a very high threshold for a single
observation of the Gaussian random field.
Furthermore, for $u_p=\Phi^{-1}(p)$, the scaling index
in~\eqref{eq:C1form} is therefore given by $\beta=1/2$.

We now investigate the behaviour of the scaling index $\beta$
of Gaussian random fields under marginal transformations.
Let us consider $X(s)=g(W(s))$, where $W$ is a Gaussian
random field with standard normal margins
satisfying Assumption~\ref{A1} and $g:\mathbb R\rightarrow\mathbb R$
is applied component-wise, is
of class $C^2$, satisfies $g'(x)>0$ for all
$x\in\mathbb R$, and is invertible on its image.
In this setting, the threshold becomes
$u_p=g(\Phi^{-1}(p))$.
In particular, the excursion sets coincide at each
probability level:
\[
E_p^X
=
\{s\in\mathbb R^2:W(s)\geq\Phi^{-1}(p)\}
=
E_p^W.
\]
With this marginal transformation, the tube equation becomes
$\text{Tube}(g,\rho)
=\{x\in\mathbb R:x\geq g^{-1}(u)-\rho\}$.
Furthermore, considering $X$ and $W$ to be arbitrary location
values in $\{X(s)\}_{s\in S}$ and $\{W(s)\}_{s\in S}$,
we have the following expansion centred about~$\rho$:
\begin{align*}
    \mathbb P\left(W\in\text{Tube}(g,\rho)\right)
    =
    \mathbb P\left(g^{-1}(X)\geq g^{-1}(u)-\rho\right) &=
    \mathbb P\left(W\geq g^{-1}(u)-\rho\right)\\ & =
    1-\Phi(g^{-1}(u))
    +\rho\varphi(g^{-1}(u))+O(\rho^2).
\end{align*}
Using the definition of $u_p$, the scaling function
can be chosen as
\begin{align*}
    \sigma(p)
    &=
    \frac{\Phi^{-1}(p)}{\sqrt{2}}
    \sim\frac{1}{\sqrt{2}}
    \frac{\mathcal M_1(p)}{\mathcal M_0(p)}=
    \frac{1}{\sqrt{2}}
    \frac{\varphi(g^{-1}(u_p))}
         {1-\Phi(g^{-1}(u_p))}=
    \frac{1}{\sqrt{2}}
    \frac{\varphi(\Phi^{-1}(p))}
         {1-\Phi(\Phi^{-1}(p))},
\end{align*}
as $p\rightarrow1$.
Therefore, the scaling function $\sigma(p)$ of Gaussian
random fields remains unchanged under marginal transformation,
and the scaling index remains $\beta=1/2$.
What is changed in this setting is the marginal quantile
function used for thresholding.
For example, when $g=\Phi$, we simply have $u_p=p$.
\smallskip

\textbf{Standardised chi-squared random fields.}
Let
$X(s)=(2\kappa)^{-1/2}(\sum_{j=1}^\kappa W_j(s)^2-\kappa)$,
for $s\in\mathbb{R}^2$ and $\kappa\in\mathbb{N}^*$,
where $W_1,\ldots,W_\kappa$ are independent copies of a
stationary isotropic standard Gaussian random field satisfying
$\operatorname{Var}(\nabla W_1(0))=\lambda_2 I_2$, with
$\lambda_2>0$.
Then, by the findings in \cite{bierme2018lipschitz},
Section 2.2.2, we obtain
\begin{align}\label{eq:chisq-C1ast-u}
    C_1^{\ast}(E^X_p)
    =&\frac{\sqrt{\pi\lambda_2}}
    {2^{\frac12(\kappa+1)}\Gamma\left(\frac{\kappa}{2}\right)}
    (\kappa+\sqrt{2\kappa}u_p)^{(\kappa-1)/2}
    \exp\left\{-\frac{1}{2}(\kappa+\sqrt{2\kappa}u_p)\right\}
    \notag\\
    \sim&
    \frac{\sqrt{\pi\lambda_2}e^{-\kappa/2}
    (2\kappa)^{\frac14(\kappa-1)}}
    {2^{\frac12(\kappa+1)}\Gamma\left(\frac{\kappa}{2}\right)}
    u_p^{(\kappa-1)/2}
    \exp\left\{-\sqrt{\kappa/2}u_p\right\}
\end{align}
as $p\to1$.
By stationarity, we have $C_2^\ast(E_p^X)=1-p$.
To inspect the ratio in Theorem~\ref{thm:LKCratio},
we need the form of $u_p$ in terms of $p$ as $p\to1$.
To do so, note that $u_p$ defines the solution to
\begin{align*}
    1-p
    =&\ \mathbb{P}\left(\chi_\kappa^2\geq \kappa+\sqrt{2\kappa}u_p\right) =\ \frac{
    \Gamma\left(\frac{\kappa}{2},
    \frac{\kappa}{2}+\sqrt{\frac{\kappa}{2}}u_p\right)}
    {\Gamma\left(\frac{\kappa}{2}\right)},
\end{align*}
where $\Gamma(s,t)$ is the upper incomplete gamma function.
Using the asymptotic expansion in \citeauthor{NIST:DLMF},
Equation 8.11.2, results in
\begin{align*}
    1-p
    &\sim
    \frac{1}{\Gamma(\frac{\kappa}{2})}
    \left(\frac{\kappa}{2}+\sqrt{\frac{\kappa}{2}}u_p\right)^{\frac{\kappa}{2}-1}
    \exp\left\{-\left(
    \frac{\kappa}{2}+\sqrt{\frac{\kappa}{2}}u_p\right)\right\}\sim
    \frac{
    \left(\frac{\kappa}{2}\right)^{\frac12(\frac{\kappa}{2}-1)}
    e^{-\kappa/2}}
    {\Gamma(\frac{\kappa}{2})}
    u_p^{\frac{\kappa}{2}-1}
    \exp\left\{-\sqrt{\frac{\kappa}{2}}u_p\right\}
\end{align*}
as $p\to1$.
Taking negative logarithms and using
$\ln u_p=o(u_p)$, we obtain
\begin{equation}\label{eq:chisq-up}
    u_p\sim-\sqrt{\frac{2}{\kappa}}\ln(1-p)
\end{equation}
as $p\to1$.
Because $C_2^\ast(E_p^X)=\mathbb{P}(X(s)>u_p)$,
note that
\begin{equation}\label{eq:chisq-C2ast-u}
    C_2^\ast(E_p^X)
    \sim
    \frac{
    \left(\frac{\kappa}{2}\right)^{\frac12(\frac{\kappa}{2}-1)}
    e^{-\kappa/2}}
    {\Gamma(\frac{\kappa}{2})}
    u_p^{\frac{\kappa}{2}-1}
    \exp\left\{-\sqrt{\frac{\kappa}{2}}u_p\right\}.
\end{equation}
Dividing \eqref{eq:chisq-C1ast-u} by
\eqref{eq:chisq-C2ast-u}
and using \eqref{eq:chisq-up} results in
\begin{align*}
    \frac{C_1^\ast(E_p^X)}{C_2^\ast(E_p^X)}
    \sim
    \frac{\sqrt{\pi\lambda_2}}{2^{5/4}}\kappa^{1/4}
    u_p^{1/2}\sim
    \frac{\sqrt{\pi\lambda_2}}{2}
    \{-\ln(1-p)\}^{1/2}
\end{align*}
as $p\to1$.
Therefore, we follow the form in
\eqref{eq:LKC-ratio-assumption} with
\[
    c_X=\frac{\sqrt{\pi\lambda_2}}{2},
    \qquad
    \sigma(p)=\{-\ln(1-p)\}^{1/2}.
\]
This choice of $\sigma$ depends only on $p$.
Since $\sigma(1-e^{-x})=x^{1/2}$ for $x>0$,
we have $\sigma(1-e^{-x})\in\mathrm{RV}_{1/2}$,
and therefore $\beta=1/2$.
For completeness, the refined asymptotic form of $u_p$ is
\begin{align*}
    u_p=\sqrt{\frac{2}{\kappa}}\Biggl[
    -\ln(1-p)
    +\left(\frac{\kappa}{2}-1\right)\ln\{-\ln(1-p)\}
    -\frac{\kappa}{2}
    -\ln\Gamma\left(\frac{\kappa}{2}\right)
    +o(1)
    \Biggr].
\end{align*}
Taking exponentials and collecting constants, we have
\[
    \exp\left\{-\sqrt{\frac{\kappa}{2}}u_p\right\}
    =
    e^{\kappa/2}\Gamma\left(\frac{\kappa}{2}\right)
    (1-p)\{-\ln(1-p)\}^{-(\frac{\kappa}{2}-1)}
    (1+o(1)).
\]
Putting this into $C_1^\ast(E_p^X)$ and using
\eqref{eq:chisq-up}, we obtain
$C_1^\ast(E_p^X)
    \sim
   \frac{\sqrt{\pi\lambda_2}}{2}
    (1-p)\{-\ln(1-p)\}^{1/2}.$ 
We therefore follow the form of \eqref{eq:C1form} with
the same constant
$c_X=\sqrt{\pi\lambda_2}/2$
and $\beta=1/2$.
\smallskip
 
\textbf{Random fields of the form \boldmath{$X(\cdot)=g(\Lambda,W(\cdot))$}.}
Suppose we are now concerned with Gaussian random fields mixed
with an independent random variable $\Lambda$ in the form
$X(s)=g(\Lambda,W(s))$.
Here, $W$ is a stationary isotropic standard Gaussian random field,
with $\operatorname{Var}(\nabla W(0))=\lambda_2 I_2$ and
$\lambda_2>0$.
We assume that, for almost every $\lambda$, the function
$w\mapsto g(\lambda,w)$ is strictly increasing and admits an
inverse at the levels considered.
In this construction, \cite{di2024spatial} showed that,
under the regularity assumptions above,
the expected half-perimeter and area are given by
\begin{equation}\label{eq:ExpLKC1locscalemix} 
    \mathbb{E}\left[\mathcal{L}_1(E^X_p\cap S)\right]
    =
    \mathcal{L}_1(S)
    \mathbb{E}\left[\rho_0(h(u_p,\Lambda))\right]
    +\frac{\pi}{2}\sqrt{\lambda_2}\mathcal{L}_2(S)
    \mathbb{E}\left[\rho_1(h(u_p,\Lambda))\right] 
\end{equation}
and
$   \mathbb{E}\left[\mathcal{L}_2(E^X_p\cap S)\right]
    =
    \mathcal{L}_2(S)
    \mathbb{E}\left[\rho_0(h(u_p,\Lambda))\right],
$ 
where
$h(u,\lambda)=[g(\lambda,\cdot)]^{-1}(u)$
and
$\mathbb{E}\left[\rho_j(
h(u_p,\Lambda))\right]$,
$j=0,1$,
define the \emph{normalised Minkowski functionals},
with $\rho_0(z)=1-\Phi(z)$ and
$\rho_1(z)=(2\pi)^{-1/2}\varphi(z)$. The half-perimeter and area densities are given by
\begin{equation}\label{eq:c1ast-minkowski}
    C_1^\ast(E^X_p)
    =
    \lim_{S\nearrow\mathbb{R}^2}
    \frac{\mathbb{E}\left[\mathcal{L}_1(E^X_p\cap S)\right]}
    {\mathcal{L}_2(S)}
    =
    \frac{\pi}{2}\sqrt{\lambda_2}
    \mathbb{E}\left[\rho_1(h(u_p,\Lambda))\right]
\end{equation}
and
\[
    C_2^\ast(E^X_p)
    =
    \lim_{S\nearrow\mathbb{R}^2}
    \frac{\mathbb{E}\left[\mathcal{L}_2(E^X_p\cap S)\right]}
    {\mathcal{L}_2(S)}
    =
    \mathbb{E}\left[\rho_0(h(u_p,\Lambda))\right].
\]
The limit defining $C_1^\ast(E^X_p)$ is taken along
VH-growing observation windows satisfying
$\mathcal{L}_1(S)/\mathcal{L}_2(S)\to0$.
Thus, the first term in \eqref{eq:ExpLKC1locscalemix},
after division by $\mathcal{L}_2(S)$, vanishes.
To obtain the half-perimeter density, it therefore suffices
to derive $\mathbb{E}[\rho_1(h(u_p,\Lambda))]$.
The resulting half-perimeter and area density ratio reduces~to
\[
    \frac{C_1^\ast(E^X_p)}{C_2^\ast(E^X_p)}
    =
    \frac{\pi}{2}\sqrt{\lambda_2}\
    \frac{\mathbb{E}\left[\rho_1(h(u_p,\Lambda))\right]}
    {\mathbb{E}\left[\rho_0(h(u_p,\Lambda))\right]}.
\]
For scale and location mixtures,
\cite{di2024spatial} showed the following useful property
for computing the Minkowski functionals of order 0 and 1:
\begin{alignat}{2}
    &\mathbb{E}\left[\rho_0(h(u_p,\Lambda))\right]
    &&=
    \frac{1}{\sqrt{2\pi}}
    \int_D e^{-z^2/2}
    \bar{F}_{\Lambda}(h(u_p,z))\,dz,
    \label{Rho0}\\
    &\mathbb{E}\left[\rho_1(h(u_p,\Lambda))\right]
    &&=
    \frac{1}{2\pi}
    \int_D z e^{-z^2/2}
    \bar{F}_{\Lambda}(h(u_p,z))\,dz,
    \label{Rho1}
\end{alignat}
where $\bar F_\Lambda=1-F_\Lambda$,
$h(u_p,z)=u_p/z$ and
$D=(0,\infty)$
for scale mixtures of the form $\Lambda W(s)$
with $\Lambda>0$ almost surely,  $u_p>0$, $h(u_p,z)=u_p-z$, and $D=\mathbb{R}$
for location mixtures of the form $\Lambda+W(s)$.

When the marginal distribution is known, it may occasionally
be more useful to use the fact that, under stationarity,
\[
    C_2^\ast(E^X_p)
    =
    \mathbb{E}\left[\rho_0(h(u_p,\Lambda))\right]
    =
    \mathbb{P}\left(X(s)>u_p\right)
    =1-p
\]
for an arbitrary $s\in S$.
The following paragraphs are devoted to specific cases of
the $\Lambda$ distribution, in which the calculations can
be carried out explicitly.
\smallskip

\textit{Laplace random fields:} 
Let $X(\cdot)=\sqrt{\Lambda}W(\cdot)$, with $\Lambda$
an exponential random variable with rate parameter $1/2$,
\emph{i.e.},
$1-F_{\Lambda}(x)=e^{-x/2}$
for $x\geq0$ \citep{Opitz.2016}.
Applying \eqref{Rho1} with mixing variable $\sqrt{\Lambda}$
and using
$\mathbb{P}(\sqrt{\Lambda}>x)=e^{-x^2/2}$ for $x\geq0$,
the first-order Minkowski functional is given by
\begin{align*}
    \mathbb{E}\left[\rho_1(u_p/\sqrt{\Lambda})\right]
     =
    -\int_0^\infty
    \rho_1^\prime(z)
    \exp\left\{-\frac{u_p^2}{2z^2}\right\}\,dz=
    \frac{1}{2\pi}
    \int_0^\infty
    &\exp\left\{-t-\frac{u_p^2}{4t}\right\}\,dt\\
    &=
    \frac{1}{2\pi}u_pK_1(u_p)\sim
    \frac{1}{2\sqrt{2\pi}}u_p^{1/2}e^{-u_p},
\end{align*}
as $p\to1$, where the representation involving the modified
Bessel function of the second kind, $K_1(u_p)$, follows from
Equation (10.32.10) of \citeauthor{NIST:DLMF}, and the
asymptotic equivalence follows from Equation (10.40.2)
of \citeauthor{NIST:DLMF}. The resulting half-perimeter density is given by
$C_1^\ast(E^X_p)
    =
    \frac{\sqrt{\lambda_2}}{4}u_pK_1(u_p)
    \sim
    \frac14\sqrt{\frac{\pi\lambda_2}{2}}
    u_p^{1/2}e^{-u_p}
$ as $p\to1$.
Because the margins are standard Laplace-distributed,
for $p>1/2$ we have
$    C_2^\ast(E^X_p)
    =
    \frac12 e^{-u_p}
    =1-p.
$ 
Consequently,
    $u_p=F_L^{-1}(p)
    =-\ln\{2(1-p)\}
    \sim-\ln(1-p)$, for  $p\to1$. The half-perimeter and area density ratio therefore satisfies
\begin{align*}
    \frac{C_1^\ast(E^X_p)}{C_2^\ast(E^X_p)}
    &\sim
    \frac12\sqrt{\frac{\pi\lambda_2}{2}}u_p^{1/2} 
    \frac12\sqrt{\frac{\pi\lambda_2}{2}}
    (-\ln(1-p))^{1/2}
\end{align*}
as $p\to1$.
Therefore, we follow the form of
\eqref{eq:LKC-ratio-assumption} with $   c_X=\frac12\sqrt{\frac{\pi\lambda_2}{2}}$, 
    $\sigma(p)=\{-\ln(1-p)\}^{1/2}$.
This scaling function depends only on $p$.
Since $\sigma(1-e^{-x})=x^{1/2}$ for $x>0$,
we have $\sigma(1-e^{-x})\in\mathrm{RV}_{1/2}$,
and the scaling index is $\beta=1/2$.  Finally,
\[
    C_1^\ast(E^X_p)
    \sim\frac{c_X}{2}
    u_p^{1/2}e^{-u_p}
    =
    c_X(1-p)u_p^{1/2}
    \sim
    c_X(1-p)\{-\ln(1-p)\}^{1/2},
\]
which gives \eqref{eq:C1form} with
the same constant $c_X$
and scaling index $\beta=1/2$.
\smallskip

\textit{Elliptical Student random fields:}
Consider the stochastic representation of elliptical Student
random fields in \cite{Ma2013},
$
    X(\cdot)=\Lambda W(\cdot)
$,
where $\Lambda\overset{\mathrm{d}}{=}\sqrt{k/\chi^2_k}$.
Here, $\Lambda$ is heavy-tailed, with regular variation index
$-k$, and we have
\[
    \bar{F}_{\Lambda}(h(u_p,z))
    =
    \Gamma(k/2)^{-1}
    \gamma\left(
    \frac{k}{2},
    \frac{kz^2}{2u_p^2}
    \right),
\]
where $\gamma(\cdot,\cdot)$ is the lower incomplete gamma function. Using \eqref{Rho1}, the Minkowski functional of order 1 is
\begin{align*}
    \mathbb{E}\left[
    \rho_1\left(u_p/\Lambda\right)\right]
    &=
    \frac{1}{2\pi\Gamma(k/2)}
    \int_0^\infty
    ze^{-z^2/2}
    \gamma\left(
    \frac{k}{2},
    \frac{kz^2}{2u_p^2}
    \right)\,dz=
    \frac{1}{2\pi\Gamma(k/2)}
    \int_0^\infty
    e^{-t}
    \gamma\left(
    \frac{k}{2},
    \frac{kt}{u_p^2}
    \right)\,dt\\
    &=
    \frac{1}{2\pi\Gamma(k/2)}
    \int_0^\infty
    \int_{u_p^2x/k}^\infty
    e^{-t}x^{k/2-1}e^{-x}\,dt\,dx=
    \frac{1}{2\pi}
    \left(1+\frac{u_p^2}{k}\right)^{-k/2} \sim
    \frac{1}{2\pi}
    k^{k/2}u_p^{-k},
\end{align*}
as $p\to1$.

Using the fact that the margins are Student with $k$ degrees
of freedom \citep{Ma2013}, we have
\[
    C_2^\ast(E^X_p)
    \sim
    \frac{\Gamma((k+1)/2)}
    {\sqrt{\pi}\Gamma(k/2)}
    k^{k/2-1}
    u_p^{-k}
\]
as $p\to1$.
Examining the half-perimeter and area density ratio,
\begin{align*}
    \frac{C_1^\ast(E^X_p)}{C_2^\ast(E^X_p)}
    &\sim
    \frac{\pi}{2}\sqrt{\lambda_2}
    \frac{
    \frac{1}{2\pi}k^{k/2}u_p^{-k}}
    {
    \frac{\Gamma((k+1)/2)}
    {\sqrt{\pi}\Gamma(k/2)}
    k^{k/2-1}
    u_p^{-k}} =
    \frac{\sqrt{\pi\lambda_2}}{2}
    \frac{\Gamma(1+k/2)}{\Gamma((k+1)/2)}
\end{align*}
as $p\to1$. We follow the form of \eqref{eq:LKC-ratio-assumption} with
\[
    c_X=
    \frac{\sqrt{\pi\lambda_2}}{2}
    \frac{\Gamma(1+k/2)}{\Gamma((k+1)/2)},
    \qquad
    \sigma(p)=1.
\]
In this case, we recover Equation \eqref{eq:C1form} with $\beta=0$.
\smallskip

\textit{Sub-Gaussian random fields:} Suppose $\Lambda$ is such that $\mathbb{E}[e^{-t\Lambda}]=e^{-t^{\varrho/2}}$, with $0<\varrho<2$. The \emph{sub-Gaussian} random field in \cite{adler2010excursion} is obtained by $X(\cdot)=\sqrt{\Lambda}W(\cdot)$. Using this definition, \cite{adler2010excursion} obtain the expected LKC of order 0,
$$
    \mathbb{E}[\mathcal{L}_0(E^X_p\cap S)] \sim u_p^{-\varrho}\Gamma(1-\varrho/2)^{-1}\sum_{j=0}^2 \lambda_2^{j/2}\mathcal{L}_j(S)c_j(\varrho),
$$
as $p\rightarrow1$, where $c_0(\varrho)$, $c_1(\varrho)$, and $c_2(\varrho)$ are constants independent of $p$ which, under the above normalisation, are given by $c_0(\varrho)=2^{\varrho/2-1}\Gamma((\varrho+1)/2)/\sqrt{\pi}$, $c_1(\varrho)=2^{\varrho/2-1}\Gamma(1+\varrho/2)/\pi$, and $c_2(\varrho)=\varrho c_0(\varrho)/(2\pi)$.
Using the Gaussian Kinematic Formula (see Equation~(23) of \cite{di2024spatial}), one can read off the Minkowski functionals
$
    \mathbb{E}[\rho_j(u_p/\sqrt{\Lambda})] \sim u_p^{-\varrho}\Gamma(1-\varrho/2)^{-1}c_j(\varrho)
$ 
as $p\rightarrow1$ for $j=0,1,2$.
The ratio of half-perimeter and area densities is therefore obtained by
$
    \frac{C_1^\ast(E^X_p)}{C_2^\ast(E^X_p)}\sim\frac{\pi}{2}\sqrt{\lambda_2}\frac{c_1(\varrho)}{c_0(\varrho)}
$ 
as $p\rightarrow1$.
Substituting these values, we obtain
$$
    \frac{C_1^\ast(E^X_p)}{C_2^\ast(E^X_p)}\sim\frac{\sqrt{\pi\lambda_2}}{2}\frac{\Gamma(1+\varrho/2)}{\Gamma((\varrho+1)/2)}
$$
as $p\rightarrow1$.
Therefore, we follow the form of \eqref{eq:LKC-ratio-assumption} with $c_X=\frac{\sqrt{\pi\lambda_2}}{2}\frac{\Gamma(1+\varrho/2)}{\Gamma((\varrho+1)/2)}$ and $\sigma(p)=1$. As a result, the scaling index is given by $\beta=0$. 

\textit{Gaussian-based threshold-stable processes:} Suppose that $X^{\textit{scal}}(s)=\Lambda W(s)$ for $\Lambda\in\text{RV}_{\infty}(\varrho)$, or $X^{\textit{loc}}(s)=\Lambda + W(s)$ for $e^{\Lambda}\in\text{RV}_{\infty}(\varrho)$.   
\cite{di2024spatial} discussed the importance of these mixture models, as both contain the threshold stability property often desired when modelling extreme events. Furthermore, \cite{cotsakis2024spatial} studied the LER of $X^{\textit{scal}}$, with simulations done in the special case when $\Lambda$ is a Pareto random variable with scale parameter 1 and shape parameter 2, resulting in $\varrho=2$.

Suppose $L^{\Lambda}$ is the slowly-varying function associated with either of the mixture random fields.
Beginning with the scale mixture $X^{\textit{scal}}$, we obtain the zero and first-order Minkowski functionals
\begin{align*}
    \mathbb{E}\left[\rho_0\left({u_p}/{\Lambda}\right)\right] =& \frac{1}{\sqrt{2\pi}}\int\limits_{0}^{\infty}e^{-\frac{1}{2}z^2}L^{\Lambda}(u_p/z)(u_p/z)^{-\varrho}dz\sim \frac{1}{\sqrt{2\pi}}L^{\Lambda}(u_p) u_p^{-\varrho}\int\limits_{0}^{\infty}z^\varrho e^{-\frac{1}{2}z^2}dz\\
    =&  \frac{1}{\sqrt{2\pi}}2^{(\varrho-1)/2}\Gamma\left(\frac{\varrho+1}{2}\right)L^{\Lambda}(u_p) u_p^{-\varrho},\\
    \mathbb{E}\left[\rho_1\left({u_p}/{\Lambda}\right)\right] =& \frac{1}{2\pi}\int\limits_{0}^{\infty}ze^{-\frac{1}{2}z^2}L^{\Lambda}(u_p/z)(u_p/z)^{-\varrho}dz\sim \frac{1}{2\pi} L^{\Lambda}(u_p) u_p^{-\varrho}\int\limits_{0}^{\infty}z^{\varrho+1} e^{-\frac{1}{2}z^2}dz\\
    =& \frac{1}{2\pi}2^{\varrho/2}\Gamma\left(1+\varrho/2\right)L^{\Lambda}(u_p) u_p^{-\varrho},
\end{align*}
where asymptotic equivalences are as $p\rightarrow1$.
As a result, for  $p\rightarrow1$, the half-perimeter and area density ratio is given by
$$
 \frac{C_1^{\ast}(E^X_p)}{C_2^{\ast}(E^X_p)}\sim \frac{1}{2}\pi^{1/2}\frac{\Gamma(1+\varrho/2)}{\Gamma((1+\varrho)/2)}\sqrt{\lambda_2}.
$$

For the location mixture model $X^{\textit{loc}}$, we obtain the zero and first-order Minkowski functionals
\begin{align*}
    \mathbb{E}\left[\rho_0\left({u_p}-{\Lambda}\right)\right] =& \frac{1}{\sqrt{2\pi}}\int\limits_{0}^{\infty}e^{-\frac{1}{2}z^2}L^{\Lambda}(e^{u_p-z})e^{-\varrho(u_p-z)}dz\sim \frac{1}{\sqrt{2\pi}}L^{\Lambda}(e^{u_p}) e^{-\varrho u_p}\int\limits_{0}^{\infty} e^{\varrho z-\frac{1}{2}z^2}dz\\
    =&  e^{\varrho^2 /2}L^{\Lambda}(e^{u_p}) e^{-\varrho u_p},\\
    \mathbb{E}\left[\rho_1\left({u_p}-{\Lambda}\right)\right] =& \frac{1}{2\pi}\int\limits_{0}^{\infty}ze^{-\frac{1}{2}z^2}L^{\Lambda}(e^{u_p-z})e^{-\varrho(u_p-z)}dz\sim \frac{1}{2\pi} L^{\Lambda}(e^{u_p}) e^{-\varrho u_p}\int\limits_{0}^{\infty} ze^{\varrho z-\frac{1}{2}z^2}dz\\
    =& \frac{\varrho}{\sqrt{2\pi}} e^{\varrho^2 /2}L^{\Lambda}(e^{u_p}) e^{-\varrho u_p},
\end{align*}
where asymptotic equivalences are as $p\rightarrow1$.
As a result, the half-perimeter and area density ratio is given~by
$
 \frac{C_1^{\ast}(E^X_p)}{C_2^{\ast}(E^X_p)}\sim \frac{1}{2}\sqrt{\frac{\pi\lambda_2}{2}}\varrho
$ 
as $p\rightarrow1$. 
Therefore $\sigma(p)=1$ and $\beta=0$. 
Furthermore, we follow the form of \eqref{eq:C1form} with $c_{X^{scal}}=\frac{1}{2}\pi^{1/2}\frac{\Gamma(1+\varrho/2)}{\Gamma((1+\varrho)/2)}\sqrt{\lambda_2}$ and $c_{X^{loc}}=\frac{1}{2}\sqrt{\frac{\pi\lambda_2}{2}}\varrho$, with $\beta=0$ in both cases.
Recalling the special case of the Pareto-Gaussian scale mixture in \cite{cotsakis2024spatial}, we recover $c_{X^{scal}}=\sqrt{\lambda_2}$.
\smallskip

\textbf{Max-stable random fields.} 
Suppose that the random field is stationary and isotropic with the following construction
\begin{equation*}\label{eq:random-disks}
    X(s)=\max_{i\geq 1}\xi_i f_i(s)
\end{equation*}
where $\{\xi_i\}_{i\geq 1}$ are points of a Poisson point process with intensity $\xi^{-2}d\xi$, and $\{f_i\}_{i\geq 1}$ is a collection of independent copies of a non-negative spectral random field $f$, independent of $\{\xi_i\}_{i\geq 1}$, each satisfying $\mathbb{E}[f(s)]=1$ for every $s\in\mathbb{R}^2$.  By this construction,  the marginal distribution  $X(0)$ is unit Fr\'{e}chet, and there exists a so-called \emph{exponent function} (or \emph{exponent measure}) $V$ such that
$\PP(X(0)\leq x_1,X(h)\leq x_2)=\exp\{-V(x_1,x_2;h)\}$ for $\left\|h\right\|>0$. The function $V$ is $-1$-homogeneous, and defines the \emph{extremal coefficient} $\theta(h)=V(1,1;h)$ (see, for example, \citeay{davison2012statistical}).
Beginning with the half-perimeter density,  \cite{galerne2011computation} showed that the perimeter of two-dimensional excursion sets can be expressed in terms of \emph{directional crossing probabilities}. For a spatial lag $h\in\mathbb{R}^2$ and threshold probability $p$, the directional crossing probability is without loss of generality (by stationarity) the probability of crossing the boundary pf $E_p^X$ when travelling from the origin to location $h$, and is defined  by
\begin{align*}
    c_p(h) = \, & \PP\left[\left(\left\{0\in E_p^X\right\} \cap \left\{h\notin E_p^X\right\} \right)\cup \left(\left\{0\notin E_p^X\right\} \cap \left\{h\in E_p^X\right\}\right)\right]\\
    =\, & 2\, \PP\left(X(0)\geq u_p,X(h)\leq u_p\right) = 2\left(e^{-u_p^{-1}} - e^{-V(u_p,u_p;h)}\right)\\
    =\, & 2\,\left(e^{-u_p^{-1}} - e^{-u_p^{-1}\theta(h)}\right) = 2e^{-u_p^{-1}}\left(1 - e^{-u_p^{-1}\left(\theta(h)-1\right)}\right).
\end{align*}
With this, the \emph{specific directional variation} can be defined as the amount of boundaries of $E_p^X$ encountered when travelling in the direction $u\in\mathbb{S}^1:=\left\{v\in\mathbb{R}^2\,:\,\left\|v\right\|=1\right\}$, and is obtained by taking the limit $\lim_{\, r\downarrow 0}r^{-1}c_p(ru)$.
With this, Proposition 8 of \cite{galerne2011computation}  (see also Section 2.2 of \citeay{villa2023mean}) can be used to obtain the half-perimeter of $E_p^X$ contained in $S$;
\begin{align*}
    \mathbb{E}\left[\mathcal{L}_1(E_p^X,S)\right] =& \frac{1}{8}\mathcal{L}_2(S)\int\limits_{\mathbb{S}^1} \mathbb{E}\left[\lim_{r\downarrow 0}r^{-1}c_p(ru)\right] \mathcal{H}^1(du) = \frac{1}{8}\mathcal{L}_2(S) \int\limits_{\mathbb{S}^1}\mathcal{H}^1(du) \mathbb{E}\left[\lim_{r\downarrow 0}r^{-1}c_p(r\tilde{u})\right] \\
    =& \frac{\pi}{4}\mathcal{L}_2(S)\lim_{r\downarrow 0}r^{-1}c_p(r\tilde{u})
\end{align*}
for general $\tilde{u}\in\mathbb{S}^1$.
Removing the expected limit term from the integral is justified by isotropy, while the expectation is redundant because $c_p(h)$ is deterministic. Dividing by $\mathcal{L}_2(S)$ and substituting the directional crossing probabilities for max-stable processes gives
\begin{align*}
    C_1^\ast(E_p^X) =& \frac{\pi}{2}e^{-u_p^{-1}}\lim_{r\downarrow 0}r^{-1}\left(1 - e^{-u_p^{-1}\left(\theta(r\tilde{u})-1\right)}\right) = \frac{\pi}{2}c_{\theta}u_p^{-1}e^{-u_p^{-1}},
\end{align*}
where $c_{\theta}:=\lim_{\,r\downarrow 0}r^{-1}\left(\theta(r\tilde{u})-1\right)$, provided the limit exists and belongs to $(0,\infty)$.

There are several examples of max-stable random fields with valid constants $c_{\theta}$. Take, for example, the so-called Smith model \citep{smith1990max}, constructed using $X(s)=\max_{i\geq 1} \xi_i f_i(s;s_i,\sigma^2 I_2)$, where $f_i$ are bivariate Gaussian densities with randomly-chosen centres $s_i$ and identical covariance matrices $\sigma^2 I_2$. Here, the pairs $\{(\xi_i,s_i)\}_{i\geq 1}$ form a Poisson point process on $(0,\infty)\times\mathbb{R}^2$ with intensity $\xi^{-2}\,d\xi\,dz$. In this construction, the extremal coefficient is given by $\theta(h)=2\Phi(\left\|h\right\|/2\sigma)$. Considering the asymptotic expansion of $\Phi(r/(2\sigma))$ as $r\to0$ leads to $c_{\theta}=1/(\sigma\sqrt{2\pi})$.

By stationarity, the area density is found through the standard Fr\'{e}chet marginal distribution;
$$
    C_2^\ast(E_p^X) = \PP(X(0)\geq u_p) = 1-e^{-u_p^{-1}}.
$$
As a result, the quantile function is obtained by setting $1-p=C_2^\ast(E_p^X)$ and solving for $u_p$, resulting in $u_p=-1/\log p\sim (1-p)^{-1}$ as $p\to 1$. As a result, we obtain the limiting form of $C_1^\ast(E_p^X)\sim (\pi/2)c_\theta(1-p)$ as $p\to 1$, and we may write the half-perimeter and area density ratios as
$$
    \frac{C_1^\ast(E_p^X)}{C_2^\ast(E_p^X)} \sim \frac{\pi}{2}c_{\theta}
$$
as $p\to 1$. We therefore follow the form of \eqref{eq:LKC-ratio-assumption} with $c_X=(\pi/2)c_\theta$ and $\sigma(p)=1$, resulting in scaling index $\beta=0$.

\section{Detailed calculations for Section \ref{sec:COUNTERexamples}}\label{supp:calculationsBIS}

In this section, we study cases in which the half-perimeter density cannot be expressed in the form~\eqref{eq:C1form}, so that the scaling index $\beta$ cannot be obtained through this representation. \smallskip

\textbf{Student random fields.} Let $X(s)=\sqrt{k-2}{W_0(s)}/\sqrt{\sum_{j=1}^k W_j(s)^2}$, where $k\geq3$ is an integer and $W_0,W_1,\ldots,W_k$ are independent copies of $W$.
\cite{bierme2018lipschitz} (Section 2.2.3) derived the half-perimeter and area density ratio
\begin{align*}
    \frac{C_1^\ast(E_p^X)}{C_2^\ast(E_p^X)} &= \frac{\frac14\sqrt{\lambda_2}\left(1+u_p^2/(k-2)\right)^{(1-k)/2}}{\mathbb{P}\left(T_k>\sqrt{\frac{k}{k-2}}u_p\right)}.
\end{align*}
To simplify the denominator, take advantage of the Mills ratio for the Student distribution \citep{soms1976asymptotic}, which results in the approximation
\begin{equation}\label{eq:StudentMills}
    \mathbb{P}(T_k>x)\sim k^{-1}x f_{T_k}(x)\sim\frac{k^{k/2-1}\Gamma\left(\frac{k+1}{2}\right)}{\sqrt{\pi}\Gamma\left(\frac{k}{2}\right)}x^{-k}
\end{equation}
as $x\rightarrow\infty$.
Using this approximation in the denominator and approximating the numerator gives
\begin{align*}
    \frac{C_1^\ast(E_p^X)}{C_2^\ast(E_p^X)} &\sim \frac{\frac14\sqrt{\lambda_2}(k-2)^{-(1-k)/2}u_p^{1-k}}{\frac{\Gamma\left(\frac{k+1}{2}\right)}{k\sqrt{\pi}\Gamma\left(\frac{k}{2}\right)}(k-2)^{k/2}u_p^{-k}} = \frac{\sqrt{\pi\lambda_2}}4\frac{k}{\sqrt{k-2}}\frac{\Gamma\left(\frac{k}{2}\right)}{\Gamma\left(\frac{k+1}{2}\right)}u_p
\end{align*}
as $p\rightarrow1$.
Therefore, we follow the form in \eqref{eq:LKC-ratio-assumption} as $p\rightarrow1$ with
$
    c_X=\frac{\sqrt{\pi\lambda_2}}4\frac{k}{\sqrt{k-2}}\frac{\Gamma\left(\frac{k}{2}\right)}{\Gamma\left(\frac{k+1}{2}\right)}
$
and $\sigma(p)=u_p$.

As a result of \eqref{eq:StudentMills}, we have \[F^{-1}_{T_k}(p)\sim\sqrt{k}\left\{\frac{\Gamma((k+1)/2)}{k\sqrt{\pi}\Gamma(k/2)}\right\}^{1/k}(1-p)^{-1/k}\] and therefore \[u_p=\sqrt{(k-2)/k}\,F^{-1}_{T_k}(p)\sim\sqrt{k-2}\left\{\frac{\Gamma((k+1)/2)}{k\sqrt{\pi}\Gamma(k/2)}\right\}^{1/k}(1-p)^{-1/k}\].
It follows that \[\sigma(1-e^{-x})\sim\sqrt{k-2}\left\{\frac{\Gamma((k+1)/2)}{k\sqrt{\pi}\Gamma(k/2)}\right\}^{1/k}e^{x/k}\].
It follows that $\sigma(1-e^{-x})$ grows exponentially with $x$ and is therefore rapidly varying rather than regularly varying with a finite index.
Consequently, the half-perimeter density does not follow the asymptotic form in Equation~\eqref{eq:C1form}.
\smallskip

\textbf{Heavy-tailed and Gaussian additive mixture.}
Suppose that $X(s)=\Lambda+W(s)$ for $\Lambda\in\text{RV}_{\infty}(\varrho)$ independent of $W$, with $\varrho>0$.
Assume moreover that $\Lambda$ admits a density $f_\Lambda$ satisfying
\[
    f_\Lambda(u)\sim\varrho L_\Lambda(u)u^{-(\varrho+1)},\qquad u\to\infty.
\]
Beginning with the half-perimeter density, we have the first-order Minkowski functional
\begin{align*}
    \mathbb{E}[\rho_1(u_p-\Lambda)] &= \frac{1}{2\pi}\int_{\mathbb{R}}ze^{-\frac12 z^2}\bar F_\Lambda(u_p-z)\,dz = \frac{1}{\sqrt{2\pi}}\int_{\mathbb{R}}\varphi(z)f_\Lambda(u_p-z)\,dz \sim \frac{\varrho}{\sqrt{2\pi}}L_\Lambda(u_p)u_p^{-(\varrho+1)},
\end{align*}
as $p\to1$. The last equivalence follows from $(f_\Lambda*\varphi)(u)\sim f_\Lambda(u)$ as $u\to\infty$, since $f_\Lambda$ is regularly varying and $\varphi(u)=o(f_\Lambda(u))$.
This gives the half-perimeter density
$$
    C_1^\ast(E_p^X)\sim\frac{\sqrt{2\pi\lambda_2}}{4}\varrho L_\Lambda(u_p)u_p^{-(\varrho+1)}
$$
as $p\rightarrow1$.
By stationarity, we also have the area density
$$
    C_2^\ast(E_p^X)=1-p\sim L_\Lambda(u_p)u_p^{-\varrho}
$$
as $p\rightarrow1$.
The half-perimeter and area density ratio is therefore given by
\begin{equation}\label{eq:HTGaussLocLKCratio}
    \frac{C_1^\ast(E_p^X)}{C_2^\ast(E_p^X)}\sim\frac{\sqrt{2\pi\lambda_2}}{4}\varrho u_p^{-1}
\end{equation}
as $p\rightarrow1$, and we follow the form of \eqref{eq:LKC-ratio-assumption} with $c_X=\frac{\sqrt{2\pi\lambda_2}}{4}\varrho$ and $\sigma(p)=u_p^{-1}$. Composing the scaling function with the standard exponential distribution function gives $\sigma(1-e^{-x})\to0$, $x\to\infty$.
More precisely, regular variation of the marginal tail implies
$\sigma(1-e^{-x})=\exp\{-x/\varrho+o(x)\},\qquad x\to\infty.
$Thus, $x\mapsto\sigma(1-e^{-x})$ is rapidly varying with index $-\infty$, rather than regularly varying with a finite negative index.
Consequently, the half-perimeter density does not follow the asymptotic form in Equation~\eqref{eq:C1form}.
\smallskip

\section{The scaling properties of the shot-noise random field}\label{supp:shot-noise}

We consider an archetypal example of a non-continuous field, the shot-noise random field, defined as \cite{BD16}. Suppose $\Pi$ is a stationary Poisson point process on $\mathbb{R}^2\times\mathbb{R}_+$ with intensity $\nu\textit{Leb}_{\mathbb{R}^2}\, dF_R$, and define, for $s \in \R^2$,  \begin{equation}\label{ShotNOISEfield}
X(s)=\sum_{(s_i,r_i)\in\Pi}\mathbbm{1}_{r_i D}(s-s_i),
\end{equation}
where $D$ is the unit disc in $ \R^2$.  Despite the loss of differentiability of this field which does not fall within Assumption \ref{A1}, Proposition 2.9 in  \cite{bierme2018lipschitz}  shows that it is still possible to define the expected LKCs for the excursion set of $X$, due to its locally positive reach. 
Suppose the first and second moments of the radial variable $R$ are finite, and define $\alpha:=\pi\mathbb{E}[R^2]$ and $\rho:=2\pi\mathbb{E}[R]$.
Section 2.3 of \cite{bierme2018lipschitz} derived the half-perimeter and area density ratio, giving
\begin{align*} 
    \frac{C_1^{\ast}(E^X_p)}{C_2^{\ast}(E^X_p)} =& \frac{\nu\rho}{2}\frac{\frac{(\nu\alpha)^{\left\lfloor u_p\right\rfloor}}{\left\lfloor u_p\right\rfloor!}}{\sum_{n> u_p}\frac{(\nu\alpha)^n}{n!}} = \frac{\nu\rho}{2} \frac{\frac{(\nu\alpha)^{\left\lfloor u_p\right\rfloor}}{\left\lfloor u_p\right\rfloor!}}{\frac{(\nu\alpha)^{\left\lfloor u_p\right\rfloor+1}}{\left(\left\lfloor u_p\right\rfloor+1\right)!} + \frac{(\nu\alpha)^{\left\lfloor u_p\right\rfloor+2}}{\left(\left\lfloor u_p\right\rfloor+2\right)!}+\frac{(\nu\alpha)^{\left\lfloor u_p\right\rfloor+3}}{\left(\left\lfloor u_p\right\rfloor+3\right)!}+\dots}\\
    =&  \frac{\nu\rho}{2} \frac{1}{\frac{\nu\alpha}{\left\lfloor u_p\right\rfloor+1}\left[1 + \frac{\nu\alpha}{\left(\left\lfloor u_p\right\rfloor+2\right)}+\frac{(\nu\alpha)^{2}}{\left(\left\lfloor u_p\right\rfloor+2\right)\left(\left\lfloor u_p\right\rfloor+3\right)}+\dots\right]}= \frac{\rho}{2\alpha}\left(\left\lfloor u_p\right\rfloor+1\right)\left(1+o(1)\right) \sim \frac{\rho}{2\alpha} u_p,
\end{align*}
as $p\rightarrow1$.
\cite{bierme2018lipschitz} note that the marginal distribution $X(0)\sim\text{Poisson}(\nu\alpha)$; however, notice here that the asymptotic behaviour above is independent of $\nu$.
Putting this together, we follow the form in \eqref{eq:LKC-ratio-assumption} as $p\rightarrow1$ with 
$
    c_X = {\rho}/{(2\alpha)}
$
and $\sigma(p)=u_p$. 

In order to solve for $\beta$, we use the fact that the margins are $\text{Poisson}(\nu\alpha)$. 
We inspect the function
$$
    h(x):=F_{\text{Pois.}(\nu\alpha)}^{-1}(1-e^{-x})
$$
where $F_{\text{Pois.}(\nu\alpha)}^{-1}(p)=\min \left\{n\,:\,\PP(X\leq n)\geq p\right\}$. To see how this is regularly varying, we first study the distribution function. Using the probability mass function as a lower bound, and Chernoff bound as an upper bound, we have the inequality
$$
    e^{-\nu\alpha}\frac{(\nu\alpha)^{\left\lfloor u_p\right\rfloor}}{\left\lfloor u_p\right\rfloor!}\leq\PP\left(X(s)\geq u_p\right) \leq e^{-tu_p}e^{\nu\alpha(e^t-1)}.
$$
Using Stirling's approximation on the left-hand side, and minimizing the right-hand expression with respect to $t>0$, we recover
\begin{equation}\label{eq:shotnoise-squeeze}
    \frac{1}{\sqrt{2\pi}}u_p^{-1/2}\exp\left\{-u_p\ln\left(\frac{u_p}{\nu\alpha}\right)+u_p-\nu\alpha\right\}\leq \PP\left(X(s)\geq u_p\right) \leq \exp\left\{u_p-\nu\alpha-u_p\ln\left(\frac{u_p}{\nu\alpha}\right)\right\}.
\end{equation}
Taking negative logarithms and using the definition of $u_p$, we have
\begin{equation}\label{eq:shotnoise-squeeze2}
    -u_p+\nu\alpha+u_p\ln\left(\frac{u_p}{\nu\alpha}\right)\leq -\ln(1-p) \leq \frac{1}{2}\ln(2\pi) +\frac{1}{2}\ln u_p +u_p\ln\left(\frac{u_p}{\nu\alpha}\right)-u_p+\nu\alpha.
\end{equation}
Notice that both the left and right-hand sides of \eqref{eq:shotnoise-squeeze2} are asymptotically equivalent to $u_p\ln u_p$ as $p\rightarrow1$. Therefore, by a squeeze-theorem-type argument, we have, 
as $p\rightarrow1$, 
\begin{align*}
    -\ln(1-p) \sim u_p\ln u_p.
\end{align*} 
In order to solve for $u_p$, we make use of Lambert's $W$ function, resulting in
$$
   u_p \sim \exp\left\{W\left( -\ln(1-p)\right)\right\}
$$
as $p\to 1$. Using the property $W(x) e^{W(x)}=x$, we have, as $p\rightarrow1$, 
$$
    u_p \sim \frac{-\ln(1-p)}{W\left(-\ln(1-p)\right)}\sim \frac{-\ln(1-p)}{\ln\left(-\ln(1-p)\right)}
$$
  (see \citea{NIST:DLMF}, equation 4.13.1). Therefore,
$$
    \sigma(p)=u_p\sim \frac{-\ln(1-p)}{\ln(-\ln(1-p))}
$$
as $p\rightarrow1$. This results in $\sigma(1-e^{-x})\sim x/\ln x\in\text{RV}_1$, and the scaling index $\beta=1$.

In addition, note that the half-perimeter density does not follow the form in \eqref{eq:C1form}. To see this, use Stirling's approximation to observe that
\begin{align*}
    C_1^\ast(E_p^X) =& \frac{1}{\sqrt{2\pi}}\frac{(\nu\alpha)^{\left\lfloor u_p\right\rfloor}}{\left\lfloor u_p\right\rfloor!}\sim \frac{1}{\sqrt{2\pi u_p}}\exp\left\{u_p\left[1+\ln\left(\frac{\nu\alpha}{u_p}\right)\right]\right\}
\end{align*}
as $p\rightarrow1$. Because we can write
$$
    u_p=\frac{-\ln(1-p)}{\ln\left(-\ln(1-p)\right)} + o(1),
$$
we have that 
$$
    \ln u_p \sim \ln(-\ln(1-p))-\ln\ln(-\ln(1-p)) + o(1)
$$
as $p\rightarrow1$. Therefore, 
\begin{align*}
    u_p &\left[1+\ln\left(\frac{\nu\alpha}{u_p}\right)\right] =  \frac{-\ln(1-p)}{\ln\left(-\ln(1-p)\right)}\left[1+\ln(\nu\alpha)-\ln(-\ln(1-p))+\ln\ln(-\ln(1-p))+o(1)\right] \\
    =&\frac{-\ln(1-p)}{\ln\left(-\ln(1-p)\right)} -\frac{\ln(\nu\alpha)\ln(1-p)}{\ln\left(-\ln(1-p)\right)} + \ln(1-p)-\frac{\ln(1-p)\ln\ln(-\ln(1-p)))}{\ln\left(-\ln(1-p)\right)} + o(1).
\end{align*}
Putting this into $C_1^\ast(E_p^X)$, we have the form
\begin{align*}
    C_1^\ast(E_p^X)\sim& \frac{1}{\sqrt{2\pi}}(1-p)\left[\frac{-\ln(1-p)}{\ln\left(-\ln(1-p)\right)}\right]^{-1/2} \\
    &\times\exp\left\{\frac{-\ln(1-p)}{\ln\left(-\ln(1-p)\right)}\left[1+\ln(\nu\alpha)+\ln\ln(-\ln(1-p))+o(1)\right]\right\}
\end{align*}
as $p\rightarrow1$, and not that of \eqref{eq:C1form}.

\section{On the scaling index and tail dependence}\label{supp:beta-dependence}
Values of $\beta$ relate to the scaling index and to extremal dependence. To see this, recall that, under the representation of Corollary \ref{cor:scaling},  $\sigma(1-e^{-tx})\sim t^\beta \sigma(1-e^{-x})$ for $x\rightarrow\infty$. Loosely speaking, set $\tilde{p}=1-e^{-tx}$ and fix $p=1-e^{-x}$ for a large fixed value $x$. Allowing $t\rightarrow\infty$, or $\tilde{p}\rightarrow1$, we have,  that 
$$
    \sigma(\tilde{p})\rightarrow\begin{cases}\infty &,\,\beta>0\\
    c<\infty &,\,\beta=0\\
    0&,\,\beta<0\end{cases}.
$$
Under the representation \eqref{eq:C1form}, a larger $\beta>0$ implies that $\sigma(p)\to\infty$ more rapidly, corresponding to increasingly localised conditional excursion sets as $p\to1$. By Proposition~\ref{prop:scaling-AI}, $\sigma(p)\to\infty$ implies asymptotic independence. Conversely, $\beta<0$ implies that $\sigma(p)\to0$, corresponding to increasingly extensive conditional excursion sets at high levels. However, the value of $\beta$ alone does not provide an ordering of the strength of extremal dependence.
\smallskip

It is interesting to note that the setting of $\sigma(p)\rightarrow c\in[1,\infty)$ does not exclude the possibility of $\chi_p(0,s)=0$ for certain choices of $s\in S$, and $\chi_p(0,s)>0$ for other choices of $s\in S\setminus\{0\}$. 
An example of a random field satisfying \ref{A1} with scaling index $\beta\leq0$ but is not asymptotically dependent is given by
$$
    X(t_1,t_2)=\Lambda\cos(t_1+\Theta)+ W(t_1,t_2),
$$
where $\Lambda\sim\text{Pareto}(1,1)$, $\Theta\sim\text{Unif}(0,2\pi)$, and $W$ is a stationary Gaussian random field.

\begin{figure}[H]
    \centering
    \includegraphics[height=110pt, trim={20 40 0 40},clip]{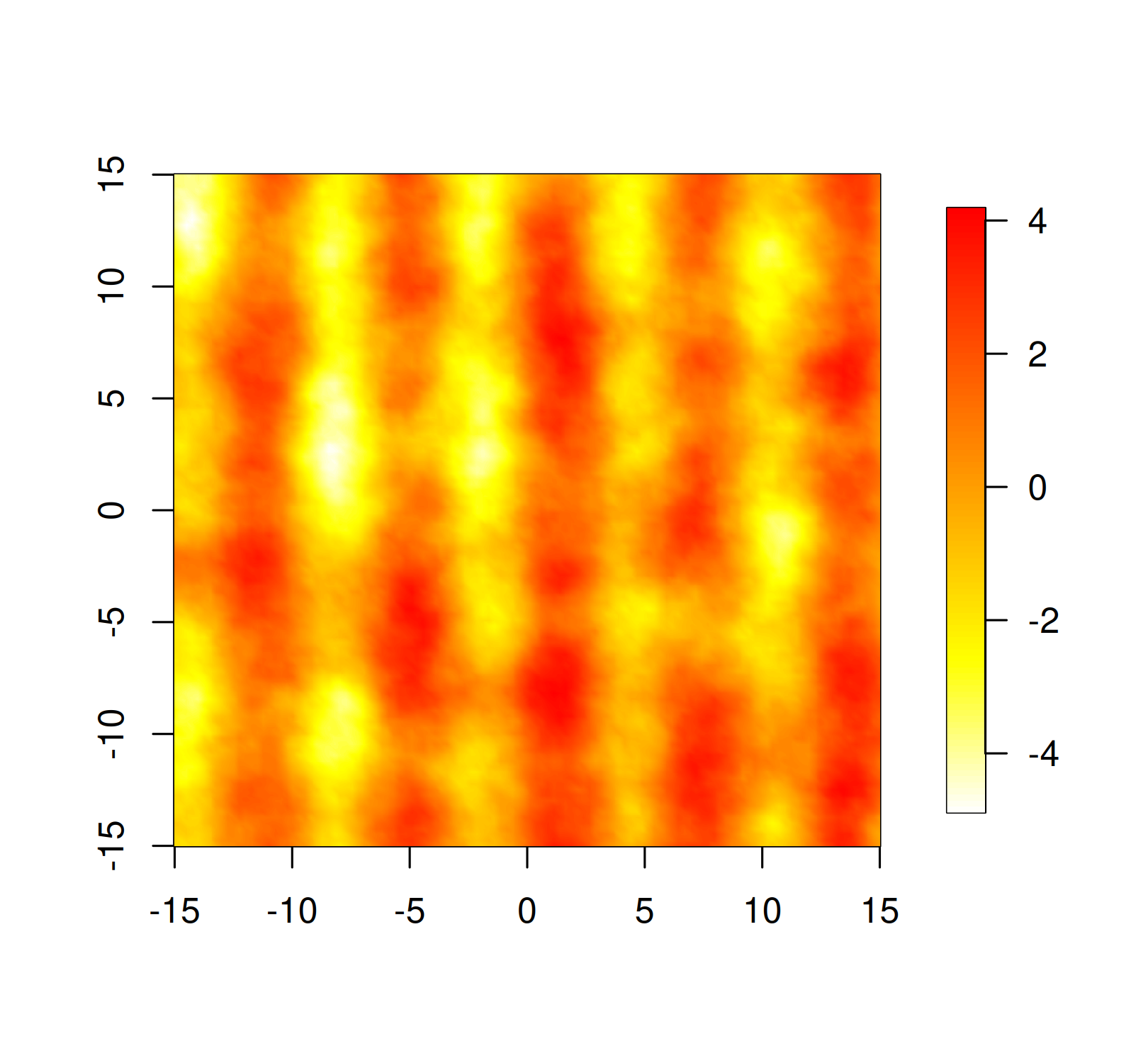}
    \includegraphics[height=110pt, trim={20 40 0 40},clip]{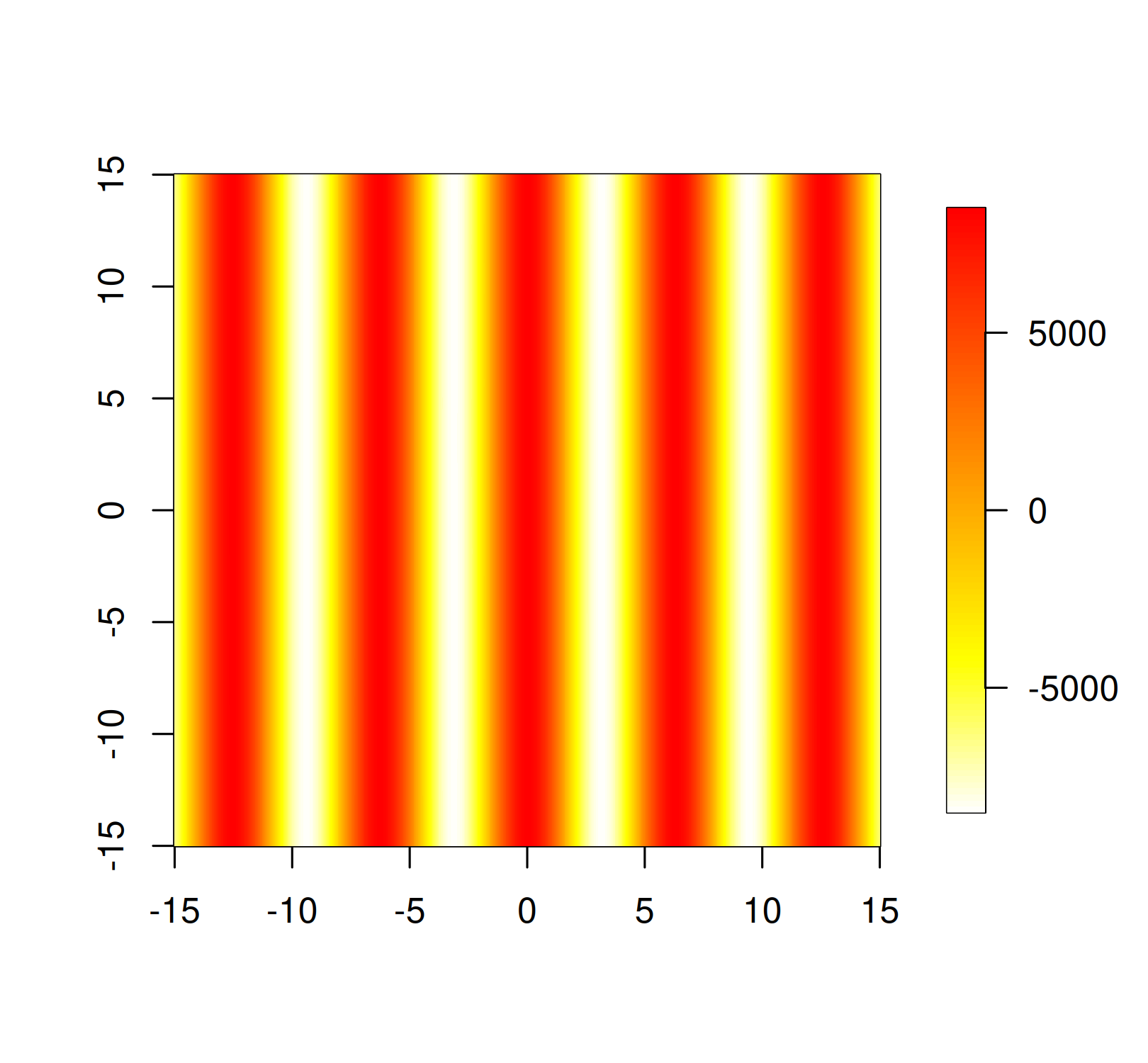}
    \caption{Observations of the random field $X(t_1,t_2)=R\cos(t_1+\Theta)+ W(t_1,t_2)$ conditioned on the event $X(0,0)>u_p$ for $p=0.50$ (left) and $p=0.9999$ (right).}
    \label{fig:HTcosUnif_Gauss_cond}
\end{figure}

Note that in the right tail, the Pareto component dominates the behaviour of $X(t_1,t_2)$. This can be seen by examining the survival function. For a threshold $u$,
\begin{align*}
    \PP(X(t_1,&t_2)>u) \sim \ \PP(\Lambda\cos(t_1+\Theta)>u)\int\limits_{0}^{2\pi}\PP\left(\Lambda>\frac{u_p}{\cos(t_1+\theta)}\right)\mathbbm{1}_{(0,\infty)}(\cos(t_1+\theta))\frac{1}{2\pi}d\theta\\
    =& \frac{1}{2\pi u}\int\limits_{0}^{2\pi}\cos(t_1+\theta)\mathbbm{1}_{(0,\infty)}(\cos(t_1+\theta))d\theta = \frac{1}{\pi u}.
\end{align*}
Above, the first asymptotic equality is as $u\rightarrow\infty$, the second line holds by the law of total probability, and the third line holds by the fact that $\Lambda\sim\text{Pareto}(1,1)$.
A direct consequence of the above is that $u_p\sim \pi^{-1}(1-p)^{-1}$ as $p\rightarrow1$, and the area density is given by $C_2^\ast(E_p^X)\sim\pi^{-1} u_p^{-1}$ as $p\rightarrow1$.
Figure~\ref{fig:HTcosUnif_Gauss_cond} shows observations above the $p=0.50$ and $p=0.9999$ marginal quantile at the origin $(t_1,t_2)=(0,0)$. The quantile $u_{0.50}=0$ is due to symmetry of each of the additive components of $X(t_1,t_2)$, and $u_{0.9999}=3183$ is an approximate result based on the heavy-tailed Pareto component. 

In order to find the scaling index, we do as before and find the ratio of the half-perimeter and area densities. For the half-perimeter density, we  follow the procedure for Gaussian location and scale mixtures outlined in Supplementary Material~\ref{supp:calculations} and obtain
\begin{align*}
   C_1^\ast(E_p^X) =   & \frac{\pi}{2}\sqrt{\lambda_2}\mathbb{E}\left[\rho_1\left(u_p-\Lambda\cos(t_1+\Theta)\right)\right]\\
    =&\frac{1}{4}\sqrt{\lambda_2}\int\limits_{-\infty}^\infty z e^{-\frac{1}{2}z^2}\PP\left(\Lambda\cos(t_1+\Theta)>u_p-z\right)dz
    \sim\frac{1}{4\pi}\sqrt{\lambda_2}\int\limits_{-\infty}^\infty z (u_p-z)^{-1}e^{-\frac{1}{2}z^2}dz\\
    \sim& \frac{1}{4\pi}\sqrt{\lambda_2}u_p^{-1}\int\limits_{-\infty}^\infty z (1-u_p^{-1}z)^{-1}e^{-\frac{1}{2}z^2}dz=\frac{1}{2\sqrt{2\pi}}\sqrt{\lambda_2}u_p^{-2},
\end{align*}
where all asymptotic equivalences hold for $p\rightarrow1$. As a result, we follow the form of \eqref{eq:LKC-ratio-assumption} with $c_X=\frac{1}{2\sqrt{2\pi}}\sqrt{\lambda_2}$ and $\sigma(p)=u_p^{-1}$. 
As a result, we follow the form of~\eqref{eq:LKC-ratio-assumption} with
$c_X=\frac{1}{2\sqrt{2\pi}}\sqrt{\lambda_2},
$ $\sigma(p)=u_p^{-1}.$ Since $u_p\sim{\pi(1-p)}^{-1}$, we have $\sigma(1-e^{-x})\sim\pi e^{-x},$
which is rapidly decreasing rather than regularly varying with a finite index. This example therefore falls outside the finite-$\beta$ framework of Equation~\eqref{eq:C1form}, while illustrating that a non-diverging spatial scaling does not by itself imply asymptotic dependence.
It remains to establish non-asymptotic dependence. To do so, note that the point $X(2\pi,0)\approx \Lambda$ when $X(0,0)>u_p$ for $p\rightarrow1$, but $X(\pi,0)\approx -\Lambda$ when $X(0,0)>u_p$ for $p\rightarrow1$. So, 
there exists locations $s=(t_1,t_2)$ of the form $(\pi+2\pi n,0)$ for $n=0,1,2,\dots$ such that $\lim_{p\rightarrow1}\chi(0,s)=0$.

\bigskip 
If the continuity assumption on sample paths of $X$ is relaxed, a concrete example of this is given in the max-stable random field constructed from random disks in \cite{schlather2002models}.
The random field is constructed in the following manner;
$$
    X(s)=\max_{i\geq 1} \xi_i \mathbbm{1}_{B(s_i,r_i)}(s),
$$
where $B(s,r)$ is a ball centred at site $s$ with radius $r>0$, and $\xi_i$ is a standard Fr\'{e}chet observation.
A direct result of this definition is that the marginal distribution follows a standard Fr\'{e}chet distribution. 
\begin{figure}[H]
    \centering    \includegraphics[height=110pt, trim={20 40 0 40},clip]{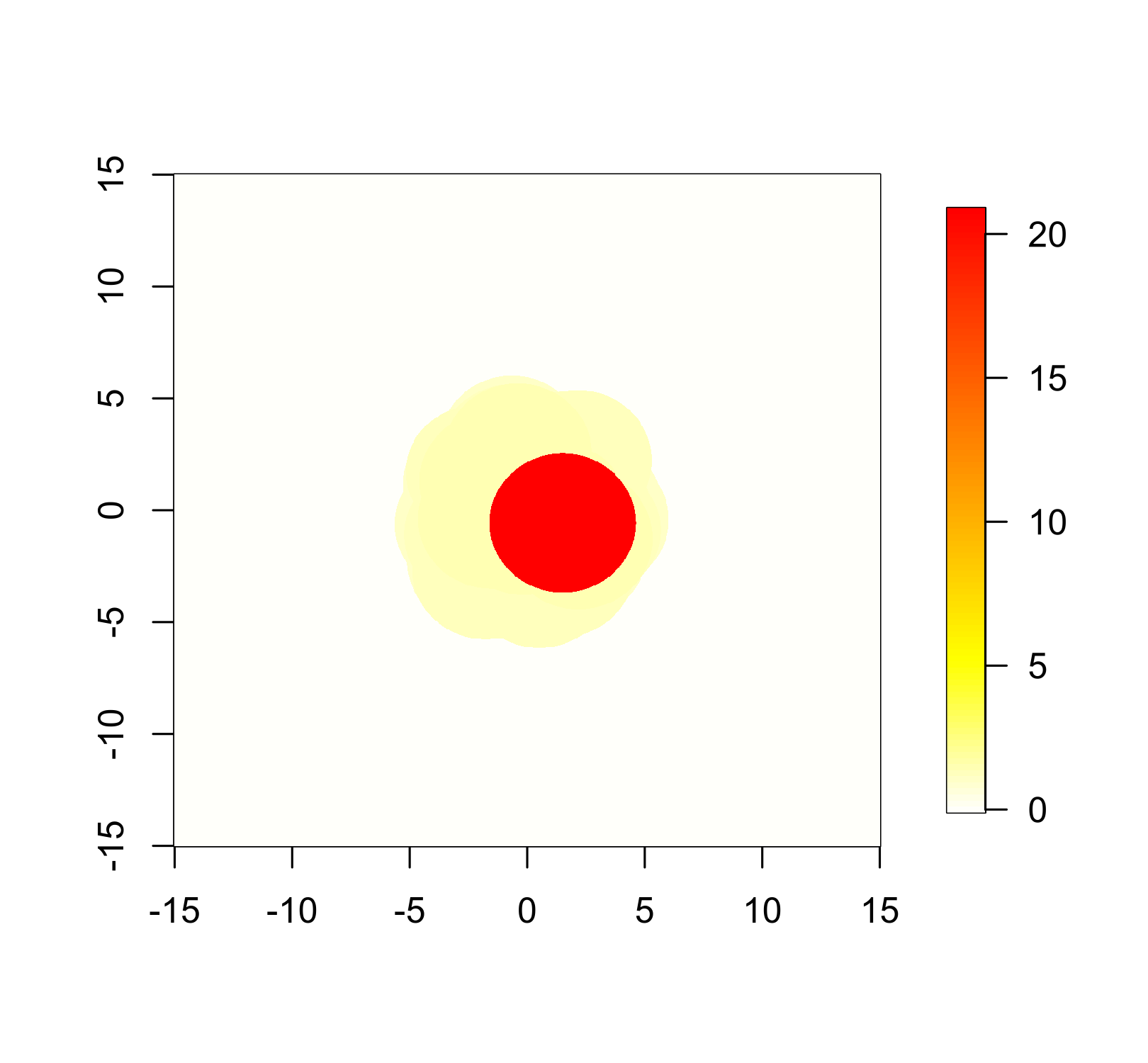}
    \includegraphics[height=110pt, trim={20 40 0 40},clip]{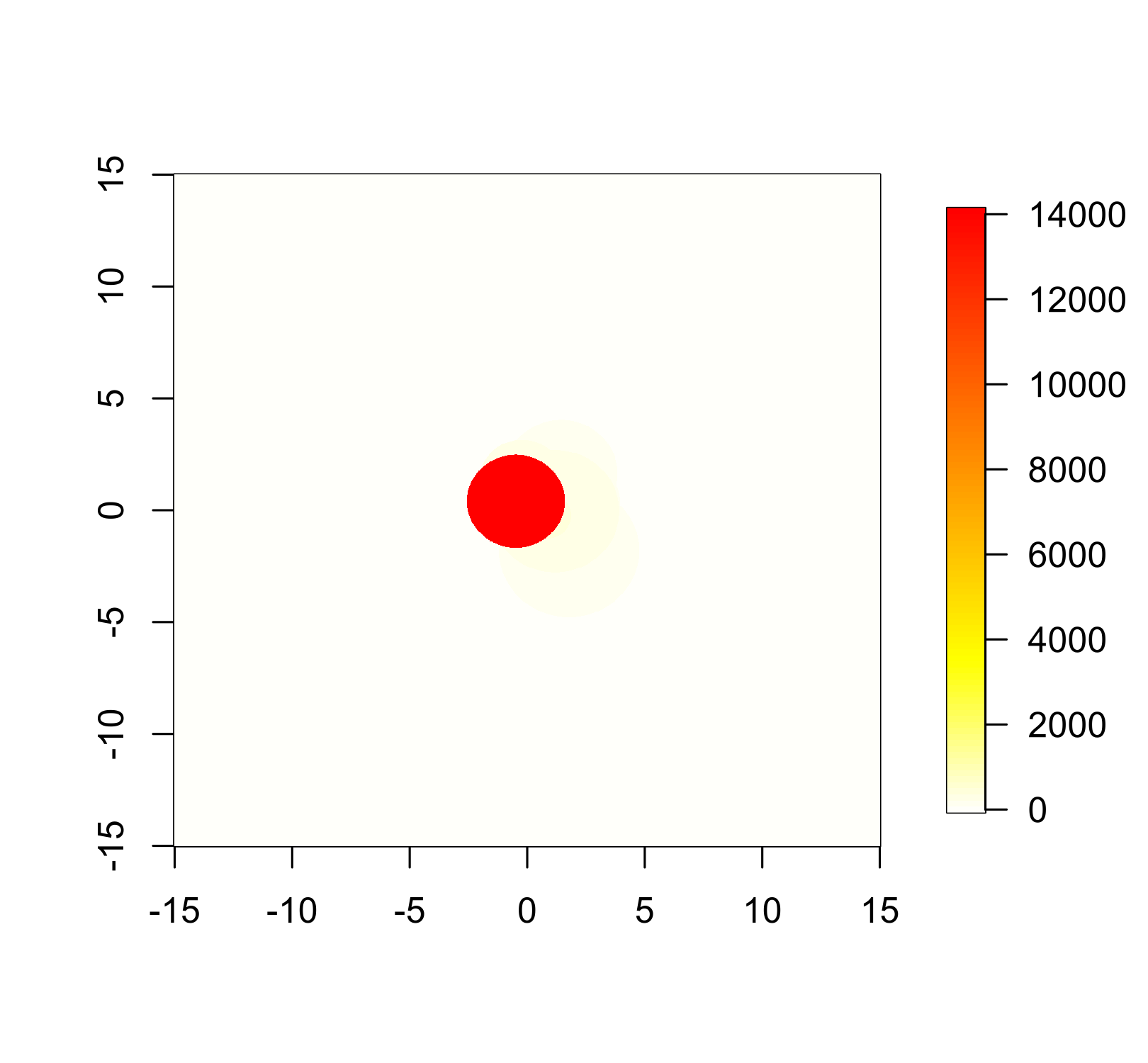}
    \caption{Observations of \cite{schlather2002models}'s  max-stable random disk field conditioned on the event $X(0)>u_p$ for $p=0.50$ (left) and $p=0.9999$ (right). The radial distribution is given by $F_R(r)=1-\left(1-\theta r^2\right)^{1/2}$ with $\theta=0.10$.}
    \label{fig:randomDisks_cond}
\end{figure}
We have already established in Supplementary Material~\ref{supp:calculationsBIS} that the scaling function $\sigma(p)$ is constant in $p$; therefore, we fall under the null hypothesis that $\beta\leq0$.  Let $0$ denote the site at the origin, and suppose the disks' radial component has distribution function $F_R(r)=1-(1-\theta r^2)^{1/2}$ with $\theta>0$.
Figure~\ref{fig:randomDisks_cond} shows observations above the $p=0.50$ and $p=0.9999$ marginal quantile at the origin for this particular random field with $\theta=0.10$.
By this definition, radial values lie in the interval $r\in[0,{1}/{\sqrt{\theta}}]$. Therefore, a disk covers both the origin and a location $s\in S$ if $|s|\leq {2}/{\sqrt{\theta}}$.
As a result, the pairwise coefficient of tail dependence has limiting form
\begin{align*}
	\chi_p(0,s)
    \rightarrow
    \begin{cases}
    c>0&;\,\,|s|\leq\frac{2}{\sqrt{\theta}}\\
    0&;\,\,|s|>\frac{2}{\sqrt{\theta}}
    \end{cases}
\end{align*}
as $p\rightarrow1$.
Therefore, while this max-stable random field does indeed satisfy the null hypothesis that $\beta\leq0$, it is not asymptotically dependent.

\section{On the tail behaviour of the scale mixture component and its effect on $\beta$}\label{supp:scale-mixtures}

Below, we describe the effect  that the tail behaviour $\Lambda$ in the scale mixture $X(s)=\Lambda W(s)$ has on the scaling index $\beta$, and therefore on the asymptotic dependence properties of $X$.

\begin{theorem}\label{thm:gauss-scale-mixture-beta}
    Let $\{{W}(s)\}_{s\in S}$ be a stationary Gaussian random field, $\Lambda$ be scalar random variable independent of $W$, $c>0$, $\gamma\in\mathbb{R}$,$\delta>0$, and $\varrho>0$. Then the following hold,
    \begin{enumerate}
        \item\label{lab:polynomial-decay} If $\PP(\Lambda>\lambda)\sim c\lambda^{-\varrho}$ as $\lambda\rightarrow
        \infty$, then $\beta=0$.
        \item\label{lab:exponential-decay} If $\PP(\Lambda>\lambda)\sim c\lambda^{\gamma}\exp\{-\delta\lambda^\varrho\}$ as $\lambda\rightarrow
        \infty$, then $\beta=1/2$. 
    \end{enumerate}
\end{theorem}
\begin{proof}[Proof of Theorem \ref{thm:gauss-scale-mixture-beta}]
    Beginning with case~\ref{lab:polynomial-decay}, suppose that $\PP(\Lambda>\lambda)\sim c\lambda^{-\varrho}$. 
    Computing the Minkowski functionals \eqref{Rho0} 
  and \eqref{Rho1} results in \[        \mathbb{E}\left[\rho_0(u_p/\Lambda)\right] \sim \frac{c}{\sqrt{2\pi}}u_p^{-\varrho}\int\limits_0^\infty z^\varrho e^{-\frac{1}{2}z^2}dz = \frac{c}{\sqrt{2\pi}}2^{(\varrho-1)/2}\Gamma((\varrho+1)/2)u_p^{-\varrho}\] 
    and   \[       \mathbb{E}\left[\rho_1(u_p/\Lambda)\right] \sim \frac{c}{{2\pi}}u_p^{-\varrho}\int\limits_0^\infty z^{\varrho+1} e^{-\frac{1}{2}z^2}dz = \frac{c}{{2\pi}}2^{\varrho/2}\Gamma(\varrho/2+1)u_p^{-\varrho}
 \]
    as $p\rightarrow1$. Using the form \eqref{eq:LKC-ratio-assumption} in Theorem~\ref{thm:LKCratio} gives $\sigma(p)=1$, and therefore $\beta=0$. 

    On the other hand, if $\PP(\Lambda>\lambda)\sim c\lambda^{\gamma}\exp\{-\delta\lambda^\varrho\}$ as $\lambda\rightarrow\infty$, then Laplace's method leads to the following asymptotic form of the Minkowski functionals \eqref{Rho0} and \eqref{Rho1},
    \begin{align*}
        \mathbb{E}\left[\rho_0(u_p/\Lambda)\right] &\sim \frac{c}{\sqrt{2\pi}}u_p^{\gamma}\int\limits_0^\infty z^{-\gamma}\exp\left\{-\frac{1}{2}z^2-\delta u_p^\varrho z^{-\varrho}\right\}dz \\
        &\sim  c(\varrho+2)^{-1/2}(\varrho\delta)^{-\gamma/(\varrho+2)}u_p^{-\frac{\varrho\gamma}{\varrho+2}-\gamma}\exp\left\{-\left(\frac{\varrho+2}{2\varrho}\right)(\varrho\delta)^{2/(\varrho+2)}u_p^{2\varrho/(\varrho+2)}\right\}
    \end{align*}
    and
    \begin{align*}
        \mathbb{E}\left[\rho_1(u_p/\Lambda)\right] &\sim \frac{c}{{2\pi}}u_p^\gamma\int\limits_0^\infty z^{1-\gamma}  \exp\left\{-\frac{1}{2}z^2-\delta u_p^\varrho z^{-\varrho}\right\}dz \\
        &\sim \frac{c}{\sqrt{2\pi}}(\varrho+2)^{-1/2}(\varrho\delta)^{(1-\gamma)/(\varrho+2)}u_p^{-\frac{\varrho(1-\gamma)}{\varrho+2}-\gamma}\exp\left\{-\left(\frac{\varrho+2}{2\varrho}\right)(\varrho\delta)^{2/(\varrho+2)}u_p^{2\varrho/(\varrho+2)}\right\}
    \end{align*}
    as $p\rightarrow1$.
    Using the form \eqref{eq:LKC-ratio-assumption} in Theorem~\ref{thm:LKCratio} gives $\sigma(p)=u_p^{\varrho/(\varrho+2)}$, where $u_p$ is the marginal quantile function of $\Lambda W(s)$. In the proof of Theorem 2 of \cite{huser2017GSMs}, it is established that, up to proportionality, $u_p\sim\left(-\ln(1-p)\right)^{(\varrho+2)/(2\varrho)}$ as $p\rightarrow1$, resulting in $\beta=1/2$. Hence the result. 
\end{proof}
\begin{remark*}
    Note that Theorems 2 and 3 in \cite{huser2017GSMs} prove that $\Lambda$ of condition~\ref{lab:polynomial-decay} leads to asymptotic dependence, and that $\Lambda$ of condition~\ref{lab:exponential-decay} leads to asymptotic independence. This is consistent with the value $\beta=0$ obtained under condition~\ref{lab:polynomial-decay}; however, we point out the interesting behaviour that, no matter the value of $\varrho$, scale mixtures satisfying condition~\ref{lab:exponential-decay} all lead to $\beta=1/2$.
\end{remark*}

\section{Additional  figures}\label{supp:SSfigs}

\begin{figure}[htpb]
    \centering
    \includegraphics[width=0.47\linewidth,trim={0 0 0 0},clip]{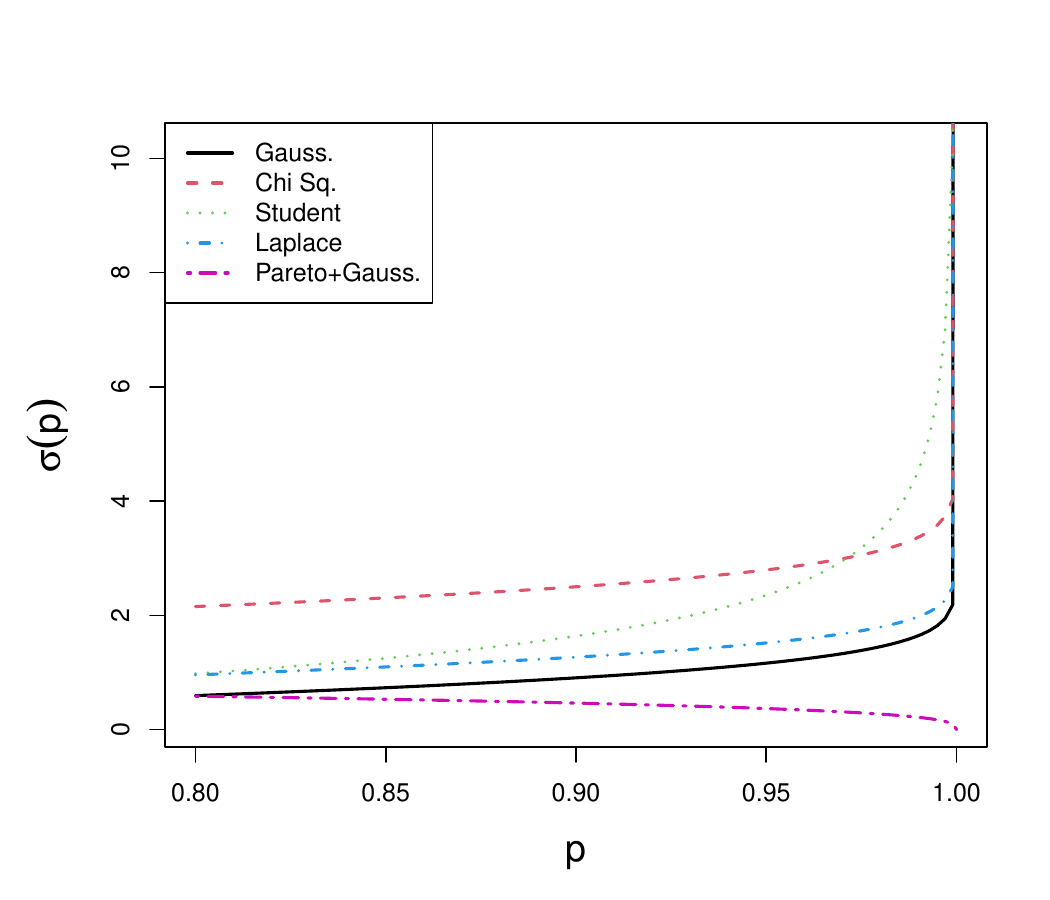}
    \includegraphics[width=0.47\linewidth,trim={0 0 0 0},clip]{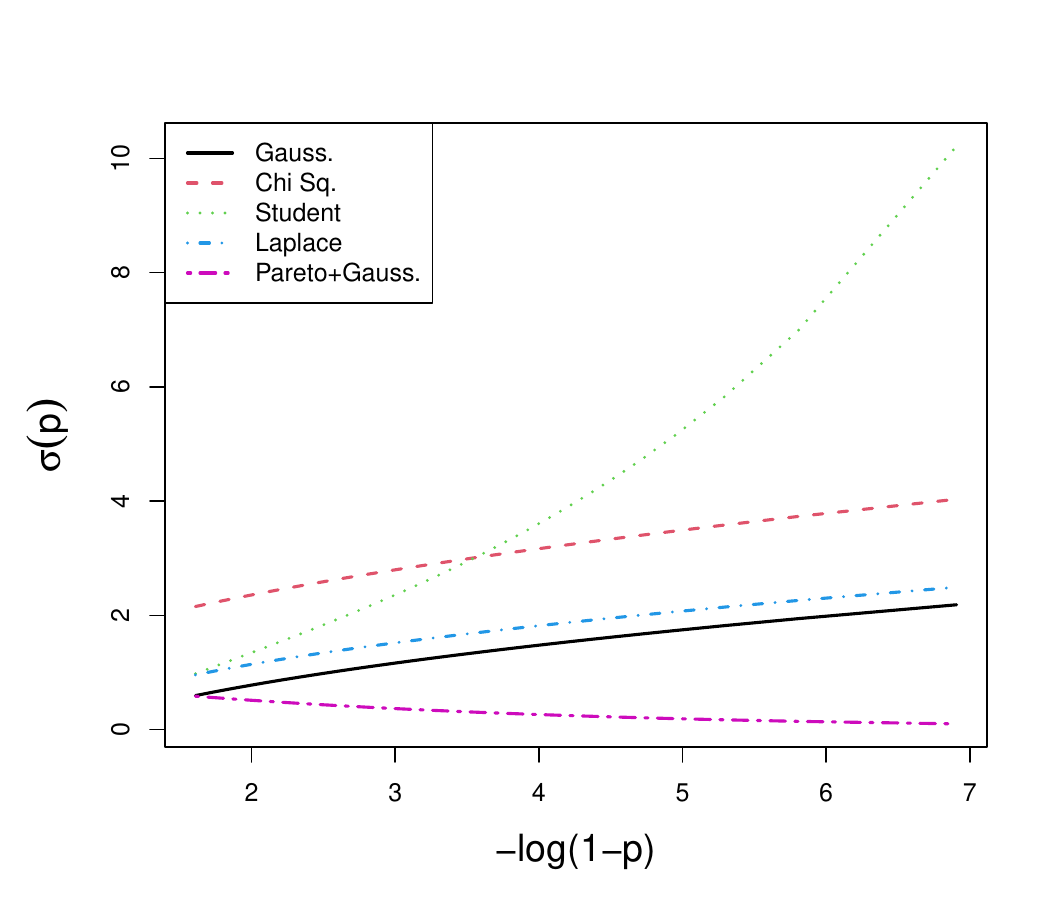}
    \vspace{-0.4cm}
    \caption{Values of the non-constant scaling functions $\sigma$ for random fields considered in Sections~\ref{sec:examples} and~\ref{sec:COUNTERexamples} on the original scale (left) and the $-\ln(1-p)$ scale (right).}
    \label{fig:sigmaplot}
\end{figure}

This section presents additional figures from the numerical simulation study, providing further illustrations of the results discussed in the main manuscript. Firstly, several of the functions $\sigma$ obtained in Sections~\ref{sec:examples} and~\ref{sec:COUNTERexamples} are illustrated in Figure~\ref{fig:sigmaplot}. 

\begin{figure}[ht!]
     \centering
     \includegraphics[width=0.32\linewidth, trim=0 10 0 50, clip]{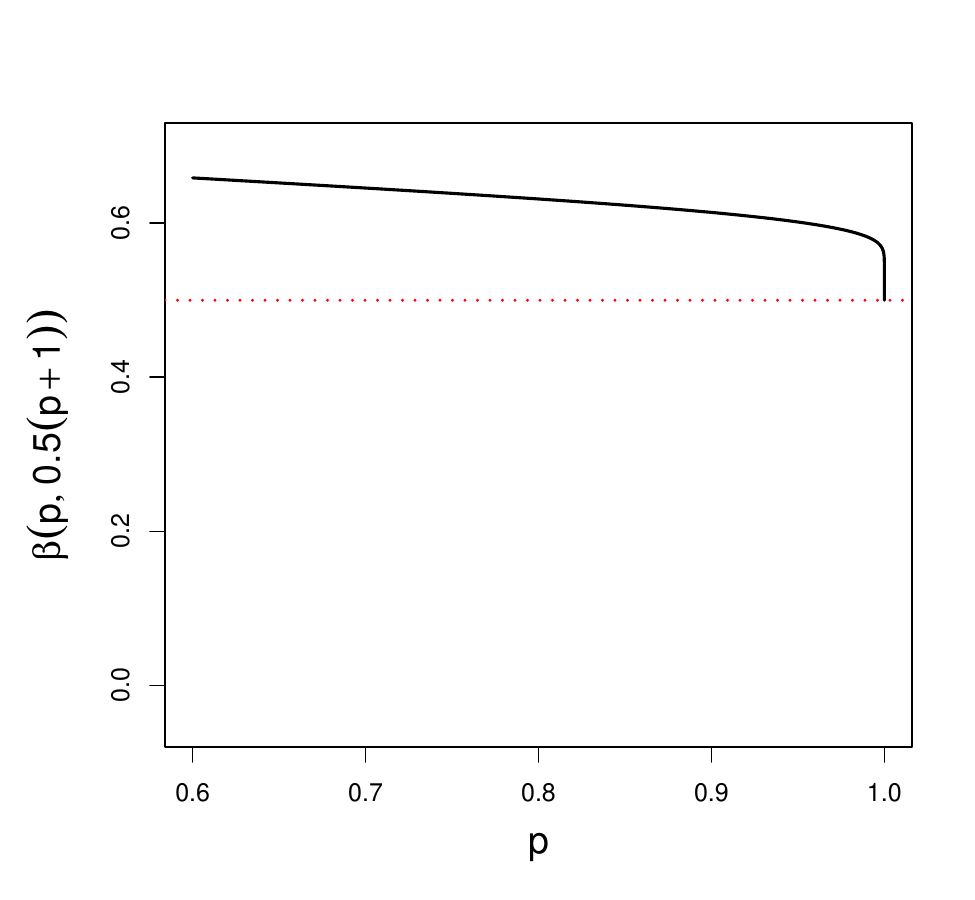}
    \includegraphics[width=0.32\linewidth, trim=0 10 0 50, clip]{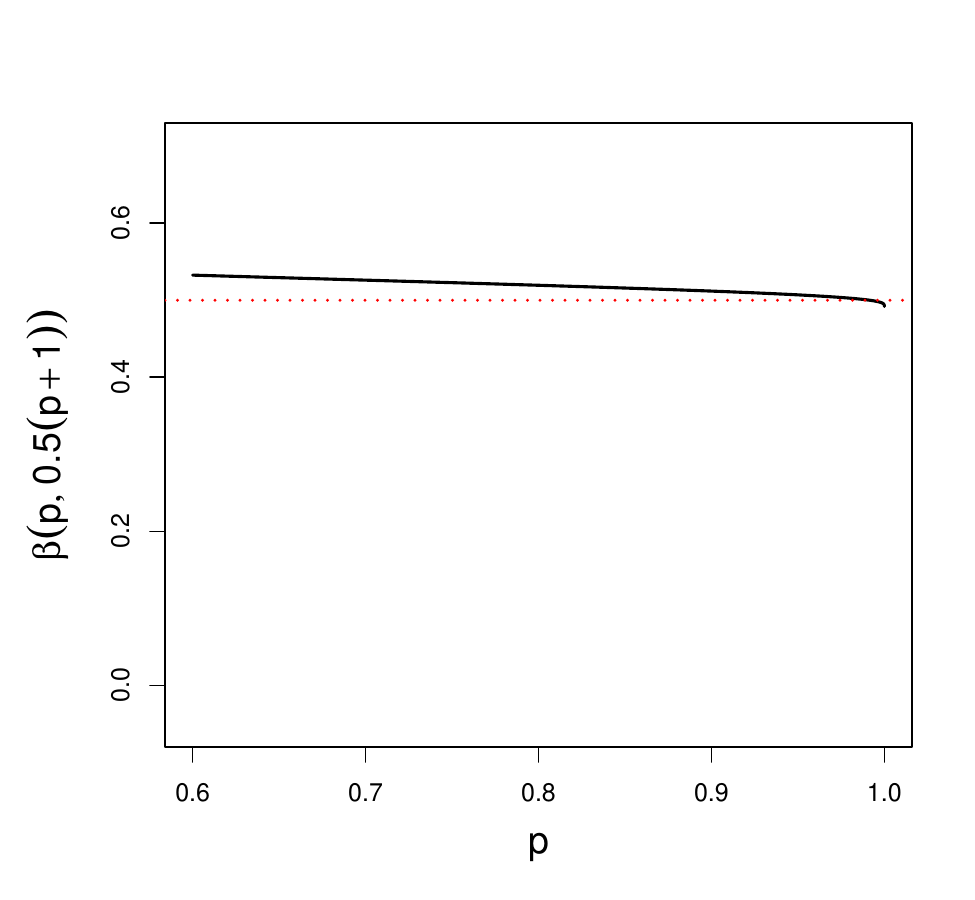}
    \includegraphics[width=0.32\linewidth, trim=0 10 0 50, clip]{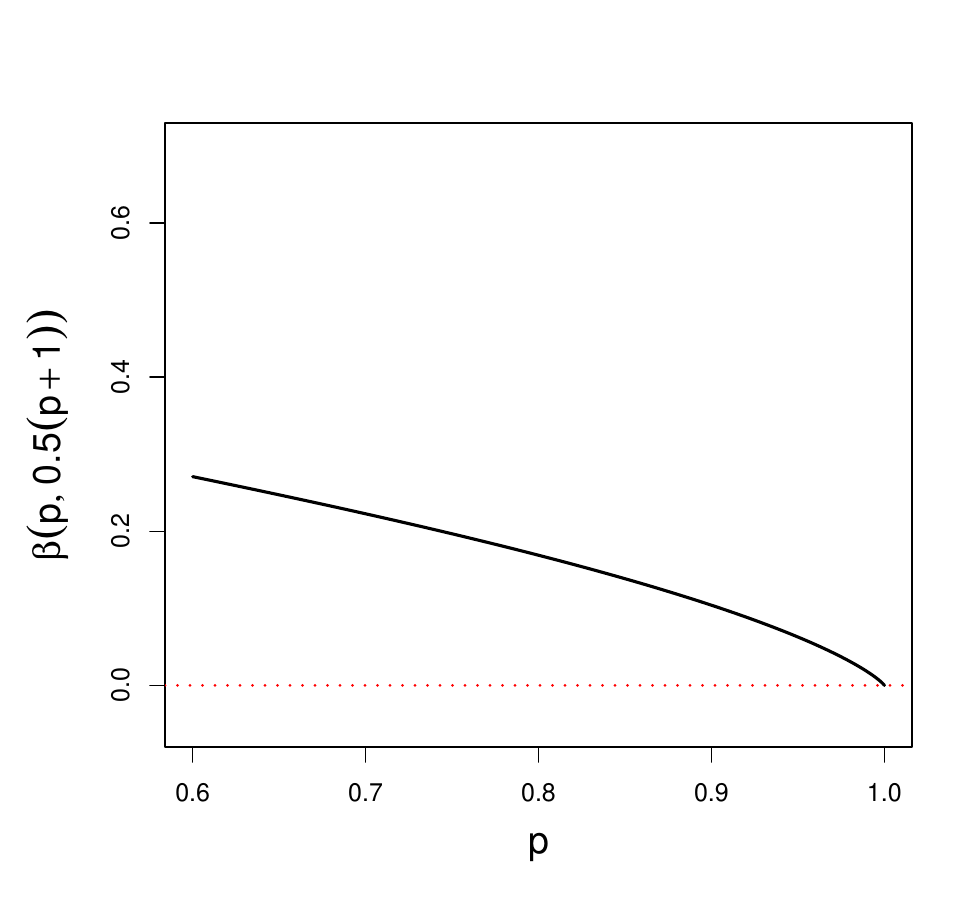}
     \caption{Values of 
 $\beta(p, (p+1)/2)$ (solid line) for the Gaussian, standardised chi-squared with $\kappa=3$ degrees of freedom, and Smith models (left to right), with limiting values $\beta$ indicated by the dotted line.}
     \label{fig:betap1p2}
 \end{figure}

In Figure \ref{fig:betap1p2}  we display the values of  
 $\beta(p, (p+1)/2)$ along with $\beta$ for several choices of  $p$ for the  Gaussian, standardised chi-squared, Smith models. Information on these models can be found in Section \ref{sec:examples}, and note that for the anisotropic Gaussian field with the curve is exactly identical to that of the isotropic Gaussian field. 
 
\begin{figure}[ht!]
    \centering
    \includegraphics[width=0.4\linewidth]{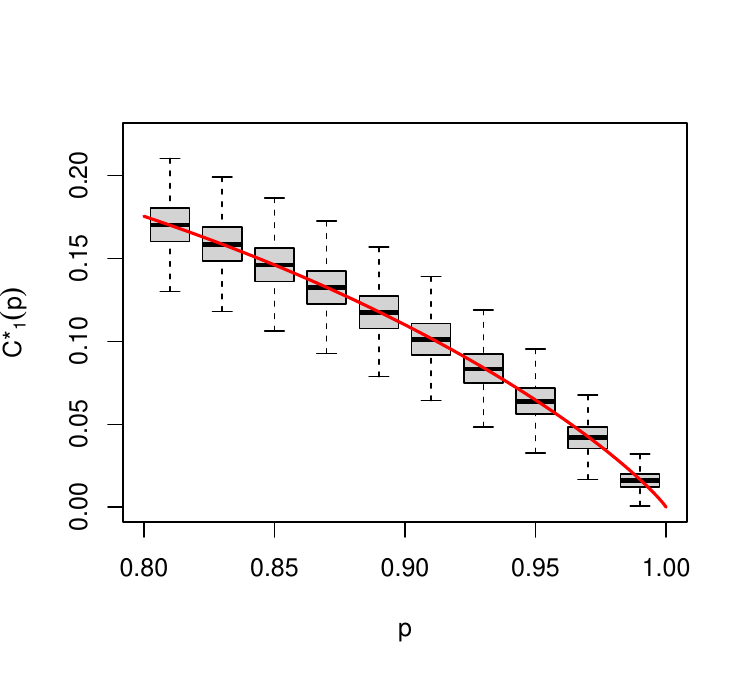}
    \includegraphics[width=0.4\linewidth]{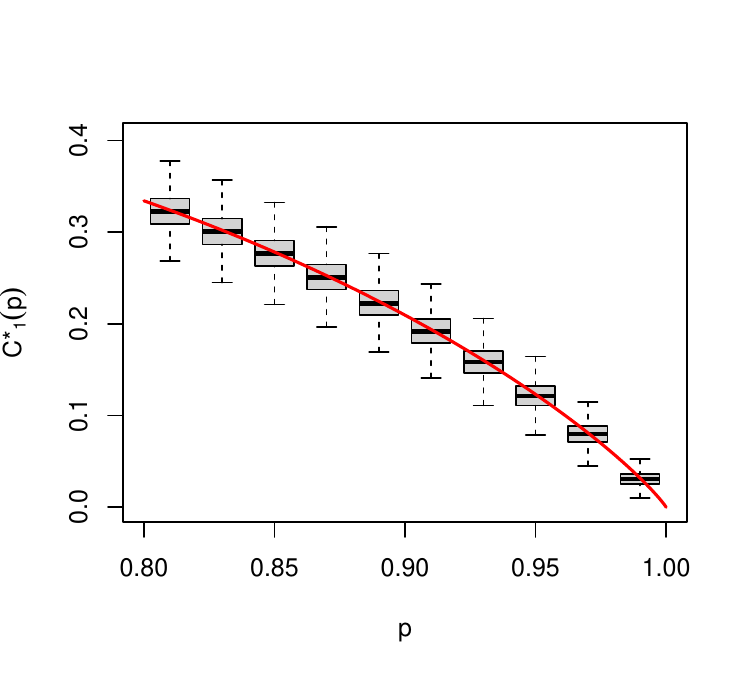}
    
    \includegraphics[width=0.4\linewidth]{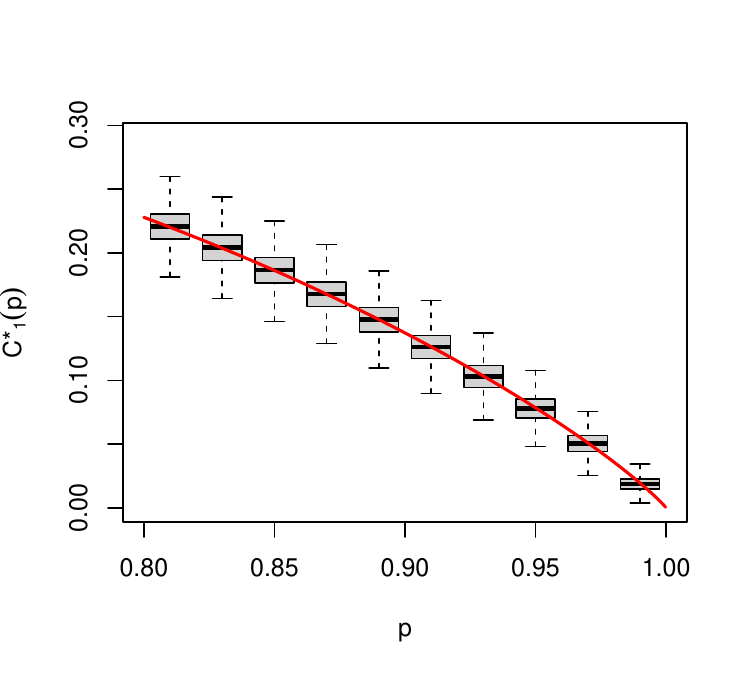}
    \includegraphics[width=0.4\linewidth]{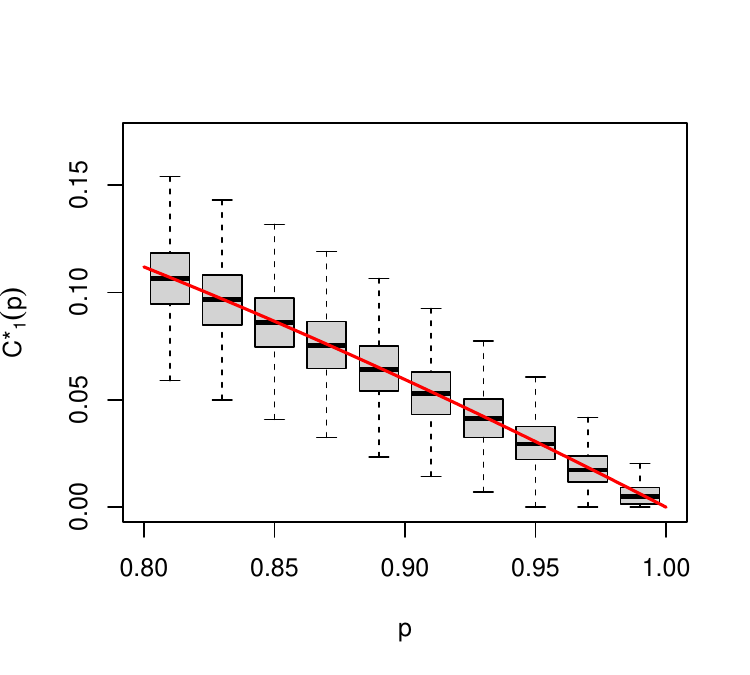}
    \vspace{-0.43cm}
    \caption{Boxplots of the estimated half-perimeter density $\hat{C}^{(m,\delta,S_n)}_1(X,p)$ at a mesh of $p$ values. The random fields  considered are Gaussian (top-left), anisotropic Gaussian (top-right), standardized chi-squared (bottom-left), and the Smith model (bottom-right). The true  half-perimeter  densities $C_1^\ast(E_p^X)$ are given by the solid red line.}
    \label{fig:C1boxplots-supp}
\end{figure}

\begin{figure}[ht!]
    \centering
    \includegraphics[width=0.32\linewidth, trim=0 10 0 50, clip]{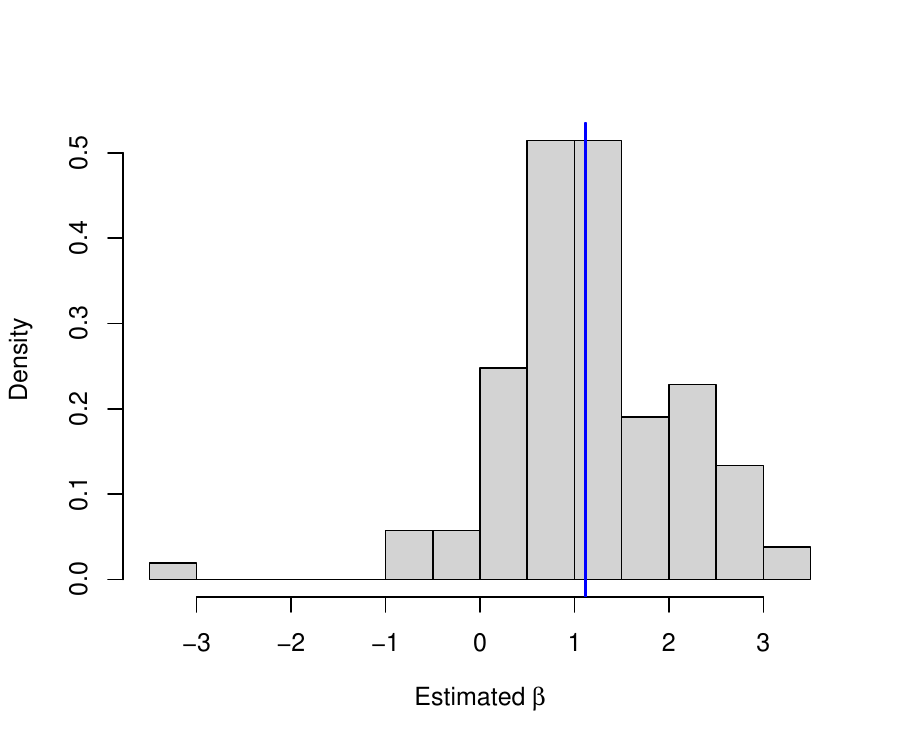}
    \includegraphics[width=0.32\linewidth, trim=0 10 0 50, clip]{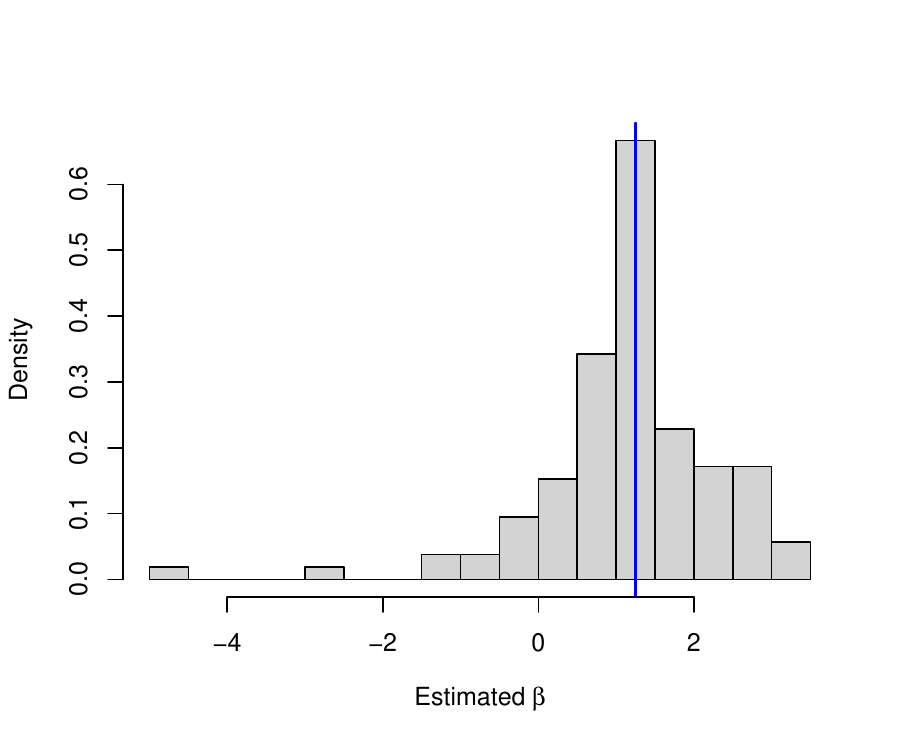}
    \includegraphics[width=0.32\linewidth, trim=0 10 0 50, clip]{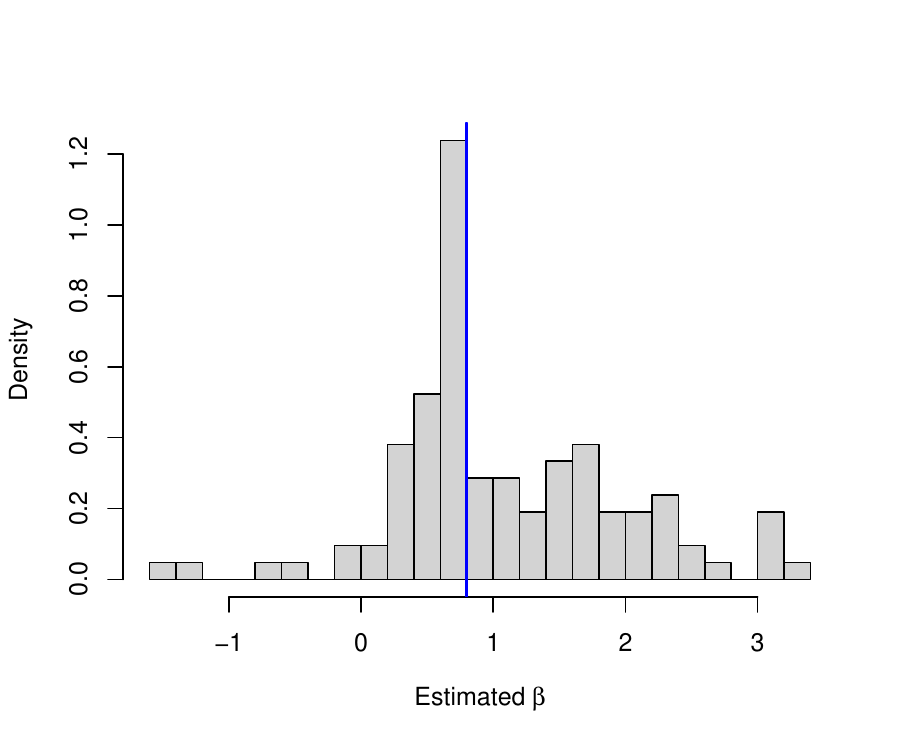}

    \caption{Histogram of estimates $\hat \beta(s_0,p_1,p_2;X,m,\delta,S_n^{(1)},S_n^{(2)})$  for pairs of thresholds $(p_1,p_2)$, with $s_0$ denoting the locations of interest of Paris, Bourges, and Valence (left to right). The solid blue vertical line denotes the median value.}
    \label{fig:beta-hist-comephore}
\end{figure}

\newpage
\section{Empirical assessment of Assumptions on Random Fields}\label{supp:assumption-check}

 \textit{Diagnostic for the condition in Equation~\eqref{densityLER}.}
In the statement of Theorem~\ref{thm:LKCratio},
 the density condition requires that
$\big|f_{\tilde R_0^{(p)}}(r)-f_{\tilde R_0^{(p)}}(0)\big|$
becomes arbitrarily small as $r\to0$, uniformly for
$p$ sufficiently close to $1$. To  investigate this numerically, we obtain
1000 observations  of rescaled LER values $\tilde{R}_{0}^{(p)}$
 from excursion sets conditioned on an exceedance
at the origin, and obtain kernel density estimates (KDE)
of the density $f_{\tilde{R}_{0}^{(p)}}$, denoted $\hat{f}_p$,
using automated bandwidth selection and a boundary correction
at zero.
This is repeated for a mesh of values $p\in[0.95,1)$ on a $[-25,25]^2$ grid
$S^\delta$ with grid spacing $\delta=0.9804$.

\begin{figure}[ht!]
    \centering       
    \includegraphics[width=0.32\linewidth]{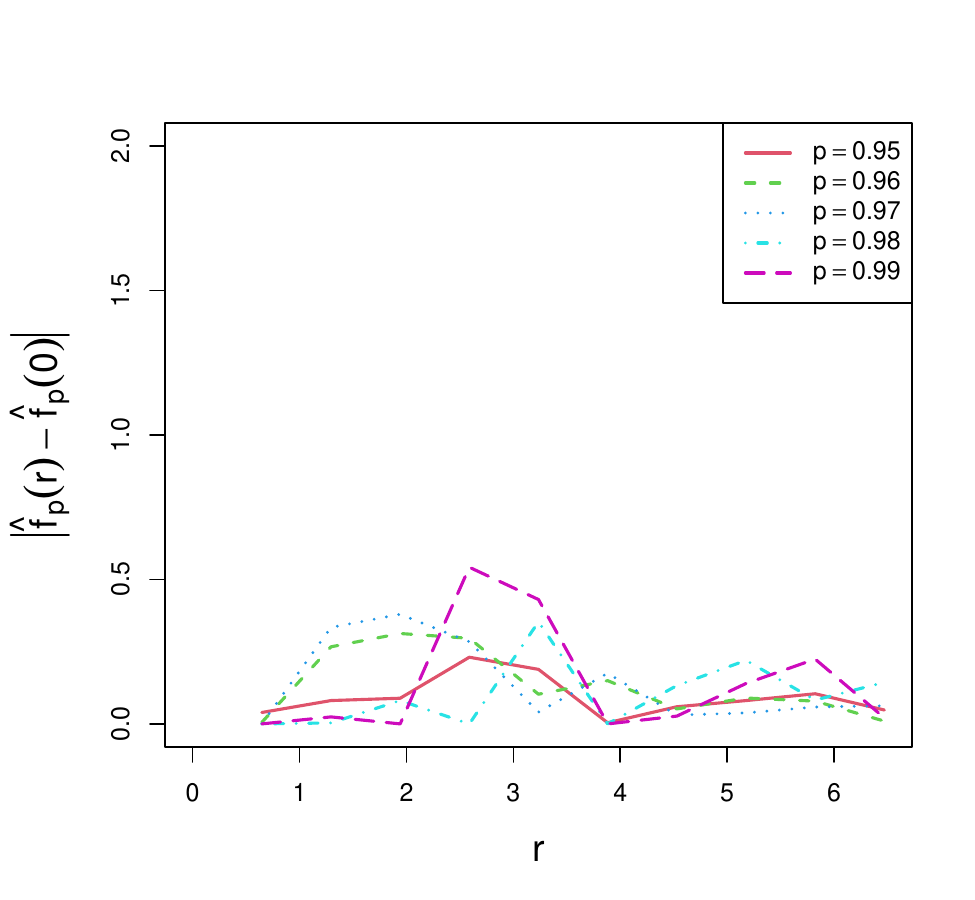}
   \includegraphics[width=0.32\linewidth]{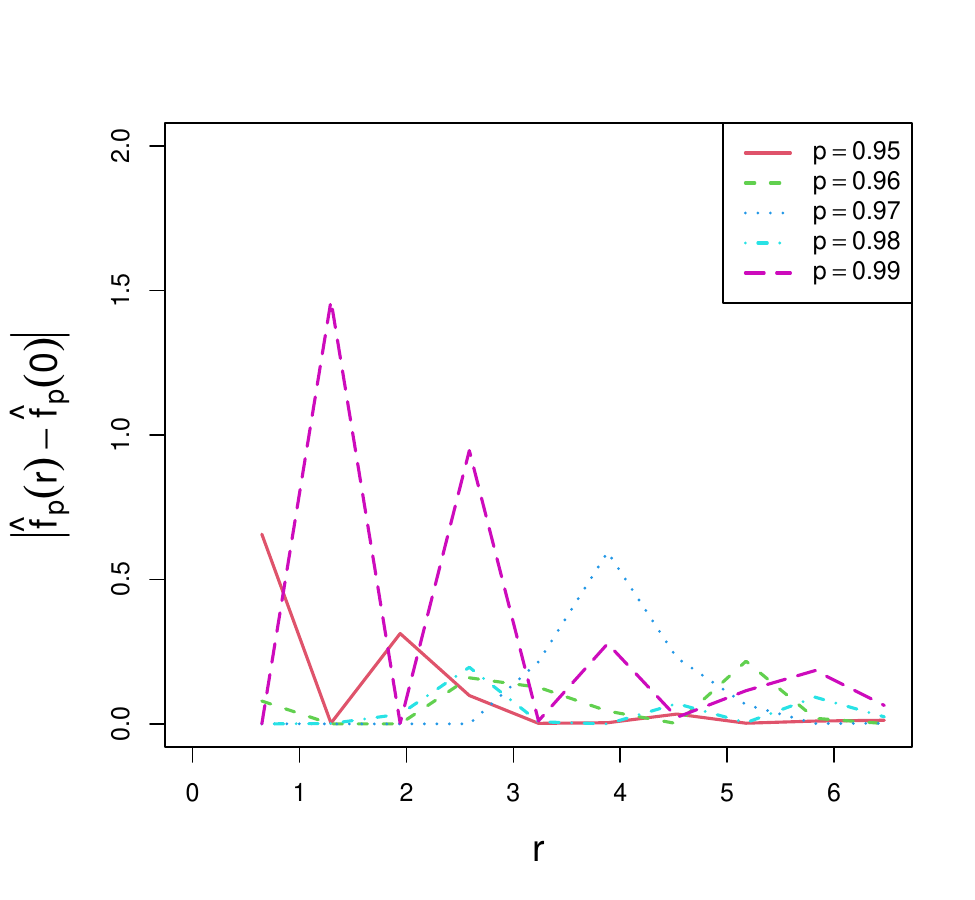} 
   \includegraphics[width=0.32\linewidth]{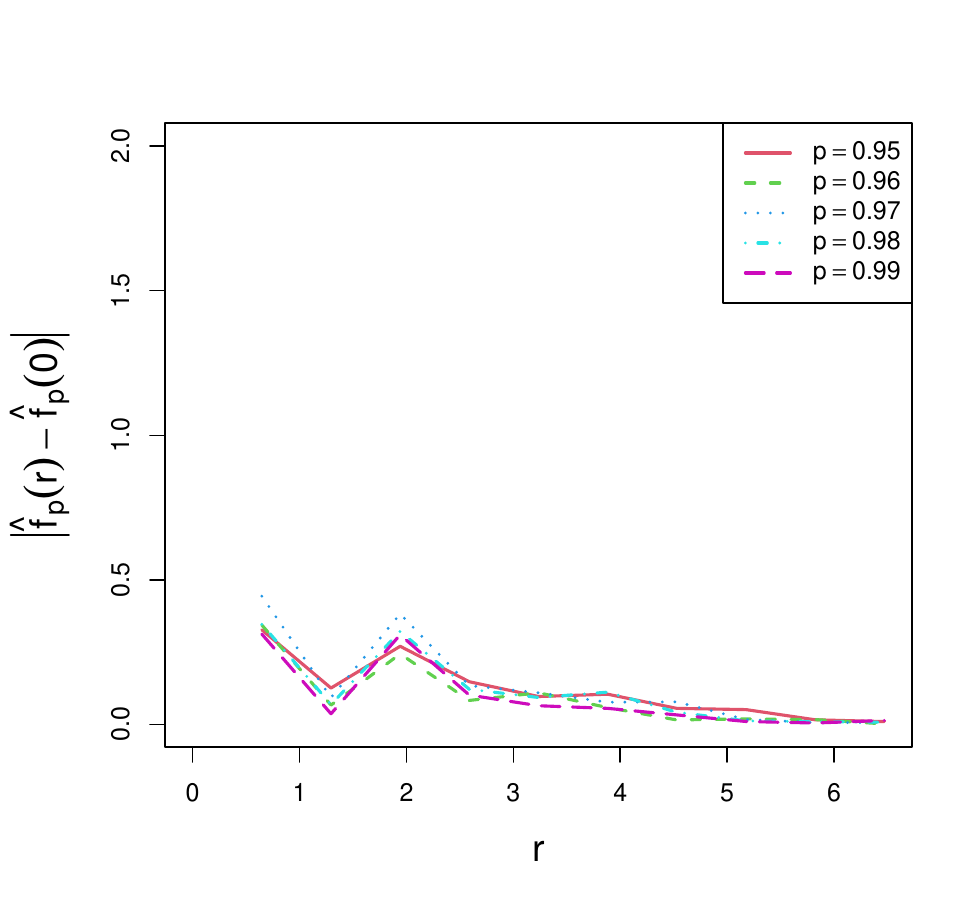}
   \vspace{-0.5cm}
    \caption{Absolute differences
$\left|\hat f_p(r)-\hat f_p(0)\right|$ between KDEs of
rescaled LER densities evaluated at $r>0$ and at zero,
with distances expressed in rescaled spatial units.
Each line corresponds to a different threshold level $p$. 
The random fields considered are the Gaussian,
standardised chi-squared, and Smith models (left to right).}
    \label{fig:Assumption1-study}
\end{figure}

Using the KDE-based estimate $\hat{f}_p$, we plot
$\left|\hat f_p(r)-\hat f_p(0)\right|$ against values of $r$
corresponding to 1 to 10 pixels away from the origin,
 expressed in rescaled spatial units, 
using observations from Gaussian, standardised chi-squared,
and Smith random fields.
The respective scaling functions $\sigma(p)$ are those
displayed in Table~\ref{tab:beta-vals} in the manuscript.
 Figure~\ref{fig:Assumption1-study} compares
the estimated density variations at the selected distances
and threshold levels in order to  provide an exploratory
diagnostic of the condition in Equation~\eqref{densityLER}.

\begin{figure}[ht!]
    \centering
    \includegraphics[width=0.32\linewidth]{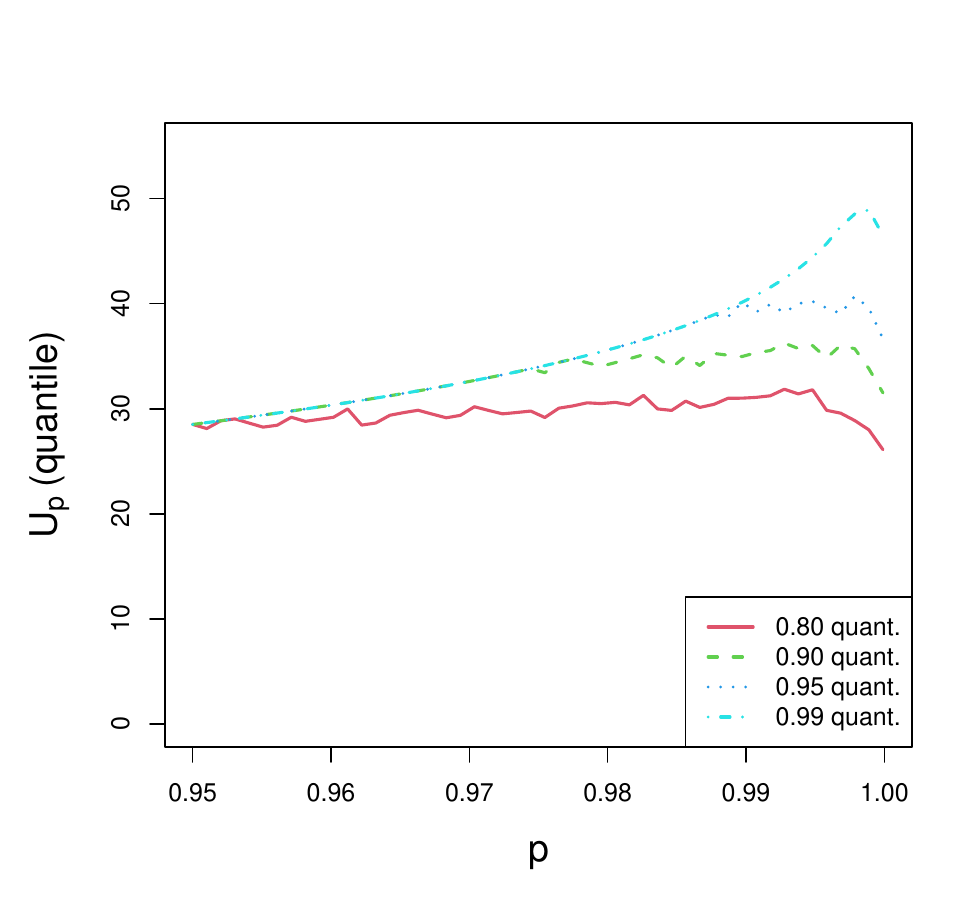}
    \includegraphics[width=0.32\linewidth]{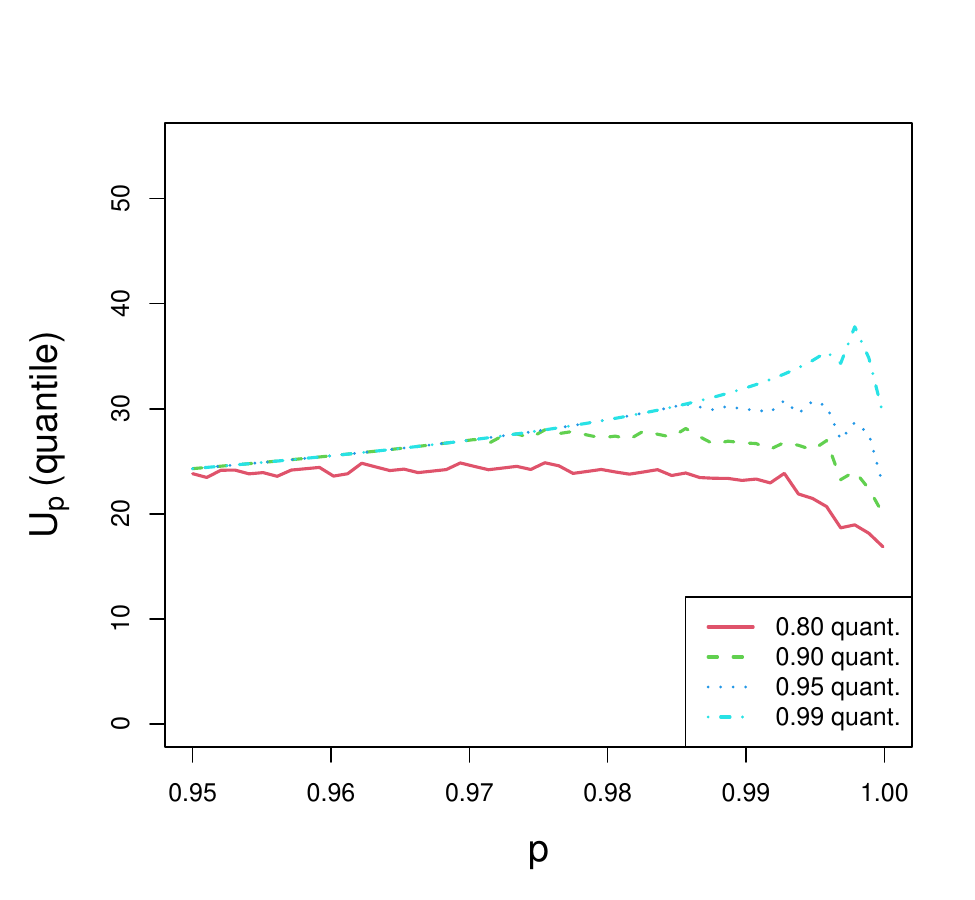}
    \includegraphics[width=0.32\linewidth]{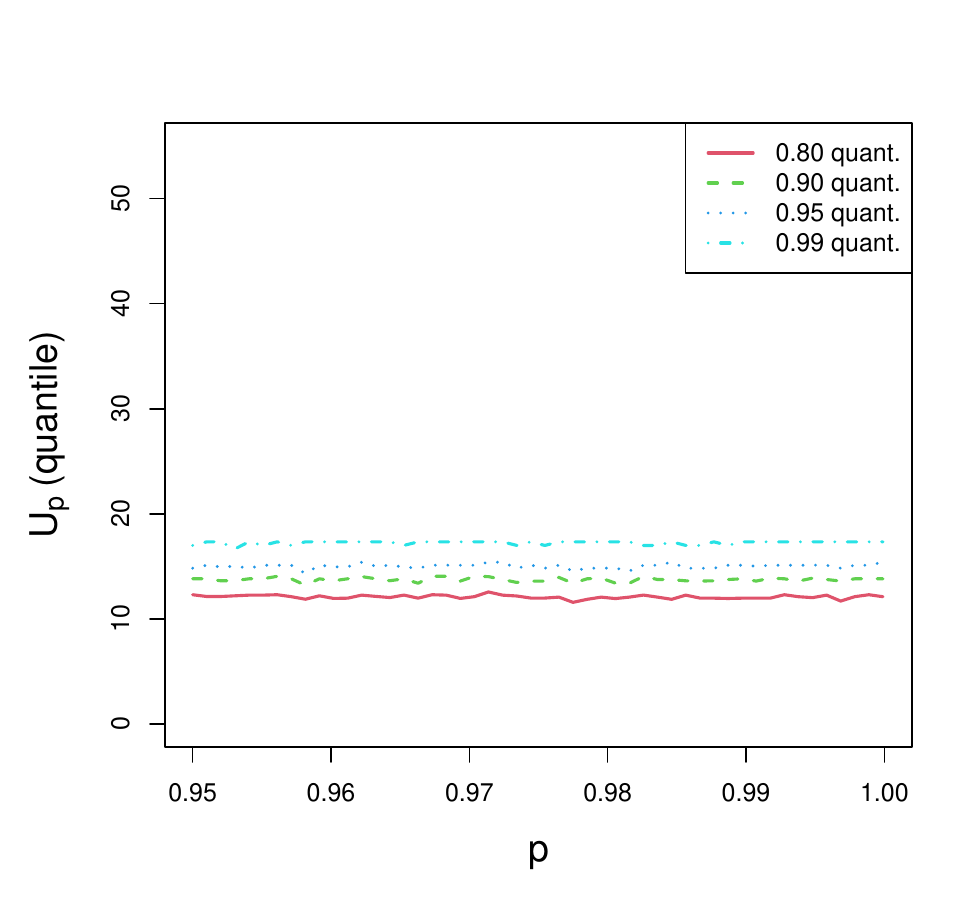}
    \vspace{-0.5cm}
    \caption{Empirical quantiles of
   grid approximations to
    $U_{p,[-25,25]^2}$ against the threshold level $p$,
    with each line indicating a different quantile level.
    The random fields considered are the Gaussian,
    standardised chi-squared, and Smith models (left to right).}
    \label{fig:Proposition1-study}
\end{figure}

\textit{Diagnostic for the condition in Equation~\eqref{ConfinementCondition}.}
The condition in the statement of Proposition~\ref{prop:scaling-AI}
 requires that, within every fixed compact window,
the rescaled conditional excursion sets remain bounded in probability
as $p\to1$. 
To investigate this numerically, we fix $s_0=0$
and simulate 1000 conditional excursion sets at each
threshold level, given that the field exceeds its $p$-quantile
at the origin. For each simulation, we approximate the rescaled
maximum excursion distance within the fixed window $[-25,25]^2$:
$$
U_{p,[-25,25]^2}
=
\sigma(p)
\sup_{s\in E_p^X(0)\cap[-25,25]^2}\|s\|.
$$ 
We do this by including all excursion components
(i.e., not just the connected component containing 
the origin).
This is repeated for a mesh of threshold levels
$p\in[0.95,1)$, using observations generated from
the Gaussian, standardised chi-squared, and Smith random fields on a $[-25,25]^2$ grid $S^\delta$
with grid spacing $\delta=0.9804$.
In Figure~\ref{fig:Proposition1-study}, we display the
0.80, 0.90, 0.95, and 0.99
 empirical quantiles of
these grid approximations to 
$U_{p,[-25,25]^2}$.
These plots provide a finite-threshold diagnostic
for the selected window; they do not establish the confinement
condition, which concerns every compact window and thresholds
arbitrarily close to one. For the Smith model,
$\sigma(p)=1$, so $U_{p,[-25,25]^2}\leq25\sqrt{2}$ deterministically,
and the confinement condition is automatically satisfied.

\section{On the expected perimeter of conditional excursion sets for large domains}\label{supp:cond-perims}

In this section, we provide the sufficient conditions
for the ratio of the conditional and unconditional expected perimeters of excursion sets to converge to one
as the domain tends to $\mathbb{R}^2$. Begin by recalling the expression obtained for the expected perimeter of
$E_p^X$ when restricted to
the domain
$S\subset\mathbb{R}^2$
given by Rice's formula (see, for example,
the Kac--Rice framework in
\citeay{AT07}, \citeay{ArmentanoAzaisLeon2025}, Theorem 2.1),
$$
\mathbb{E}\left[
\mathcal{H}_1(S \cap \partial E^X_p)
\right]
=
\int_S f_{X(s)}(u_p)
\mathbb{E}\left[
\left\|\nabla X(s)\right\|
\middle| X(s)=u_p
\right]\,ds.
$$

Supposing the random field defining $X(s)$ is stationary
(which is covered by our Assumption~\ref{A1}),
then we have
\begin{align*}
\mathbb{E}\left[
\mathcal{H}_1(S \cap \partial E^X_p)
\right]
=&
f_{X(s_0)}(u_p)
\mathbb{E}\left[
\left\|\nabla X(s_0)\right\|
\middle|X(s_0)=u_p
\right]\mathcal{L}_2(S),
\end{align*}
for some fixed location $s_0\in S$. We further assume that $f_{X(s_0)}(u_p)>0$ and that \[m_p:=
\mathbb{E}\left[
\|\nabla X(s_0)\|
\middle|X(s_0)=u_p
\right]\in(0,\infty).\] Using the same location $s_0$, the conditional  expected perimeter can also be
obtained
by using Rice's formula:
\begin{align*}
&\mathbb{E}\left[
\mathcal{H}_1(S \cap \partial E^X_p)\middle|X(s_0)>u_p\right]\\
&= \int_S
f_{X(s)\mid\{X(s_0)>u_p\}}(u_p)
\mathbb{E}\left[
\left\|\nabla X(s)\right\|
\middle| X(s)=u_p,X(s_0)>u_p
\right]ds\\
&=
\int_S f_{X(s)}(u_p)
\frac{
\PP\left(X(s_0)>u_p\middle|X(s)=u_p\right)
}{
\PP\left(X(s_0)>u_p\right)
}
\mathbb{E}\left[
\left\|\nabla X(s)\right\|
\middle| X(s)=u_p,X(s_0)>u_p
\right]ds\\
& =
\int_S f_{X(s)}(u_p)(1-p)^{-1}
\PP\left(X(s_0)>u_p\middle|X(s)=u_p\right)
\mathbb{E}\left[
\left\|\nabla X(s)\right\|
\middle| X(s)=u_p,X(s_0)>u_p
\right]ds.
\end{align*}

More generally, the weighted Rice formula
gives the equivalent expression
\begin{align*}
&
\mathbb{E}\left[
\mathcal{H}_1(S \cap \partial E^X_p)
\middle|X(s_0)>u_p
\right]
=\frac{1}{1-p}\int_S f_{X(s)}(u_p)
\mathbb{E}\left[
\|\nabla X(s)\|
\mathbbm{1}_{(u_p,\infty)}(X(s_0))
\middle| X(s)=u_p
\right]ds.
\end{align*}

Define $L_p(S)$ to be the ratio of the conditional
and unconditional expectations at threshold level $p$
on the 
bounded
domain $S$.
\begin{align*}
L_p(S)
&:=
\frac{
\mathbb{E}\left[
\mathcal{H}_1(S \cap \partial E^X_p)
\middle|X(s_0)>u_p
\right]
}{
\mathbb{E}\left[
\mathcal{H}_1(S \cap \partial E^X_p)
\right]
}\\
&=
\mathcal{L}_2(S)^{-1}
\int_S
\frac{
\PP\left(X(s_0)>u_p\middle|X(s)=u_p\right)
\mathbb{E}\left[
\left\|\nabla X(s)\right\|
\middle| X(s)=u_p,X(s_0)>u_p
\right]
}{
(1-p)
\mathbb{E}\left[
\left\|\nabla X(s_0)\right\|
\middle|X(s_0)=u_p
\right]
}ds.
\end{align*}
where by stationarity we have that
$f_{X(s)}(u_p)/f_{X(s_0)}(u_p)=1$.
Equivalently,
\begin{equation*}
L_p(S) = \frac{1}{\mathcal{L}_2(S)}
\int_S G_p(s)\,ds, \quad \mbox{ where }\quad G_p(s):=
\frac{\mathbb{E}\left[
\|\nabla X(s)\|
\mathbbm{1}_{(u_p,\infty)}(X(s_0))
\middle|X(s)=u_p
\right]}{(1-p)m_p}. 
\end{equation*}
With this, and using the limit as $h:=\|s-s_0\|\to\infty$
in arbitrary spatial directions, the following are sufficient conditions for
$\lim_{S\nearrow\mathbb{R}^2}L_p(S)=1$:
\begin{enumerate}
\item[(i)]
As $h$ increases, the conditional exceedance probability
converges to its marginal counterpart,
i.e.,
\begin{equation*}
\lim_{\|s-s_0\|\to\infty}
\PP\left(
X(s_0)>u_p
\middle|X(s)=u_p
\right)
=
\PP\left(X(s_0)>u_p\right)
=1-p.
\label{eq:supp-cond1}
\end{equation*}
\item[(ii)]
As $h$ increases, conditioning on the exceedance at $s_0$
has an asymptotically negligible effect on the
conditional mean gradient norm at $s$,
\emph{i.e.},
\begin{equation*}
\begin{aligned}
&\lim_{\|s-s_0\|\to\infty}
\mathbb{E}\left[
\left\|\nabla X(s)\right\|
\middle|X(s)=u_p,X(s_0)>u_p
\right]=
\mathbb{E}\left[
\left\|\nabla X(s)\right\|
\middle|X(s)=u_p
\right]
=m_p.
\end{aligned}
\label{eq:supp-cond2}
\end{equation*}
\end{enumerate}
Indeed, conditions (i) and (ii) imply that
$\lim_{\|s-s_0\|\to\infty}G_p(s)=1$. Moreover, since
$0\leq\mathbbm{1}_{(u_p,\infty)}(X(s_0))\leq1$,
stationarity yields $0\leq G_p(s)\leq\frac{1}{1-p}$,  
for almost every $s\in\mathbb{R}^2$. Let $(S_n)_{n\geq1}$ be a sequence of bounded
domains containing $s_0$ such that
$\mathcal{L}_2(S_n)\to\infty$.
For every $\varepsilon>0$, there exists $R>0$
such that $
|G_p(s)-1|<\varepsilon$ whenever $\|s-s_0\|>R$.  
Denoting by $B_R(s_0)$ the ball of radius $R$
centred at $s_0$, we obtain
\begin{align*}
|L_p(S_n)-1|
&\leq
\frac{1}{\mathcal{L}_2(S_n)}
\int_{S_n}|G_p(s)-1|\,ds \leq
\left(1+\frac{1}{1-p}\right)
\frac{\mathcal{L}_2(B_R(s_0))}
{\mathcal{L}_2(S_n)}
+\varepsilon.
\end{align*}
Letting $n\to\infty$ and then
$\varepsilon\downarrow0$, we conclude that $\lim_{n\to\infty}L_p(S_n)=1$.

\clearpage
\bibliographystyle{plainnat}
\bibliography{paper-ref}

\end{document}